\documentstyle[preprint,aps,eqsecnum,epsfig]{revtex}
\tighten
\begin{document}
\draft
\title{\bf OUT OF EQUILIBRIUM FIELDS IN SELFCONSISTENT INFLATIONARY
DYNAMICS. DENSITY FLUCTUATIONS}
\author{{\bf D. Boyanovsky$^{(a)}$ and  H. J. de Vega$^{(b)}$, }}
\address
{(a)Department of Physics and Astronomy, University of
Pittsburgh, Pittsburgh, PA 15260, USA \\
 (b)  LPTHE, Universit\'e Pierre et Marie Curie (Paris VI) 
et Denis Diderot  (Paris VII), Tour 16, 1er. \'etage, 4, Place Jussieu
75252 Paris, Cedex 05, France}
\date{May 2000}
\maketitle
\begin{abstract}
The physics during the inflationary stage of the universe is of
quantum nature involving extremely high energy densities. Moreover, it
is out of equilibrium on a fastly expanding dynamical geometry.
We present in these lectures non-perturbative out of equilibrium field
theoretical methods in cosmological universes. We then study 
the non-linear dynamics of quantum fields in matter and radiation
dominated FRW  and de Sitter universes. For a variety of initial
conditions, we compute the evolution of the quantum inflaton field,
its large quantum fluctuations and the equation of state. We investigate
the explosive particle production due to spinodal instabilities
and parametric amplification in FRW and de Sitter universes with and
without symmetry breaking. We show how the particle production
is  sensitive to the expansion of the universe.  
In the large $N$ limit for symmetry breaking scenarios, we determine
generic late time fields behavior for any flat FRW and de Sitter cosmology. 
We show that the amplitude of the quantum fluctuations fall off in FRW
with the square of the scale factor while the order parameter
approaches a minimum of the potential in the same manner. We  present
a complete and numerically accessible renormalization scheme for the
equation of motion and the energy momentum tensor in flat
cosmologies. 
Furthermore, we consider an $ O(N) $ inflaton model coupled
self-consistently to gravity in the 
semiclassical approximation, where the field is subject to `new inflation'
type initial conditions. We study the dynamics  {\bf self-consistently and
non-perturbatively} with non-equilibrium field theory methods in
the large $ N $ limit. We find that spinodal instabilities drive the growth
of non-perturbatively large quantum fluctuations which shut off the
inflationary growth of the scale factor. We find that 
a very specific combination of  these large quantum
fluctuations plus the inflaton zero mode assemble into a new effective field.
This new field behaves classically and it is the object which actually
rolls down. The metric perturbations during inflation are computed
using this effective field and the Bardeen variable for superhorizon modes
during inflation.  We compute the amplitude and index for the spectrum of
scalar density and tensor perturbations and argue that in all models of this
type the spinodal instabilities are responsible for a `red' spectrum of
primordial scalar density perturbations.  
\end{abstract}

\section{Introduction and Motivation}

Inflationary cosmology has come of age. From its beginnings as a solution to
the horizon, flatness,  entropy and monopole problems\cite{guth}, it
has grown into the main contender for the explanation of the source of
primordial fluctuations giving rise to large scale
structure\cite{defe}. There is evidence from the 
measurements of temperature anisotropies in the cosmic microwave background
radiation (CMBR) that the scale invariant power spectrum predicted by generic
inflationary models is consistent with
observations\cite{kolb,turner,lyth1} and we can expect further and
more exacting tests of the predictions of inflation  
when the MAP and PLANCK missions are flown. In particular, if the
fluctuations that are responsible for the temperature anisotropies of
the CMB truly originate from 
quantum fluctuations during inflation, determinations of the spectrum of scalar
and tensor perturbations will constrain inflationary models based on particle
physics scenarios and probably will validate or rule out specific
proposals\cite{turner,lyth2}. 

The tasks for inflationary universe researchers are then two-fold. First,
models of inflation must be constructed on the basis of 
a realistic particle physics model. This is in contrast
to the current situation where most, if not all acceptable inflationary models
are ad-hoc in nature, with fields and potentials put in for the sole purpose of
generating an inflationary epoch. Second, and equally important, the
quantum dynamics of inflation must be understood. This is 
extremely important, especially in light of the fact that it is {\it
exactly} this 
quantum behavior that is supposed to give rise to the primordial metric
perturbations which presumably have imprinted themselves in the CMBR. This
latter problem is the focus of this review. 

The inflaton must be treated as a {\it non-equilibrium} quantum field . The
simplest way to see this comes from the requirement of having small enough
metric perturbation amplitudes which in turn requires that the quartic self
coupling 
$ \lambda $ of the inflaton be extremely small, typically of order $ \sim
10^{-12} $. Such a small coupling cannot establish local thermodynamic
equilibrium (LTE) for {\it all} field modes; typically the long wavelength
modes will respond too slowly to be able to enter LTE. In fact, the
superhorizon sized modes will be out of the region of causal contact and
cannot thermalize. We see then that if we
want to gain a deeper understanding of inflation, non-equilibrium tools must be
developed. Such tools exist and have now been developed to the point that they
can give quantitative answers to these questions in 
cosmology\cite{barrabajo} -\cite{frw3},\cite{ctp,hu,motola,largen}. 
These methods permit us to follow the {\bf dynamics} of quantum fields
in situations where the energy  
density is non-perturbatively large ($ \sim 1/\lambda $). That is, they allow
the computation of the time evolution of non-stationary states and of
non-thermal  density matrices.

Our programme on non-equilibrium dynamics of quantum field theory, started in
 1992\cite{barrabajo}, is naturally poised to provide a framework to
 study these  problems. The larger goal of the program is to study the
 dynamics of  non-equilibrium processes from a fundamental
 field-theoretical description, by solving  the dynamical 
equations of motion of the underlying four dimensional quantum field
 theory for physically relevant problems:  the early universe
 dynamics, high energy particle collisions, phase transitions out of
 equilibrium,  symmetry breaking and  dissipative processes.  

The focus of our work is to describe the quantum field dynamics when
the  energy density is {\bf high}. That is, a large number of particles per
volume $ m^{-3} $, where $ m $ is the typical mass scale in the
theory. Usual S-matrix calculations apply in the opposite limit of low
energy density and since they only provide information on {\em in}
$\rightarrow$ {\em out} matrix elements,  are unsuitable for calculations of
expectation values. 

In high  energy density situations such as in the early universe,
the particle propagator (or Green function) depends on the particle
distribution in momenta in a nontrivial way. This makes the 
quantum dynamics intrinsically nonlinear and calls to the use of
self-consistent non-perturbative approaches as the large $N$ limit,
Hartree and self-consistent one-loop approximations. 

There are basically three different levels to study the early universe
dynamics:
\begin{enumerate}

\item To work out the nonlinear dynamics of quantum fields in
Minkowski spacetime. By non-linear dynamics we understand to solve the
quantum equations of motion {\bf  including} the quantum back-reaction
quantitatively\cite{barrabajo} -\cite{mink},
\cite{kaiser,baacke,motola,largen}.
This level is in fact appropriate to describe high
energy particle collisions \cite{dcc}. 

\item To work out the nonlinear dynamics of quantum fields in fixed
cosmological backgrounds\cite{frw2,De Sitter}. New phenomena arise
then compared  with 1.  showing that a Minkowski analysis is not
quantitatively precise for expanding universes. 

\item A self-consistent treatment of the quantum fields and the
cosmological background\cite{din,ramsey}. That is, the metric is
obtained dynamically from the quantum fields (matter source)
propagating in the that metric.  

\end{enumerate}

We shall successively present the three levels of study. The first
stage was reviewed in the 1996 Chalonge School \cite{mink}.
The second level is the subject of
secs. VI and VII. We study the parametric  and spinodal
resonances both in FRW and de Sitter backgrounds
wide range of initial conditions both in FRW and de Sitter backgrounds
\cite{frw,De Sitter}. [Parametric resonance appears in chaotic
inflationary scenarios for unbroken symmetry whereas spinodal
unstabilities show up in new inflation scenarios with broken symmetry].
Both types of  unstabilities shut-off through the
non-linear quantum evolution as described in secs.  VI and VII
\cite{frw,De Sitter} both analytically and numerically. 
We follow the equation of state of the quantum matter during the evolution
and analyze its properties. 

The third stage of 
our approach is to apply non-equilibrium quantum field theory techniques to the
situation of a scalar field coupled to {\it semiclassical} gravity, where the
source of the gravitational field is the expectation value of the stress energy
tensor in the relevant, dynamically changing, quantum state. In this way
we can go beyond the standard analyses\cite{linde2,vilenkin,stein,guthpi} which
treat the background as fixed and do not consider the non-linear
quantum field dynamics. 

In all cases 1. - 3. , the quantum fields energy- momentum tensor is
covariantly conserved both at the regularized as well as the
renormalized levels \cite{us1} - \cite{din}.

We  mainly consider for the stage 3.   new inflation scenarios where a
scalar field 
$\phi$ evolves under the action of a typical symmetry breaking potential. The
initial conditions will be taken so that the initial value of the order 
parameter is near the top of the potential
(the disordered state) with essentially zero time derivative. 
What we find is that the existence of spinodal instabilities, i.e. the fact
that eventually (in an expanding universe) all modes will act as if they have a
{\it negative} mass squared, drives the quantum fluctuations to grow {\it
non-perturbatively} large. We have the picture of an initial wave-function or
density matrix peaked near the unstable state and then spreading until it
samples the stable vacua. Since these vacua are non-perturbatively far from
the initial state (typically $\sim m\slash \sqrt{\lambda}$, where $m$ is the
mass scale of the field and $\lambda$ the quartic self-coupling), the spinodal
instabilities will persist until the quantum fluctuations, as encoded in the
equal time two-point function $\langle \Phi(\vec{x}, t)^2 \rangle$, grow to
${\cal O}( m^2\slash \lambda$).

This growth eventually shuts off the inflationary behavior of the scale factor
as well as the growth of the quantum fluctuations (this last also happens
in Minkowski spacetime \cite{us1,mink}).

The scenario envisaged here is that of a quenched or super-cooled phase 
transition where the order parameter is zero or very small. Therefore one is 
led to ask: 

a) What is rolling down?. 

b) Since the quantum fluctuations are non-perturbatively large ( $ \sim  
1/\lambda $), will not they modify drastically the FRW dynamics?.

c) How can one extract (small?) metric perturbations from non-perturbatively 
large field fluctuations?

We address the questions a)-c) as well as  other issues in sec. IX. 

We choose such type of new inflationary scenario because the issue of
large quantum fluctuations is particularly dramatic there. However,
our methods do apply to any inflationary scenario as  chaotic,
extended and hybrid inflation.

\section{Non-Equilibrium Quantum Field Theory, Semiclassical Gravity and
Inflation}

We present here the framework of the non-equilibrium closed time path
formalism. For a more complete discussion, the reader is referred to
\cite{us1}-\cite{datan}. 

The time evolution of a system is determined in the Schr\"odinger picture
by the functional Liouville equation
\begin{equation}\label{liouville}
i\frac{\partial \rho (t)}{\partial t} = [H(t),\rho(t)],
\end{equation}
where $\rho$ is the density matrix and we allow for an explicitly time
dependent Hamiltonian as is necessary to treat quantum fields in a time
dependent background.  Formally, the solutions to this equation for the time
evolving density matrix are given by the time evolution operator, $U(t,t^{'})$,
in the form
\begin{equation}
\rho(t) = U(t,t_0)\rho(t_0)U^{-1}(t,t_0).
\label{rhoevol}
\end{equation}
The quantity $\rho(t_0)$ determines the initial condition for the evolution. We
choose this initial condition to describe a state of local equilibrium in
conformal time, which is also identified with the conformal adiabatic vacuum
for short wavelengths. 

Given the evolution of the density matrix (\ref{rhoevol}), ensemble averages
of operators are given by the expression (again in the Schr\"odinger picture)
\begin{equation}
\langle{\cal O}(t)\rangle = 
\frac{Tr[U(t_0,t){\cal O}U(t,t^{'})U(t^{'},t_0)\rho(t_0)]}{Tr\rho(t_0)},
\label{expect}
\end{equation}
where we have inserted the identity, $U(t,t^{'})U(t^{'},t)$ with $t^{'}$ an
arbitrary time which will be taken to infinity. The state is first evolved
forward from the initial time $t_0$ to $t$ when the operator is inserted.  We
then evolve this state forward to time $t^{'}$ and back again to the initial
time\cite{us1,frw}.

We shall study the inflationary dynamics in a spatially flat
Friedmann-Robertson-Walker background with scale factor $a(t)$ and
line element: 
\begin{equation}\label{metric}
ds^2 = dt^2 - a^2(t)\; d\vec{x}^2.
\end{equation}
Our Lagrangian density has the form
\begin{equation}
{\cal L} =\sqrt{-g}\left[ \frac12 \nabla_\mu\Phi\nabla^\mu\Phi -
V(\Phi)\right]\; . \label{lagrangian}
\end{equation}
Our approach can be generalized to open as well as closed cosmologies.

Our program incorporates the non-equilibrium
behavior of the quantum fields involved in inflation into a framework
where the geometry (gravity) is dynamical and is  
treated  self consistently. We do this
via the use of semiclassical gravity\cite{birrell} where we say that
the metric is classical and determined through  the Einstein equations 
using the expectation value of the stress energy tensor $\langle
T_{\mu \nu} \rangle$. Such
 expectation value is taken in the dynamically determined state
described by the density matrix $\rho(t)$. This dynamical problem can be
described schematically as follows: 

\begin{enumerate}

\item{The dynamics of the scale factor $a(t)$ is driven by the 
semiclassical Einstein equations 

\begin{equation}
\frac{1}{8\pi G_R} G_{\mu \nu} + \frac{\Lambda_R}{8\pi G_R} \; g_{\mu \nu} +
\left(\rm{higher\  curvature}\right)= -\langle T_{\mu \nu}\rangle_R .
\end{equation}
Here $G_R, \Lambda_R$ are the renormalized values of Newton's constant and the
cosmological constant, respectively and $G_{\mu \nu}$ is the Einstein
tensor. The higher curvature terms must be included to absorb
ultraviolet divergences.}

\item{On the other hand, the density matrix $\rho(t)$ of the matter
(that determines $\langle T_{\mu \nu}\rangle_R$) obeys the Liouville
equation (\ref{liouville}) with $H$ is the evolution Hamiltonian,
which is dependent on the scale factor,  $a(t)$.}
\end{enumerate}

It is this set of equations we must try to solve specifying the
appropriate initial conditions.

\subsection{On the initial state: dynamics of phase transitions} 

The situations we consider are 
\begin{enumerate}
\item{the theory
admits a symmetry breaking potential and in which the field expectation value
starts its evolution near the unstable point.}
\item{The symmetry is not broken and  the field expectation value
starts its evolution at a finite distance from the absolute minimum.}
\end{enumerate}

 There is an
issue as to how the field got to have an expectation value near the
unstable point (typically at $\Phi=0$) as well as an issue concerning
the initial state of the non-zero momentum modes.

Since our background is an FRW spacetime, it is spatially homogeneous and we
can choose our state $\rho(t)$ to respect this symmetry. Starting from the full
quantum field $\Phi(\vec{x},t)$ we can extract its expectation value $
\phi(t) $ by writing:
\begin{equation}
\Phi(\vec{x}, t) = \phi(t) + \Psi(\vec{x}, t) \quad , \quad 
\phi(t) =  {\rm Tr}[\rho(t)\Phi(\vec{x}, t)]\equiv \langle
\Phi(\vec{x}, t)) \rangle .
\end{equation}
The quantity $\Psi(\vec{x}, t)$ represents the quantum fluctuations about the
zero mode $\phi(t)$ and clearly satisfies $\langle \Psi(\vec{x}, t) \rangle=0$.

We need to choose a basis to represent the density matrix. A natural choice
consistent with the translational invariance of our quantum state is that given
by the Fourier modes, in {\it comoving} momentum space, of the quantum
fluctuations $\Psi(\vec{x}, t)$:

\begin{equation}
\Psi(\vec{x}, t) = \int \frac{d^3 k}{(2 \pi)^3} \; \exp(-i\ \vec{k}
\cdot \vec{x})\; \psi_k (t) .
\end{equation}

In this language we can state our ansatz for the initial condition of the
quantum state as follows. We take the zero mode $\phi(t=0)=\phi_0,\
\dot{\phi}(t=0)=0$, where $\phi_0$ will typically be very near the
origin for broken symmetry and at a finite distance from it in the
unbroken symmetry case. The initial conditions on the the nonzero
modes $ \psi_k (t) $ will be chosen such that the initial density
matrix $ \rho(t=0) $ describes a vacuum state 
(i.e. an initial state in local thermal equilibrium at a temperature $T_i=0$).
There are some subtleties involved in this choice. First, as
explained in \cite{frw2}, in order for the density matrix to commute with the
initial Hamiltonian, we must choose the modes to be initially in the conformal
adiabatic vacuum (these statements will be made more precise below). This
choice has the added benefit of allowing for time independent renormalization
counterterms to be used in renormalizing the theory.

We are making the assumption of an initial vacuum state
in order to be able to proceed with the calculation. It would be
interesting to understand what forms of the density matrix can be used for
other more general initial conditions\cite{tsun}. 

The assumptions of an initial equilibrium vacuum state are essentially 
the same used in refs. \cite{linde2}, \cite{vilenkin} and
\cite{guthpi} in the analysis of the quantum mechanics of inflation in
a fixed de Sitter background. 

As discussed in the introduction, if we start from such an initial state,
spinodal or parametric instabilities will drive the growth of
non-perturbatively large 
quantum fluctuations. In order to deal with these, we need to be able to
perform calculations that take these large fluctuations into account. 
Although the quantitative features of the dynamics will depend on the
initial state, the qualitative features associated with spinodal or
parametric unstabilities are fairly robust for a wide choice of
initial states that describe a phase transition.

\section{The Inflaton Model and the Equations of Motion}

Having recognized the appearance of large quantum fluctuations driven
by parametric or spinodal unstabilities, we need to study the dynamics
within a non-perturbative framework. That is, a framework allowing calculations
for non-perturbatively large energy densities. 
We require that such a framework be: i) renormalizable, ii)
covariant energy conserving, iii) numerically implementable.  There are very
few schemes that fulfill all of these criteria: the large $ N $
and the Hartree approximation\cite{us1}-\cite{din}. Whereas the
Hartree approximation 
is basically a Gaussian variational approximation\cite{jackiwetal} that
in general cannot be consistently improved upon, the large $ N $ approximation
can be consistently implemented beyond leading order\cite{motola,largen}.
In addition, the presence of a large number of fields in most of the
GUT's models suggest that the large $ N $ limit will be actually a
realistic one. Moreover, for the case of broken symmetry
it has the added bonus of providing many light fields (associated with
Goldstone modes) that will permit the study of the effects of other fields
which are lighter than the inflaton on the dynamics. Thus we will
study the inflationary dynamics  within the framework of
the large $ N $ limit of a scalar theory in the vector representation of $ O(N)
$ both for unbroken and broken symmetry. In the second case we will
have a quenched phase transition. 

We assume that the universe is spatially flat with a metric given by
eq.(\ref{metric}). The matter action and Lagrangian density are given
by eq.(\ref{lagrangian}), 
\begin{equation}
S_m  =  \int d^4x\; {\cal L}_m = \int d^4x \;
a^3(t)\left[\frac{1}{2}\dot{\vec{\Phi}}^2(x)-\frac{1}{2} 
\frac{(\vec{\nabla}\vec{\Phi}(x))^2}{a^2(t)}-V(\vec{\Phi}(x))\right]
\label{action}
\end{equation}
\begin{equation}
V(\vec{\Phi})  =  \frac{m^2}2\; \vec{\Phi}^2 +
\frac{\lambda}{8N}\left(\vec{\Phi}^2 +\frac{2N m^2}{\lambda}\right)^2
+\frac12 \, \xi\; {\cal R} \;\vec{\Phi}^2  \;, \label{potential}
\end{equation}
where $ m^2 > 0 $ for unbroken symmetry and  $ m^2 < 0 $ for broken
symmetry. Here $ {\cal R}(t) $ stands for the scalar curvature
\begin{equation}
{\cal R}(t)  =  6\left(\frac{\ddot{a}(t)}{a(t)}+
\frac{\dot{a}^2(t)}{a^2(t)}\right), \label{ricciscalar}
\end{equation}
The $\xi$-coupling of $ \vec{\Phi(x)}^2 $ to the scalar curvature ${\cal
R}(t)$ has been included since  arises anyhow as a consequence of
renormalization\cite{frw}.

The gravitational sector includes the usual Einstein term in addition
to a higher order curvature term and a cosmological constant term 
which are necessary to renormalize the theory. The action for
the gravitational sector is therefore:
\begin{equation}
S_g  =  \int d^4x\; {\cal L}_g = \int d^4x \, a^3(t) \left[\frac{{\cal
R}(t)}{16\pi G}  
+ \frac{\alpha}{2}\; {\cal R}^2(t) - K\right].
\end{equation}
with $ K $ being the cosmological constant.
In principle, we also need to include the terms $R^{\mu\nu}R_{\mu\nu}$
and $R^{\alpha\beta\mu\nu}R_{\alpha\beta\mu\nu}$ as they are also terms
of fourth order in derivatives of the metric (fourth adiabatic order),
but the variations resulting from these terms turn out not to be 
independent of that of ${\cal R}^2$ in the flat
FRW cosmology we are considering.

The variation of the action $S = S_g + S_m$ with respect to the 
metric $g_{\mu\nu}$ gives us Einstein's equation
\begin{equation}
\frac{G_{\mu\nu}}{8\pi G} + \alpha \; H_{\mu\nu} + K \;  g_{\mu\nu}
= - <T_{\mu\nu}>\; ,
\label{extendEinstein}
\end{equation}
where $G_{\mu\nu}$ is the Einstein tensor given by the variation of
$\sqrt{-g}{\cal R}$, $H_{\mu\nu}$ is the higher order curvature term given
by the variation of $\sqrt{-g}{\cal R}^2$, and $T_{\mu\nu}$ is the
contribution  from the matter Lagrangian. 
 With the metric (\ref{metric}), the various components
of the curvature tensors in terms of the scale factor are:
\begin{eqnarray}
G^{0}_{0} & = & -3\, \left({\dot{a} \over a}\right)^2\; , \;
G^{\mu}_{\mu}  =  -{\cal R} = -6\left(\frac{\ddot{a}}{a}
+\frac{\dot{a}^2}{a^2}\right)\; ,\nonumber \\
H^{0}_{0} & = & -6\left(\frac{\dot{a}}{a}\; \dot{{\cal R}} + 
\frac{\dot{a}^2}{a^2}\; {\cal R} - \frac{1}{12}\; {\cal R}^2\right)\; , \;
H^{\mu}_{\mu}  =  -6\left(\ddot{{\cal R}} + 
3\; \frac{\dot{a}}{a}\; \dot{{\cal R}}\right)\; .\nonumber
\end{eqnarray}
Eventually, when we have fully renormalized the theory,
we will set $\alpha_R=0$ and keep as our only contribution to
$K_R$ a piece related to the matter fields which we shall 
incorporate into $T_{\mu\nu}$.  

The use of semiclassical gravity (i. e. neglecting graviton loops) is
justified since graviton loops are suppressed by factors $
(m/M_{Pl})^2 $ where $ m $ is at the GUT scale and hence $
(m/M_{Pl})^2 \sim 10^{-8} $. 

\section{The Large N Limit for a scalar field with an arbitrary
invariant self-interaction}

We present here the systematic derivation of the $ 1/N $ expansion for
the scalar model with an arbitrary $O(N)$-invariant self-interaction and the
scalar field in the vector representation of $ O(N)
$\cite{losalamos}-\cite{avan}. We consider Minkowski space-time. The
generalization to the cosmological space-time (\ref{metric}) is
discussed at the end of the section.    

The action and Lagrangian density are given by,
\begin{equation}\label{action2} 
S  = \int d^4x\; {\cal L} \quad ,  \quad
{\cal L}  =   \frac{1}{2}\left[\partial{\vec{\Phi}}(x)\right]^2
-N \, V\left(\frac1{N} \; \vec{\Phi}^2(x)\right)\; .
\end{equation}
This choice of the $ N $ dependence in the interaction ensures that
the large $ N $ limit exists. In particular, for a quartic potential
we choose according to eq.(\ref{potential}):

\begin{equation}\label{fi4}
V\left(\frac1{N} \; \vec{\Phi}^2(x)\right) =
\frac{\lambda}{8N^2}\left(\vec{\Phi}^2+\frac{2N m^2}{\lambda}\right)^2 \quad . 
\end{equation}

The functional integral for the model takes the form
\begin{equation}\label{zjota}
{\cal Z}[J(.)] = \int \int D\vec{\Phi} \; \exp{i\int d^4x \left[ {\cal L} +
{\vec{\Phi}}(x) . {\vec{J}}(x) \right]}
\end{equation}
where $  {\vec{J}}(x) $ is an external source introduced to generate
the correlation functions. For example, the connected two point correlation
function is expressed as
\begin{equation}\label{fun2p}
< T \Phi_i(x)\,  \Phi_k(y)> = i^{-2} \; \left. { {\delta^2} \over {
\delta J_i(x) 
\, \delta J_k(y) }} \log{\cal Z}[J(.)]\right|_{J_i(.) = 0} \; .
\end{equation}

In order to compute the large $ N $ limit it is convenient to replace
the interaction term by the following functional integral
representation \cite{avan}:
\begin{eqnarray}\label{repV}
\exp{-iN\int d^4x\;  V(\frac1{N} \vec{\Phi}^2(x)) } &=&
\int \int Dz \; \Pi_x \, \delta\left( z(x) - \frac1{N}
\vec{\Phi}^2(x)\right) \; e^{-iN \int d^4x  \,  V(z(x))} \cr \cr
 &=&\int \int Dz \; D\sigma \; e^{iN  \int d^4x  \,\left[\frac12 \sigma(x) \, 
\left( z(x) - \frac1{N} \vec{\Phi}^2(x)\right) -  V(z(x))  \right]} 
\end{eqnarray}
Inserting eq.(\ref{repV}) into eq.(\ref{zjota}),  the integration over
$ \vec{\Phi}(x) $ becomes gaussian and therefore can be exactly computed
with the result:
\begin{eqnarray}\label{gauss}
&&{\cal Z}[J(.)] =\int \int D\vec{\Phi} \;Dz \; D\sigma \; e^{i\int
d^4x \left[ \frac{1}{2}\left[\partial{\vec{\Phi}}(x)\right]^2 + \frac{N}2 
\sigma(x) \,  z(x)- \frac12\sigma(x) \, \vec{\Phi}^2(x) - N \,  V(z(x)) + 
{\vec{\Phi}}(x). {\vec{J}}(x) \right]} \cr \cr
&& =\int \int Dz \; D\sigma \; \left\{\det\left[
\partial^2 + \sigma(.)  \right]\right\}^{ -\frac{N}2}
\;
e^{iN\int d^4x \left[\frac12\sigma(x) \,  z(x)- V(z(x))\right] + \frac{i}2
\int d^4x\;   d^4y\; J_i(x) \; G(x,y,\sigma(.)) \; J_i(y)}
\end{eqnarray}
Here, $ G(x,y,\sigma(.)) $ is the inverse operator of $ \partial^2 +
\sigma(.) $. 
That is,
\begin{equation}\label{defG}
\left[ \partial^2 + \sigma(x)  \right]G(x,y,\sigma(.))=\delta(x-y)
\end{equation}

It is useful to introduce a source $ N \,  K(x) $ for the field $ z(x)
$. This will permit to generate the correlation functions of the
composite field $ \vec{\Phi}^2(x) $. The generating functional now
takes the form 
\begin{eqnarray}\label{zetap}
&&{\cal Z}[J(.), K(.) ] =\int \int Dz \; D\sigma \; \left\{\det\left[
\partial^2 + \sigma(.)  \right]\right\}^{ -\frac{N}2} \;
e^{iN\int d^4x \left[\frac12\sigma(x) \,  z(x)- V(z(x))\right]} \cr \cr 
&& e^{\frac{i}2\int d^4x\;   d^4y\; J_i(x) \; G(x,y,\sigma(.)) \; J_i(y) + i N
\int d^4x\; K(x)\;  z(x)  }  
= \int \int Dz \; D\sigma \; e^{i N {\cal S}[z(.),\sigma(.)]} \; .
\end{eqnarray} 
Here,
\begin{eqnarray} \label{accefe}
{\cal S}[z(.),\sigma(.)] &=& \frac{i}2 \log \det\left[\partial^2 +
\sigma(.)  \right] + \int d^4x \left[\frac12\sigma(x) \,  z(x)-
V(z(x)) + K(x)\;  z(x)\right]   \cr \cr 
&+&\frac1{2N}\int d^4x\;   d^4y \; G(x,y,\sigma(.)) \; J_i(x) \; J_i(y) \; .
\end{eqnarray}
Notice that the source terms in $ {\cal S} $ are both of order one
since $  K(x) $ and $  J_i(x) $ are assumed to be of order one. 

The functional derivatives with respect to the source $
K(x) $ at $ K(x) = 0 $ produce the insertions
\begin{equation} \label{inK}
{ 1 \over i}  { {\delta} \over {\delta K(x) }} \Rightarrow  N \, z(x) =
\vec{\Phi}^2(x) \; .
\end{equation}
as follows from eqs.(\ref{repV}) and (\ref{zetap}) 
That is, $ K(x) $ is the source of the composite field $ \vec{\Phi}^2(x) $.
The correlations of $ \vec{\Phi}^2(x) $ follow as functional
derivatives of $ \log {\cal Z}[J(.), K(.) ] $ with respect to $ K(x)
$. For example, we have for the two points function
$$
<T\, \vec{\Phi}^2(x) \, \vec{\Phi}^2(y) > = i^{-2} \; \left. 
{ {\delta^2} \over { \delta K(x)   
\, \delta K(y) }} \log{\cal Z}[J(.), K(.) ]\right|_{J(.) = , K(.) =0} \; .
$$

Since in eq.(\ref{zetap})  the $ N$-dependence is explicit in the
exponent, we can take the large $N$ limit by looking for the
stationary points of the action there.

Extremizing $ {\cal S}[z(.),\sigma(.)] $ with respect to the field $
\sigma(x) $ yields: 
\begin{equation} \label{ecsigma}
i\,G(x,x,\sigma(.)) + z(x) - \frac1{N}\, \left[ \int d^4y\;  G(x,y,\sigma(.))\,
J_i(y) \right]^2 = 0  
\end{equation}
Where we used that,
\begin{equation} \label{derG}
{ {\delta G(y,z,\sigma(.))}\over { \delta \sigma(x)} } = -
G(y,x,\sigma(.))\;G(x,z,\sigma(.))\; . 
\end{equation}
Eq.(\ref{derG}) can be derived as follows. Taking the functional
derivative of eq.(\ref{defG}) with respect to $ \sigma(.) $ yields,
$$
\left[ \partial^2 + \sigma(x)  \right]{ {\delta G(x,y,\sigma(.))}\over
{ \delta \sigma(z)} }=- \delta(x-z) G(x,y,\sigma(.))
$$
This is an equation for $ { {\delta G(x,y,\sigma(.))}\over
{ \delta \sigma(z)} } $ that can be solved using the inverse operator
of $ \left[ \partial^2 + \sigma(x)  \right] $ as given by
eq.(\ref{defG}). This gives eq.(\ref{derG}).

Extremizing $ {\cal S}[z(.),\sigma(.)] $ with respect to the field $
z(x) $ yields:
\begin{equation} \label{ecz}
\frac12\sigma(x) - V'\left(z(x)\right) + K(x) = 0\; .
\end{equation}

Eqs.(\ref{ecsigma})-(\ref{ecz}) define the saddle point $
\sigma(x), z(x) $ as a functional of the sources $ J(.), K(.) $. 
One has thus to make the following shift of the functional integration
variables 
\begin{equation}\label{camvar}
\sigma(x) = \sigma(x) + \xi(x) \; , \; z(x) = z(x) + w(x)
\end{equation}
where $ \xi(x), w(x) $ are the new functional integration variables.

Now we have to insert the change (\ref{camvar}) into eq.(\ref{accefe}) and
expand $ {\cal S}[z(.),\sigma(.)] $ in powers of $ \xi(x) $ and $ w(x)
$.
The zeroth order, that is $ {\cal S}[z(.),\sigma(.)] $ yields the
$ N = \infty $ limit of the model. The quadratic part in $ \xi(x) $
and $ w(x) $ provides the propagators and the higher orders provide
the vertices of the $ 1/N $ perturbation theory. 

\subsection{The Large $N$ Limit for the $[\vec{\Phi}^2]^2$ theory}

For simplicity, we will restrict ourselves to the $ \Phi^4 $ theory
with potential (\ref{fi4}). That is,
\begin{equation}\label{vcua}
V(z) = \frac{\lambda}{8}\left(z+\frac{2 m^2}{\lambda}\right)^2 
= \frac{\lambda}{8}\, z^2 + \frac{m^2}2 \, z + {{m^4}\over { 2 \lambda}}
\quad .  
\end{equation}
In this important case we have $ V'(z) = \frac{\lambda\; z}{4} +
\frac{m^2}2 $ and
we can eliminate $ z(x) $ using the saddle point equation (\ref{ecz}),
\begin{equation}\label{zetas}
z(x) = \frac{2}{\lambda} \left[ \sigma(x) + 2 K(x) - m^2 \right]\; .
\end{equation}
Equivalently, in this case we can integrate exactly over $ z(x) $ in
eqs.(\ref{repV}), (\ref{gauss}) and (\ref{zetap}) since these
functional integrals become gaussian in the field $ z $
when $ V(z) $ is given by eq.(\ref{vcua}).  

The saddle point equation (\ref{ecsigma}) for $ \sigma(x) $ becomes 
\begin{equation} \label{ecsigmb}
i\,G(x,x,\sigma(.)) + \frac{2}{\lambda} \left[ \sigma(x) + 2 K(x) - m^2
 \right]  - \frac1{N}\,\left[ \int d^4y\;  G(x,y,\sigma(.))\, J_i(y)
 \right]^2 = 0 
\end{equation}
where $ G(x,y,\sigma(.)) $ is defined by eq.(\ref{defG}).

The action $ {\cal S} $ at the saddle point takes then the form
\begin{eqnarray} \label{aefi4}
{\cal S}[\sigma(.)] &=& \frac{i}2 \log \det\left[\partial^2 +
\sigma(.)  \right] + \frac{1}{2\lambda}\int d^4x\; \left[ \sigma(x) + 2
K(x) - m^2  \right]^2  \cr \cr 
&+&\frac1{2N}\int d^4x\;   d^4y \; G(x,y,\sigma(.)) \; J_i(x) \; J_i(y) \; .
\end{eqnarray}
where we used eqs.(\ref{accefe}), (\ref{ecsigma}), (\ref{vcua}) and
(\ref{zetas}). 

Hence, from eqs.(\ref{zetap}) and (\ref{ecsigmb}) we obtain that in
the $ N = \infty $ limit 
\begin{equation} \label{limiZ}
\lim_{N \to \infty} \frac1{N} \log {\cal Z}[J(.), K(.) ] = i \, {\cal
S}[\sigma(.)] \; .
\end{equation}

Furthermore, we can easily compute the two points function of the
scalar field $ {\vec{\Phi}}(x) $ using eqs.(\ref{fun2p}), (\ref{aefi4}) 
and (\ref{limiZ}) with the result
\begin{equation} \label{propb}
< T\, \Phi_i(x)\,  \Phi_k(y)> = - i \; \delta_{i k} \;
G(x,y,\sigma(.)) + {\cal O}\left(\frac1{N} \right) \; .
\end{equation}
and $ G(x,y,\sigma(.)) $ was defined through eq.(\ref{defG}).
The $ {\vec{\Phi}}(x) $ propagator turns to be of order one in the $ N
= \infty $ limit. 

The $ O(N) $ invariance is here explicit. We find for the expectation
value of $ {\vec{\Phi}}(x) $ (one-point function) in the infinite $ N
$ limit, 
\begin{eqnarray}\label{vexfi}
< {\vec{\Phi}}(x)_i > &=& \left. i^{-1}{ {\delta} \over {\delta
J_i(x) }}  \log {\cal Z}[J(.), K(.) ]\right|_{J_i(.) = K(.) = 0} 
=\left.  \, N \, { {\delta} \over {\delta
J_i(x) }} {\cal S}[\sigma(.)]\right|_{J_i(.) = K(.) = 0} \cr \cr\cr 
&=& \int d^4y\; G(x,y,\sigma(.)) \; J_i(x) + {\cal O}\left(\frac1{N}
\right) \; .  
\end{eqnarray}
and for the composite field  $ \vec{\Phi}^2(x) $ 
\begin{eqnarray}\label{vexfi2}
\frac1{N} <\vec{\Phi}^2(x)> &=&\frac{i}{\,N}\left.  { {\delta} \over {\delta
K(x) }} \log {\cal Z}[J(.), K(.) ]\right|_{J_i(.) = K(.) = 0} \cr \cr \cr 
&=& \left.  { {\delta} \over {\delta
K(x) }} {\cal S}[\sigma(.)]\right|_{J_i(.) = K(.) = 0} = \frac2{\lambda}
\left[ \sigma(x) - m^2\right]  + {\cal O}\left(\frac1{N} \right)
\end{eqnarray}
where we used eqs.(\ref{zjota}), (\ref{inK}) and (\ref{limiZ}).

Once we set the external sources $ J_i(x) $ and $ K(x) $ equal to
zero, the expectation value of $ {\vec{\Phi}}(x) $ vanishes
as it must be due to the $ O(N) $ invariance. On the contrary, 
$ N^{-1} <\vec{\Phi}^2(x)> $ has in general a non-zero value at
zero external sources.

It must be stressed that the present derivation applies to an
{\bf arbitrary} quantum state of the theory. Indeed, the equations simplify
for the ground state due to translational invariance. In such case $
\sigma(x) $ must be a constant $\sigma_0 $ and the propagator  $
G(x,y,\sigma_0) $ takes the form,
$$
G(x,y,\sigma_0) = \int { {d^4k} \over {(2\pi)^4}} \, {{e^{ik.(x-y)}
}\over { \sigma_0 - k^2}} \; .
$$
The equal-points propagator $ G(x,x,\sigma_0) $ needs an UV
regulator. We obtain using a momentum cutoff $ \Lambda $ and Wick
rotating $ k_0 \to i \, k_0 $,
\begin{eqnarray}
&& G_{\Lambda}(x,x,\sigma_0) = {{ 2\pi^2 i } \over { (2\pi)^4}}
 \int_0^{\Lambda} {{ k^3 \; dk}\over { \sigma_0 + k^2}} 
={ i \over (4 \pi)^2} \left[ -\Lambda^2 + \sigma_0\, \log\left( {
 \Lambda^2 \over \sigma_0^2}+1\right) \right] \nonumber
\end{eqnarray}
The $\Lambda$ dependence can be absorbed into standard mass and
coupling constant renormalization. Notice that $ \mu^2 \equiv \sigma_0
$ is the physical (renormalized) mass squared of the fundamental boson
in the $N=\infty$ limit 
as we see from eq.(\ref{propb}). $m^2\equiv m^2_B $ is the bare boson
mass and $ \lambda \equiv \lambda_B $ the bare coupling constant. 

We find from eq.(\ref{ecsigmb}) the relationship between bare and
renormalized parameters,
\begin{equation}
\lambda_B = { \lambda_r \over { 1 - {\lambda_r\over 16 \pi^2} \log{
\Lambda \over \mu }}} \quad , \quad
m^2_B = - {\lambda_B \, \Lambda^2 \over 32 \pi^2} + {\mu^2 \over   1
- {\lambda_r\over 16 \pi^2} \log{ \Lambda \over \mu }}
\end{equation}
where $ \lambda_r $ stands for the renormalized coupling constant and
we dropped all contributions of order $ \Lambda^{-2} $ and higher.
 
We can now choose for simplicity $ J_i(x) = J(x) $ for $ i = 1, \ldots, N $.
We proceed now to make a Legendre transformation such that the 
field expectation values $ \phi(x) $ and $ \xi(x) $ become the
independent functional variables. We define following
eqs.(\ref{vexfi}) and (\ref{vexfi2}),
\begin{eqnarray} \label{defphi}
\phi(x) &\equiv&  \frac1{i\,N}\; { {\delta} \over {\delta J(x) }}  
\log {\cal Z}[J(.), K(.) ] = 
\int d^4y\; G(x,y,\sigma(.)) \; J(y) + {\cal O}\left(\frac1{N}
\right) \cr \cr
\rho(x) &\equiv& \frac1{i\,N} { {\delta} \over {\delta K(x) }}  
\log {\cal Z}[J(.), K(.) ] = \frac2{\lambda}
\left[ \sigma(x) - m^2+ 2 K(x) \right]  + {\cal O}\left(\frac1{N} \right)
\end{eqnarray}
In order to compute these functional derivatives, we used the chain
rule,
$$
 \left. { {\delta} \over {\delta J(x) }}\right|_{K} =
\int d^4y \;  \left.
{ {\delta \sigma(y) } \over {\delta J(x) }}\right|_{K}\;  \left.
{ {\delta} \over {\delta \sigma(y) }}\right|_{J,K} + \left.
{ {\delta} \over {\delta J(x) }}\right|_{\sigma,K} \; .
$$
The effective action functional is then given by,
\begin{equation}
\Gamma[\phi(.),\rho(.)] = \frac1{i\,N}\;\log {\cal Z}[J(.), K(.) ]  -
\int d^4x\; \left[ J(x)\, \phi(x) + K(x) \, \rho(x) \right]
\end{equation}
We get for this effective action in the infinite $N$ limit
\begin{eqnarray}
\Gamma[\phi(.),\rho(.)] &=& \frac{i}2 \log \det\left[\partial^2 + m^2 + 
\xi(.)  \right] + \frac{1}{2\lambda}\int d^4x\; \left[ \xi(x)^2
-4 K(x)^2 \right]  \cr \cr 
&-&\frac1{2}\int d^4x\;   d^4y \; G(x,y,m^2 + \xi(.)) \; J(x) \; J(y)
+ {\cal O}\left(\frac1{N} \right) \; .
\end{eqnarray}
where we used eqs.(\ref{aefi4}),(\ref{limiZ}) and (\ref{defphi}) and
we introduced the new field variable: 
\begin{equation}\label{defxi}
\xi(x) \equiv \sigma(x) - m^2 = \frac12 \, \lambda \,\rho(x) - 2 K(x)  \; .
\end{equation}
Using now eqs.(\ref{defG}) and (\ref{defphi}) we can easily express $
J(x) $ in terms of $ \phi(x) $ and $ \xi(x) $ as follows,
$$
J(x) = \left[\partial^2 + m^2 + \xi(x)\right] \phi(x) +  {\cal
O}\left(\frac1{N} \right) \; . 
$$
The effective action can then be written in terms of  $ \phi(x) $ and
$ \xi(x) $ with the result,
\begin{eqnarray}\label{Gamm}
\Gamma[\phi(.),\xi(.)] &=& \frac{i}2 \log \det\left[\partial^2 + m^2 + 
\xi(.)  \right] + \frac{1}{2\lambda}\int d^4x\; \left[ \xi(x)^2
-4 K(x)^2 \right]  \cr \cr 
&-&\frac1{2}\int d^4x\; \phi(x) \left[\partial^2 + m^2 + \xi(x)
\right]\phi(x) + {\cal O}\left(\frac1{N} \right) \; .
\end{eqnarray}

The equations of motion on the fields $ \phi(x) $ and $ \xi(x) $
follow extremizing the effective action $ \Gamma[\phi(.),\xi(.)]
$. We find from eq.(\ref{Gamm}) always for infinite $ N $,
\begin{eqnarray}
\left[\partial^2 + m^2 + \xi(x) \right]\phi(x) &=& 0 \; , \cr \cr
\frac{i}{2} G(x,x, m^2 + \xi(.)) -\frac12 \phi(x)^2 +
\frac{1}{\lambda} \, \xi(x) &=& 0 \; . 
\end{eqnarray}
Or, in a more explicit form,
\begin{eqnarray}\label{eqgen}
\left[\partial^2 + m^2 + \frac{\lambda }2 \phi(x)^2-\frac{i\lambda}{2}\;
G(x,x, m^2 + \xi(.)) \right]\phi(x) &=& 0 \; , \\ \cr
\left[\partial^2 + m^2 + \frac{\lambda }2 \phi(x)^2-\frac{i\lambda}{2}\;
G(x,x, m^2 + \xi(.)) \right]G(x,y, m^2 + \xi(.)) &=& \delta(x-y)\; .\nonumber
\end{eqnarray}
As we see, this is a non-linear and non-local set of partial
differential equations. 

Recall that $ \xi(x) $ provides the expectation value of $
<\vec{\Phi}^2(x)> $ as derived in eqs.(\ref{vexfi2}) and (\ref{defxi}),
$$
 \xi(x) = \frac{\lambda}{2 N} <\vec{\Phi}^2(x)> +  \, {\cal
 O}\left(\frac1{N} \right)\; .  
$$
The field $ \phi(x) $ is the expectation value of $ <{\vec{\Phi}}(x)_i
> $ for all values  of $ i $ as we see from eqs.(\ref{vexfi}) and
(\ref{defphi}) (up to  $ {\cal O}\left(\frac1{N}\right) $ corrections).  

The ground state correspond to $ \xi(x) = \phi(x) = 0 $. This is the
solution $ \sigma_0 $ discussed above. We are interested on solutions with
non-zero $ \xi(x) $ and $ \phi(x) $ describing excited states. 
States with non-zero $ \phi(x) $ will  not be invariant under $ O(N) $
transformations.  

\subsection{Invariance under spatial translations }

Let us now consider states which are invariant under {\bf spatial}
translations. The fields $ \xi(x) $ and $ \phi(x) $ will thus only be
functions of time. This fact considerably simplifies the general
equations (\ref{eqgen}). We can Fourier expand $ G(x,y,m^2+\xi(.)) $ as
$$
 G(x,y,m^2+\xi(.)) = \int e^{i {\vec k}.({\vec x} - {\vec y} )} \;
 g_k(t,t') \; { { d^3k} \over { (2\pi)^3 }} \; ,
$$
where the times $ t $ and  $ t' $ are associated with $ x $ and $ y $
respectively and $ g_k(t,t') $ obeys the equation

\begin{equation}\label{gfou}
\left[{{d^2}\over {dt^2} } + k^2 + m^2 + \xi(t) \right] g_k(t,t') =
\delta(t-t') \; . 
\end{equation}

This one-dimensional Green function can be expressed in terms of 
solutions of the homogeneous equation
\begin{equation}\label{eqmodo}
\left[{{d^2}\over {dt^2} } + k^2 + m^2 + \xi(t) \right]f_k(t)=0 \; .
\end{equation}
One gets,
$$
 g_k(t,t') = {{f_k^{<}(t_{<})\; f_k^{>}(t_{>})}\over {W_k}} \; ,
$$
where $ t_{<} = $ min$(t,t')$, $ t_{>} = $ max$(t,t')$ and  $ f_k^{<}(t) $
and $ f_k^{>}(t) $ are independent solutions of eq.(\ref{eqmodo}). 
$ W_k $ stands for the Wronskian between these solutions:
\begin{equation}\label{wron}
W_k \equiv W[ f_k^{<}(t),  f_k^{>}(t)] \; .
\end{equation}
For causal bondary conditions one has  $  f_k^{>}(t) =
[f_k^{<}(t)]^* $.

Before renormalization, eqs.(\ref{eqgen}) take the form
\begin{eqnarray}\label{eqdesn}
&&\left[{{d^2}\over {dt^2} }  + m^2 + \frac{\lambda }2 \phi(t)^2
-\frac{i\lambda}{2} \; G(x,x, m^2 + \xi(.)) \right]\phi(t)=0 \cr \cr
 &&\mbox{where} \quad G(x,x, m^2 + \xi(.)) = \int { { d^3k} \over {
(2\pi)^3 }}\; {{\left|f_k(t)\right|^2 }\over  {W_k}} \; .
\end{eqnarray}
and the mode functions $ f_k(t) $ obey,
\begin{equation}\label{modosk}
\left[{{d^2}\over {dt^2} } + k^2 + m^2 + \frac{\lambda }2 \phi(t)^2
-\frac{i\lambda}{2} \; G(x,x, m^2 + \xi(.)) \right]f_k(t)=0 \; .
\end{equation}
We obtained an infinite set of coupled ordinary differential equations
on $ \phi(t), \; f_k(t) $ for $ 0 \leq k < \infty $. Notice that
eqs.(\ref{eqdesn})-(\ref{modosk}) are {\bf local} on time. That is
they involve the unknown functions $ \phi(t), \; f_k(t) $ always at
time $ t $. 

All physical quantities can be computed in terms of the mode functions
$ f_k(t) , \; 0 \leq k < \infty $ and the order parameter $ \phi(t) $. 
We have described the large $ N $ approximation. Other
non-perturbative approximations are the Hartree approximations and the 
self-consistent one-loop approximation. They were considered in
ref.\cite{eri97}. 

\bigskip

In a cosmological spacetime (\ref{metric}) the large $ N $ evolution
equations take the form
\begin{eqnarray}\label{evocos}
\left[{{d^2}\over {dt^2} } + 3H(t)\frac{d}{dt}+\xi{\cal R}(t)
+ m^2 + \frac{\lambda }2 \phi(t)^2 +
\frac{\lambda}{2}\langle \pi^2(t) \rangle \; 
\right]\phi(t)&=&0 \cr \cr
\left[{{d^2}\over {dt^2} } + 3H(t)\frac{d}{dt}+\frac{k^2}{a^2(t)}
+\xi{\cal R}(t)
+ m^2 + \frac{\lambda }2 \phi(t)^2 +
\frac{\lambda}{2}\langle \pi^2(t) \rangle \;
\right]f_k(t)&=& 0 \quad , 
\end{eqnarray}
where 
$$
H(t) \equiv \frac{\dot{a}(t)}{a(t)}
$$
and we introduced the notation
\begin{equation}\label{auto}
\langle \pi^2(t) \rangle = -i  G(x,x, m^2 + \xi(.)) = \int { { d^3k} \over {
(2\pi)^3 }}\; {{\left|f_k(t)\right|^2 }\over  {i \; W_k}} \; .
\end{equation}

\subsection{\bf The Large $N$ Limit in Cosmological Spacetimes} 

Let us connect the functional formalism in previous sections with the
operatorial approach. Let us call `$ 1 $' the direction in internal
space of the expectation value of $ \Phi(x) $. We write,
$$
\vec{\Phi}(\vec x, t) = (\sigma(\vec x,t), \vec{\pi}(\vec x,t)),
$$ 
with $\vec{\pi}$ an $(N-1)$-plet, and 
\begin{equation}
\sigma(\vec x,t) = \sqrt{N}\phi(t) + \chi(\vec x,t) \; \; ; \; \; \langle
\sigma(\vec x, t) \rangle= \sqrt{N}\phi(t) \; \; ; \; \; 
\langle \chi(\vec x,t) \rangle = 0.
\label{largenzeromode} 
\end{equation}
We can now write the operator $ \vec{\pi}(\vec x, t) $ in terms
of creation and annihilation operators and mode functions that obey the
Heisenberg equations of motion
\begin{equation}\label{oper}
\vec{\pi}(\vec x, t) = \int\frac{d^3k}{(2\pi)^3}
\left[{\vec a}_k \; f_k(t) \; e^{i\vec{k}\cdot \vec x} + {\vec a}^{\dagger}_k
 \; f^*_k(t) \; e^{-i\vec{k}\cdot \vec x} \right] .
\end{equation}
where $ {\vec a}_k $ and $ {\vec a}^{\dagger}_k $ obey canonical
commutation rules. It is easy to derive the two points function
eq.(\ref{propb}) from eqs.(\ref{eqdesn}), 
(\ref{wron}) and (\ref{oper}). 

We see that since there are $N-1$ `pion' fields, contributions from the field
$\chi$ can be neglected in the $ N \to \infty $ limit as they are of
order $1/N$ with respect those of $\pi$ and $\phi$.

The equations of motion (\ref{evocos}) can be written as

\begin{eqnarray}
\ddot{\phi}(t)+3H(t) \; \dot{\phi}(t)+{\cal M}^2(t) \; \phi(t)&=&0\quad,
\label{modcer}
\cr \cr
\left[\frac{d^2}{dt^2}+3H(t)\frac{d}{dt}+\frac{k^2}{a^2(t)}+{\cal
M}^2(t) \right]f_k(t)&=& 0\quad, 
\label{modkN}
\end{eqnarray}
where
\begin{equation}
{\cal M}^2(t) =  m^2+\xi\; {\cal R}(t)+ \frac{\lambda}{2}\phi^2(t)+
\frac{\lambda}{2}\langle \pi^2(t) \rangle \; .
\label{Ngranmass}
\end{equation}
plays the role off time-dependent effective mass.

An important point to note in the large $N$ equations of motion is that the
form of the equation for the zero mode (\ref{modcer}) is the same as for the
$k=0$ mode function (\ref{modkN}). It is this property that allows
solutions of these equations in a symmetry broken scenario  to satisfy
Goldstone's theorem in {\bf out of equilibrium} situations both in
Minkowski and cosmological spacetimes \cite{mink,simrota,frw2,din}.

In this leading order in $ 1/N $ the theory becomes Gaussian, but with the
self-consistency condition (\ref{auto}).

The initial conditions on the modes $ f_k(t) $ must now be determined. 
At this stage it proves illuminating to pass to conformal time variables
in terms of the conformally rescaled fields (see \cite{frw2} and
section V for a discussion)
in which the mode functions obey an equation which is very similar to that
of harmonic oscillators with time dependent frequencies in Minkowski
space-time.  
It has been realized that different initial conditions on the mode functions
lead to different renormalization counterterms\cite{frw2};
 in particular imposing initial conditions in comoving time leads to
counterterms that depend on these initial conditions. Thus we chose to
impose initial conditions in conformal time in terms of the conformally
rescaled mode functions leading to the following choice:
$$
f^{>}_k(t) = f_k(t) \quad , \quad f^{<}_k(t) = [f_k(t)]^{*}
$$
with initial conditions in comoving time,
\begin{equation}
f_k(t_0)=\frac{1}{\sqrt{\Omega_k}}, \;\;\; 
\dot{f}_k(t_0)=\left[-\frac{\dot{a}(t_0)}{a(t_0)}-i\Omega_k\right]f_k(t_0),
\label{condini}
\end{equation}
with
\begin{equation}
\Omega_k^2 \equiv k^2 + {\cal M}^2(t_0) - \frac{{\cal R}(t_0)}{6}.
\label{frec}
\end{equation}
We thus find from eq.(\ref{wron}) that the Wronskian takes the value $
W_k = -2i $ and the quantum fluctuations of the inflaton (\ref{auto})
take the form 
\begin{equation}\label{largenfluc}
\langle \pi^2(t) \rangle = \int { { d^3k} \over {2
(2\pi)^3 }}\; \left|f_k(t)\right|^2  \; .
\end{equation}
For convenience, we have set $a(t_0)=1$ in eq.(\ref{frec}).
At this point we recognize that when ${\cal M}^2(t_0) - {\cal R}(t_0)/6 <0$ 
the above initial condition must be modified to avoid imaginary frequencies,
which are the signal of instabilities for long wavelength modes in the
broken symmetry case. Thus
we {\em define} the initial frequencies that determine the initial conditions
(\ref{condini}) as
\begin{eqnarray}
\Omega_k^2 & \equiv &  k^2 + \left|{\cal M}^2(t_0) - \frac{{\cal
R}(t_0)}{6}\right| \; \mbox{ for } 
k^2 < \left |{\cal M}^2(t_0) - \frac{{\cal R}(t_0)}{6} \right|\; ,
\label{unstcond1} \\ 
\Omega_k^2 & \equiv &  k^2 + {\cal M}^2(t_0) - \frac{{\cal R}(t_0)}{6} \;
\mbox{ for } 
k^2 \geq \left |{\cal M}^2(t_0) - \frac{{\cal R}(t_0)}{6} \right|\;
. \label{unstcond2}  
\end{eqnarray}
In the unbroken symmetry case ($ m^2 > 0 $ ) we use
eq.(\ref{unstcond2}) for all $ k $. 

As an alternative we have also used initial conditions which smoothly
interpolate from positive frequencies for the unstable modes to the adiabatic
vacuum initial conditions defined by (\ref{condini})-(\ref{frec}) for
the high $k$ modes. While the alternative choices of initial
conditions result in small quantitative differences in the results (a
few percent in quantities which depend strongly on these low-$k$
modes), all of the qualitative features we will examine are
independent of this choice. 

In the large $N$ limit we find the energy density and pressure 
density to be given by\cite{De Sitter,frw2}
\begin{eqnarray}
&&\frac{\varepsilon}{N}  =  \frac12\dot{\phi}^2 + \frac12\;m^2\phi^2 +
\frac{\lambda}{8}\;\phi^4 + \frac{m^4}{2\lambda} - \xi\;
G^{0}_{0}\;\phi^2 +  6\;\xi\;\frac{\dot{a}}{a}\;\phi\;\dot{\phi}
+  \frac12\langle\dot{\pi}^2\rangle
\nonumber \\
&& +  \frac{1}{2a^2}\langle(\nabla\pi)^2\rangle + \frac12 m^2 
\langle\pi^2\rangle 
+ \frac{\lambda}{8}[2\phi^2\;\langle\pi^2\rangle + \langle\pi^2\rangle^2] 
-\xi\; G^{0}_{0}\langle\pi^2\rangle + 
6\;\xi\;\frac{\dot{a}}{a}\;\langle\pi\dot{\pi}\rangle\; , \nonumber \\
&&\frac{\varepsilon-3p}{N}  =  -\dot{\phi}^2 + 2m^2\;\phi^2 + 
\frac{\lambda}{2}\;\phi^4 + \frac{2m^4}{\lambda} - \xi\;
G^{\mu}_{\mu}\phi^2 + 6\xi\left(\phi\ddot{\phi} + \dot{\phi}^2 +
3\frac{\dot{a}}{a}\phi\dot{\phi}\right) - \langle\dot{\pi}^2\rangle
\nonumber \\ 
&& +   \frac{1}{a^2}\langle(\nabla\pi)^2\rangle + 
2m^2\;\langle\pi^2\rangle - \xi \; G^{\mu}_{\mu}\;\langle\pi^2\rangle  
+\frac{\lambda}{2}\;[2\phi^2\;\langle\pi^2\rangle +
\langle\pi^2\rangle^2] + 
6\;\xi\left(\langle \pi\;\ddot{\pi}\rangle + \langle \dot{\pi}^2 \rangle+
3\,\frac{\dot{a}}{a}\;\langle\pi\dot{\pi}\rangle \right)\; ,\nonumber
\end{eqnarray}
where $\langle\pi^2\rangle$ is given by equation (\ref{largenfluc}) 
and we have defined the following integrals:
\begin{eqnarray}
&&\langle(\nabla\pi)^2\rangle  =  \int \frac{d^3k}{2 (2\pi)^3}\; k^2\;
|f_k(t)|^2 \quad , \quad
\langle\dot{\pi}^2\rangle =  \int \frac{d^3k}{2(2\pi)^3}\;
|\dot{f}_k(t)|^2 \;. \label{dotpsi}
\end{eqnarray}
The composite operators $\langle \pi \dot{\pi} \rangle$ and 
$\langle \pi \ddot{\pi} \rangle$ are symmetrized by
removing a normal ordering constant. $\langle \pi \ddot{\pi} \rangle$
may be rewritten using the equation of motion (\ref{modkN}):
\begin{equation}
\langle \pi \ddot{\pi} \rangle = -3\, \frac{\dot{a}}{a}\,\langle \;
\pi \dot{\pi} \rangle
-\frac{\langle(\nabla\pi)^2\rangle}{a^2}-{\cal M}^2(t)\,
\langle\pi^2\rangle \; . 
\end{equation}
It is straightforward to show that the bare energy is
covariantly conserved by using the equations of motion for the zero mode and
the mode functions. 

\section{Renormalization, Conformal time  and Initial Conditions}

Renormalization is a very subtle but important issue in gravitational
backgrounds\cite{birrell}. The fluctuation contribution
$\langle \pi^2(\vec x,t) \rangle$, the energy, and the pressure all need to be
renormalized. The renormalization aspects in curved space times have been
discussed at length in the literature\cite{birrell} and have
been extended to the large $N$ self-consistent approximations for
the non-equilibrium backreaction problem 
in\cite{De Sitter,din,frw2,largen,ramsey}. More recently, a consistent
and covariant  regularization scheme that can be implemented
numerically has been proposed\cite{baacke}.  

The ultraviolet divergences can be seen in the present framework in
the $k$-integrals over the modes in eqs.(\ref{largenfluc}) and
(\ref{dotpsi}).  To analyze these divergences it is convenient to
change variables to conformal time $ {\cal T} $ defined as
\begin{equation}
{\cal T} =  {\cal T}_0 + \int^t_{t_0} \frac{dt'}{a(t')}~~; ~~ {\cal
T}(t=t_0)={\cal T}_0 \; , 
\label{conftime}
\end{equation}
The metric becomes then
\begin{equation}
ds^2= C^2({\cal T})\; (d{\cal T}^2 - d\vec{x}^2) \; ,
\end{equation}
where $ C({\cal T})  \equiv  a(t({\cal T}))~~;~~ C({\cal T}_0)=1 $ stands for
the scale factor in conformal time.

The issue of renormalization and initial conditions is best analyzed in
conformal time which is a natural framework for adiabatic renormalization and
regularization.

Under a conformal rescaling of the field
\begin{equation}\label{camco}
\Phi(\vec x, t) = {\chi(\vec x, {\cal T}) \over  C({\cal T})} \; ,
\end{equation}
the action for a scalar field (with the obvious generalization to $ N $
components) becomes, after an integration by parts and dropping a
 surface term
\begin{equation}
S= \int d^3x \; d{\cal T} \left[\frac12 (\chi')^2-\frac12 (\vec{\nabla}\chi)^2-
{\cal{V}}(\chi)\right] \; ,
\end{equation}
with
\begin{equation}
{\cal{V}}(\chi) = C^4({\cal T})\; V\left({\chi \over C({\cal
T})}\right)-C^2({\cal T}) \; \frac{{\cal{R}}}{12} \; \chi^2\; \;  , 
\end{equation}
where $ {\cal R }= 6\; C''({\cal T})/C^3({\cal T}) $ is the Ricci scalar,
and primes stand for derivatives with respect to conformal time $
{\cal T} $.

The conformal time Hamiltonian operator, which is the generator of translations
in ${\cal T}$, is given by
\begin{equation}
H_{{\cal T}}= \int d^3x \left\{ \frac{1}{2}\Pi^2_{\chi}+\frac{1}{2}
(\vec{\nabla}\chi)^2+{\cal{V}}(\chi) \right \}\; , \label{confham}
\end{equation}
with $\Pi_{\chi}$ being the canonical momentum conjugate to $\chi$,
$\Pi_{\chi} = \chi'$. 
Separating the zero mode of the field $\chi$ 
\begin{equation}
\chi(\vec x, {\cal T}) = \chi_0({\cal T}) + \bar{\chi}(\vec x,{\cal T}),
\end{equation}
and in the large $N$  approximation  we find 
that the Hamiltonian becomes linear plus quadratic in the fluctuations, and
similar to a Minkowski space-time Hamiltonian with 
a ${\cal T}$-dependent effective mass term
given by
\begin{equation}
B({\cal T}) = C^2({\cal T}) \left[m^2+(\xi-\frac{1}{6})\; {\cal{R}}
+ \frac{\lambda}{2}\; \chi_0^2({\cal T}) + \frac{\lambda}{2}
\langle \bar{\chi}^2 \rangle\right]\; . \label{masseff}
\end{equation}
Notice that this $ B(t) $ 
naturally appears in 
the WKB expansion of the mode functions both in cosmic and conformal
time [see eq.(\ref{largekf})].

We can now follow the steps and use the results of reference\cite{frw} for the
conformal time evolution of the density matrix by setting $a(t)=1$ in the
proper equations of that reference and replacing the frequencies by
\begin{equation}
\omega^2_k({\cal T}) = \vec{k}^2 + B({\cal T})\; , \label{freqs}
\end{equation}
and the expectation value in eq.(\ref{masseff}) is obtained in this
${\cal T}$ evolved density matrix.
The time evolution of the kernels in the density matrix (see\cite{frw})
is determined by the mode functions that obey
\begin{equation}
\left[ \frac{d^2}{d{\cal T}^2}+k^2+B({\cal T})\right] F_k({\cal T})=0\; .
\label{fmodeqn}
\end{equation}
The Wronskian of these mode functions
\begin{equation}\label{wff}
{\cal{W}}(F,F^*)= F'_k F^*_k-F_k F^{*'}_k
\end{equation} 
is a constant. It is natural to impose initial conditions such that at the
initial ${\cal T}$ the density matrix describes a situation of local
thermodynamic 
equilibrium and therefore commutes with the conformal time Hamiltonian at the
initial time. This implies that the initial conditions of the mode functions
$F_k({\cal T})$ be chosen to be (see\cite{frw})
\begin{equation}
F_k({\cal T}_o)= \frac{1}{\sqrt{\omega_k({\cal T}_o)}} \; \; ; \; 
F'_k({\cal T}_o)= -i\sqrt{\omega_k({\cal T}_o)} \; \; . \label{inicond} 
\end{equation}
With such initial conditions, the Wronskian (\ref{wff}) takes the value
\begin{equation}\label{wro}
{\cal{W}}(F,F^*)= -2i \; .
\end{equation}
These initial conditions correspond to the choice of mode functions which
coincide with the first order adiabatic modes and those of the Bunch-Davies
vacuum for large momentum\cite{birrell}.  To see this clearly, we
write the solution of eq.(\ref{fmodeqn}) in the form,
\begin{equation}
D_k({\cal T}) = e^{\int^{{\cal T}}_{{\cal T}_o} R_k({\cal T}')d{\cal T}'}\; ,
\end{equation} 
with the function $ R_k({\cal T}) $ obeying the Riccati equation
\begin{equation}
R'_k + R^2_k+ k^2+B({\cal T})=0\; .
\end{equation} 
This equation posses the solution
\begin{equation}
R_k({\cal T}) = -ik+R_{0,k}({\cal T})-i\frac{R_{1,k}({\cal T})}{k}+
\frac{R_{2,k}({\cal T})}{k^2}-i\frac{R_{3,k}({\cal T})}{k^3}+
\frac{R_{4,k}({\cal T})}{k^4}+ {\cal O}\left(\frac{1}{k^5} \right)
\end{equation}
and its complex conjugate. We find for the  coefficients:
\begin{eqnarray}
&& R_{0,k} = 0 \; \; ; \; \; R_{1,k} = \frac{1}{2}\; B({\cal T}) \quad ;
R_{2,k} = -\frac{1}{2}\; R'_{1,k} \nonumber \\ &&R_{3,k} = \frac{1}{2}\left(
R'_{2,k}-R^2_{1,k} \right) \; \; ; \; \; R_{4,k} = -\frac{1}{2}\left(
R'_{3,k}+2R_{1,k}R_{2,k} \right)\; .
\end{eqnarray}
The solutions $F_k({\cal T})$ obeying the boundary conditions
(\ref{inicond}) are obtained as linear combinations of this WKB
solution and its complex conjugate 
\begin{equation}
F_k({\cal T}) = \frac{1}{2\sqrt{\omega_k({\cal T}_o)}}\left[
(1+\gamma)D_k({\cal T})+(1-\gamma)D^*_k({\cal T}) \right]\; ,
\end{equation}
where the coefficient $\gamma$ is obtained from the initial conditions. It is
straightforward to find that the real and imaginary parts are given by
\begin{equation}
\gamma_R= 1+ {\cal{O}}(1/k^4) \; \; ; \; \; \gamma_I= {\cal{O}}(1/k^3)\; .
\end{equation}
Therefore the large-$k$ mode functions satisfy the adiabatic vacuum initial
conditions\cite{birrell}. This, in fact, is the rationale for the choice of the
initial conditions (\ref{inicond}).

Following the analysis presented in \cite{frw} we find, in conformal time that
\begin{equation}
\langle \bar{\chi}^2(\vec x,{\cal T}) \rangle = \int \frac{d^3k}{2(2\pi)^3}
\; |F_k({\cal T})|^2\; .
\end{equation}
The Heisenberg field operators $\bar{\chi}(\vec x, {\cal T})$  and 
their canonical momenta $\Pi_{\chi}(\vec x, {\cal T})$ can be expanded as:
\begin{eqnarray}
&& \bar{\chi}(\vec x, {\cal T}) = \int {{d^3k}\over {\sqrt2 \, (2\pi)^{3/2}}} 
\left[ a_k F_k({\cal T})+ a^{\dagger}_{-k}F^*_k({\cal T}) \right]
 e^{i \vec k \cdot \vec x}\; , \label{heisop}\\
&& \Pi_{\chi}(\vec x, {\cal T}) = 
\int {{d^3k}\over {\sqrt2 \, (2\pi)^{3/2}}} 
\left[ a_k F'_k({\cal T})+ a^{\dagger}_{-k}F^{*'}_k({\cal T}) \right]
 e^{i \vec k \cdot \vec x}\;, \label{canheisop}
\end{eqnarray}
with the time independent creation and
annihilation operators $ a_k$ and $ a^{\dagger}_k $ 
obeying canonical commutation relations. Since the fluctuation fields
in comoving and conformal time are related by the conformal rescaling
(\ref{camco}), 
it is straightforward to see that the mode functions in comoving time are
related to those in conformal time simply as
\begin{equation}
f_k(t) = \frac{F_k({\cal T})}{C({\cal T})}.
\end{equation}
Therefore the initial conditions (\ref{inicond}) on the conformal time mode
functions imply the initial conditions for the mode functions in comoving time
are given by eq.(\ref{condini}).
 
For renormalization purposes we need the large-$k$ behavior of $|f_k(t)|^2 \; ,
|\dot{f}_k(t)|^2$, which are determined by the large-$k$ behavior of the
conformal time mode functions and its derivative. These are given
by appropriately adapting the Minkowski formulas\cite{mink} 
\begin{eqnarray}
|F_k({\cal T})|^2 &=& \frac{1}{k} + { B({\cal T}) \over 2 \, k^2} + {
 1 \over 8 \, k^5}\left[ 3 \, B^2({\cal T}) + B''({\cal T}) \right] 
+ {\cal O}\left(\frac{1}{k^7}\right) \; ,  \cr \cr
|F'_k({\cal T})|^2 &=& k + { B({\cal T}) \over 2 \, k} - {
 1 \over 8 \, k^3}\left[ B^2({\cal T}) + B''({\cal T}) \right]
+ {\cal O}\left(\frac{1}{k^5}\right)  \; . \label{largekf}
\end{eqnarray}
We note that the large $k$ behavior of the mode functions to the order
needed to 
renormalize the quadratic and logarithmic divergences is insensitive to the
initial conditions. This is not the case when the initial conditions
are imposed as described 
in\cite{frw,De Sitter}. Thus the merit in considering the initial conditions in
conformal time \cite{frw2}.

There is an important physical consequence of the choice
(\ref{inicond}) of initial
conditions, which is revealed by analyzing the evolution of the density matrix.

In the large $N$ or Hartree (also to one-loop) approximation, the
density matrix is Gaussian, and defined by a normalization factor, a
complex covariance that 
determines the diagonal matrix elements and a real covariance that determines
the mixing in the Schr\"odinger representation as discussed in
reference\cite{frw} (and references therein).

In conformal time quantization and in the Schr\"odinger representation in which
the field $\chi$ is diagonal the conformal time evolution of the density matrix
is via the conformal time Hamiltonian (\ref{confham}). The evolution equations
for the covariances is obtained from those given in reference\cite{frw} by
setting $a(t) = 1$ and using the frequencies $\omega^2_k({\cal T})=
k^2+{\cal{M}}^2({\cal T})$. In particular, by setting the covariance of the
diagonal elements (given by equation (2.20) in\cite{frw}; see also equation
(2.44) of\cite{frw}),
\begin{equation}
{\cal{A}}_k({\cal T}) = -i \frac{F^{'*}_k({\cal T})}{F^*_k({\cal T})},
\end{equation}
we find that with the initial conditions (\ref{inicond}), the
conformal time density matrix is that of local equilibrium at ${\cal T}_0$
in the sense that it commutes with the conformal time Hamiltonian. 
However, it is straightforward to see, that the comoving time density
matrix {\em does not} commute with the {\it comoving time} Hamiltonian at
the initial time $t_0$.  

An important corollary of this analysis and comparison with other initial
conditions used in comoving time is that assuming initial conditions of local
equilibrium in comoving time leads to divergences that depend on the initial
condition as discussed at length in\cite{frw}.  This dependence of the
renormalization counterterms on the initial condition was also
realized in ref.\cite{leutwyler} within the context of the CTP formulation.
Imposing the initial conditions corresponding to local thermal equilibrium in
{\em conformal} time, we see that: i) the renormalization counterterms do not
depend on the initial conditions and ii) the mode functions are identified with
those corresponding to the adiabatic vacuum for large momenta.
This is why we prefer the  initial conditions (\ref{inicond}).


For our main analysis we choose this initial temperature to be zero so that the
resulting density matrix describes a pure state, which for the large
momentum modes coincides with the conformal adiabatic vacuum. Such
zero temperature choice seems appropriate after the exponential
inflation of the universe.

{\bf Particle Number:}

We write the Fourier components of the field $\chi$ and its canonical
momentum $\Pi_{\chi}$ given by (\ref{heisop}) -(\ref{canheisop}) as:
\begin{eqnarray}
&& \bar{\chi}_k({\cal T}) = \frac{1}{\sqrt{2}}\left[
a_k \; F_k({\cal T})+ a^{\dagger}_{-k} \;F^*_k({\cal T}) \right],
  \label{heisopfu}\\
&& \Pi_{\chi,k}({\cal T}) = \frac{1}{\sqrt 2}\left[
a_k \; F'_k({\cal T})+ a^{\dagger}_{-k} \;F^{*'}_k({\cal T}) \right].
  \label{canheisopfu}
\end{eqnarray}
These (conformal time) Heisenberg operators can be written equivalently
in terms of the ${\cal T}$ dependent creation and annihilation operators
\begin{eqnarray}
&& \bar{\chi}_k({\cal T}) = \frac{1}{\sqrt{2 \omega_k({\cal T}_0)}}\left[
\tilde{a}_k({\cal T}) \; e^{-i\omega_k({\cal T}_0){\cal T}}+ 
\tilde{a}^{\dagger}_k({\cal T}) \; e^{i\omega_k({\cal T}_0){\cal T}}
 \right],
  \label{newheis}\\
&& \Pi_{\chi,k}({\cal T}) = -i\sqrt{\frac{\omega_k({\cal T}_0)}{2}}\left[
\tilde{a}_k({\cal T}) \; e^{-i\omega_k({\cal T}_0){\cal T}}-
\tilde{a}^{\dagger}_k({\cal T}) \; e^{i\omega_k({\cal T}_0){\cal T}}
 \right].
  \label{newcanheisopfu}
\end{eqnarray}
The operators $\tilde{a}_k({\cal T}) \; ; a_k({\cal T}) $ are related by a 
Bogoliubov transformation. The number of particles referred to the
initial Fock vacuum of the modes $F_k$, is given by
\begin{equation}
N_k({\cal T}) = \langle \tilde{a}^{\dagger}_k({\cal T}) 
\tilde{a}_k({\cal T}) \rangle 
= \frac{1}{4}\left[ \left|\frac{F_k({\cal T})}{F_k({\cal T}_0)}\right|^2 
+ \frac{1}{\omega^2_k({\cal T}_0)} \left| 
\frac{F'_k({\cal T})}{F_k({\cal T}_0)} \right|^2 \right] - \frac{1}{2}
\; , \label{partnumb} 
\end{equation}
 or alternatively, in terms of the comoving mode functions $f_k(t) =
F_k({\cal T})/C({\cal T})$ we find
\begin{equation}
N_k(t) = \frac{a^2(t)}{4} \left[ \left| \frac{f_k(t)}{f_k(t_0)}\right|^2
+ \frac{1}{\omega^2_k(t_0)} \left| \frac{ \dot{f}_k(t) + H(t) f_k(t)}{f_k(t_0)}
\right|^2 \right] - \frac{1}{2} \; . \label{partnumbcomo}
\end{equation}
Using the large $k$-expansion of the conformal mode functions given by
eqs. (\ref{largekf})  we find the large-$k$ behavior
of the particle number to be $N_k \buildrel {k\to \infty} \over =
{\cal{O}}(1/k^4) $, and the 
total number of particles (with reference to the initial state at
${\cal T}_0$) is therefore finite.

{\bf Renormalization.}

  We make our 
subtractions using an ultraviolet cutoff, $\Lambda a(t)$, constant in
{\bf physical coordinates}. This guarantees that the counterterms will be time
independent\cite{din}. The renormalization then proceeds much in the
same manner as in 
reference\cite{frw}; the quadratic divergences renormalize the mass 
and the logarithmic terms renormalize the quartic coupling and the coupling
to the Ricci scalar. In addition, there is a quartic divergence which
renormalizes the cosmological constant while the leading renormalizations
of Newton's constant and the higher order curvature coupling are quadratic and
logarithmic respectively.
The renormalization conditions on the mass, coupling to the Ricci
scalar and coupling constant are obtained from the requirement that the
frequencies that appear in the mode equations are finite\cite{frw}, i.e:
\begin{equation}
m^2_B+\xi_B {\cal R}(t)+\frac{ \lambda_B}{2}\phi^2(t)
+\frac{\lambda_B}{2}\langle\pi^2(t)\rangle_B=
m^2_R+\xi_R {\cal R}(t)+\frac{ \lambda_R}{2}\phi^2(t)
+\frac{\lambda_R}{2}\langle\pi^2(t)\rangle_R, \label{rencond}
\end{equation}
while the renormalizations of Newton's constant, the higher order curvature
coupling, and the cosmological constant are given by the condition of
finiteness of the semi-classical Einstein-Friedmann equation:
\begin{equation}
\frac{G^0_0}{8\pi G_B} + \alpha_B H^0_0 + K_B g^0_0 + 
\langle T^0_0 \rangle_B = \frac{G^0_0}{8\pi G_R} + 
\alpha_R H^0_0 + K_R g^0_0 + \langle T^0_0 \rangle_R \; .
\end{equation}
Finally, we arrive at the following set of renormalizations\cite{din,frw2}:
\begin{eqnarray}
&&\frac{1}{8\pi N G_R} = \frac{1}{8\pi N G_B} 
- {2 \over (4 \pi)^2} \left(\xi_R-\frac16\right)\left[\Lambda^2
-m_R^2\; \ln(\Lambda/\kappa) \right] \; , \;
\lambda_R = \lambda_B - \lambda_R\;\frac{\ln(\Lambda/\kappa)}{(4 
\pi)^2} \; ,\cr \cr 
&&\frac{\alpha_R}{N} = \frac{\alpha_B}{N} 
- \left(\xi_R-\frac16\right)^2\frac{\ln(\Lambda/\kappa)}{(4
\pi)^2}\quad , \quad 
\frac{K_R}{N} = \frac{K_B}{N} - \frac{\Lambda^4}{(4 \pi)^2}
- m_R^2\;\frac{\Lambda^2}{(4 \pi)^2}
+ \frac{m_R^4}{2}\;\frac{\ln(\Lambda/\kappa)}{(4 \pi)^2}\; , \cr \cr 
&& m_R^2 = m_B^2 + \lambda_R\;\frac{\Lambda^2}{(4 \pi)^2}
-\lambda_R\; m_R^2\;\frac{\ln(\Lambda/\kappa)}{(4 \pi)^2}\quad  , \quad 
\xi_R = \xi_B 
- \lambda_R\left(\xi_R-\frac16\right)\frac{\ln(\Lambda/\kappa)}{(4
\pi)^2} \; , \cr \cr 
&& \langle \pi^2(t) \rangle_R = \int \frac{d^3k}{2(2\pi)^3}\left\{|f_k(t)|^2-
\frac{1}{ka^2(t)} + \frac{\Theta(k-\kappa)}{2k^3}
\left[{\cal M}^2(t)-{{{\cal R}(t)}\over 6} \right] \right\} \; . \nonumber
\end{eqnarray}
Here, $\kappa$ is the renormalization point.  As expected, the logarithmic 
terms are consistent with the renormalizations found using dimensional
regularization\cite{baacke,ramsey}.  Again, we set $\alpha_R=0$ and
choose the renormalized cosmological constant such that the vacuum
energy is zero in the true vacuum.  We emphasize that while 
the regulator we have chosen does not respect the covariance of 
the theory, the renormalized energy momentum tensor defined in this 
way nevertheless retains the property of covariant conservation in the
limit when the cutoff is taken to infinity.

The logarithmic subtractions can be neglected because of the coupling
$ \lambda \leq 10^{-12} $.  Using the Planck scale as the cutoff and the
inflaton mass $m_R$ as a renormalization point, these terms are of order
$ \lambda \ln[M_{pl}/m_R] \leq 10^{-10} $, for $ M_{Pl} > m \geq 10^9
\mbox{ GeV }$. An 
equivalent statement is that for these values of the coupling and inflaton
masses, the Landau pole is well beyond the physical cutoff $M_{pl}$.
Our relative 
error in the numerical analysis is of order $10^{-8}$, therefore our numerical
study is insensitive to the logarithmic corrections. Though these corrections
are fundamentally important, numerically they can be neglected. Therefore, in
the numerical computations that follow, we will neglect logarithmic 
renormalization and subtract only
quartic and quadratic divergences in the energy and pressure, and quadratic
divergences in the fluctuation contribution. 

\subsection{Renormalized Equations of Motion for Dynamical Evolution in the 
Large $ N $ limit}

It is convenient to introduce the following dimensionless quantities
and definitions,
\begin{equation}
\tau = m_R \; t \quad ; \quad h(\tau)= \frac{H(t)}{m_R} \quad ; 
\quad q=\frac{k}{m_R} \quad \; \quad
\omega_q = \frac{\Omega_k}{m_R} \quad ; \quad g= \frac{\lambda_R}{8\pi^2} \; ,
\label{dimvars1}
\end{equation}
\begin{equation}
\eta^2(\tau) = \frac{\lambda_R}{2m^2_R} \; \phi^2(t)
\quad ; \quad  g\Sigma(\tau) = \frac{\lambda}{2m^2_R}\; \langle \pi^2(t)
\rangle_R  \quad ; \quad f_q(\tau) \equiv \sqrt{m_R} \; f_k(t) \; .
\label{dimvars3}
\end{equation}

Choosing $\xi_R=0$ (minimal coupling)  and the renormalization
 point $\kappa = |m_R|$ and setting $
 a(\tau_0)=1 $, the equations of motion become for unbroken symmetry:

\begin{eqnarray}\label{modknr}  
&&\left[\frac{d^2}{d \tau^2}+ 3h(\tau) \frac{d}{d\tau}+1+\eta^2(\tau)+
g\Sigma(\tau)\right]\eta(\tau) = 0\; , \label{modcnr} \cr \cr
& &\left[\frac{d^2}{d \tau^2}+3h(\tau)
\frac{d}{d\tau}+\frac{q^2}{a^2(\tau)}+1+\eta^2+g\Sigma(\tau)
\right]f_q(\tau)=0\;,  \\ 
& &  f_q(\tau_0)  =  \frac{1}{\sqrt{\omega_q}} \quad ; \quad 
\dot{f}_q(\tau_0)  = \left[-h(\tau_0)-i\omega_q\right]f_q(\tau_0)\; , \;
\omega_q  =   \left[q^2+1+\eta^2(\tau_0)-\frac{{\cal
R}(\tau_0)}{6m^2_R}+g\Sigma(\tau_0)\right]^{\frac{1}{2}} \; . \nonumber
\end{eqnarray}
We find for broken symmetry,
\begin{eqnarray}
&&\left[\frac{d^2}{d \tau^2}+ 3h(\tau) \frac{d}{d\tau}-1+\eta^2(\tau)+
g\Sigma(\tau)\right]\eta(\tau) = 0\; , \label{modcr}\cr \cr
& &\left[\frac{d^2}{d \tau^2}+3h(\tau)
\frac{d}{d\tau}+\frac{q^2}{a^2(\tau)}-1+\eta^2+g\Sigma(\tau)
\right]f_q(\tau)=0\; ,  \\ 
& &  f_q(\tau_0)  =  \frac{1}{\sqrt{\omega_q}} \quad ; \quad 
\dot{f}_q(\tau_0)  = \left[-h(\tau_0)-i\omega_q\right]f_q(\tau_0)\;, \nonumber \\
& & \omega_q  =  
\left[q^2-1+\eta^2(\tau_0)-\frac{{\cal
R}(\tau_0)}{6m^2_R}+g\Sigma(\tau_0)\right]^{\frac{1}{2}} \; \mbox{ for } \; q^2
> 1-\eta^2(\tau_0)+\frac{{\cal R}(\tau_0)}{6m^2_R}-g\Sigma(\tau_0)\;, \nonumber \\ 
& & \omega_q  =  
\left[q^2+1-\eta^2(\tau_0)+\frac{{\cal
R}(\tau_0)}{6m^2_R}-g\Sigma(\tau_0)\right]^{\frac{1}{2}} \; \mbox{ for } \; q^2
< 1-\eta^2(\tau_0)+\frac{{\cal R}(\tau_0)}{6m^2_R}-g\Sigma(\tau_0) \; .  \label{modkr}  
\end{eqnarray}
Here,
\begin{equation}\label{sigre}
\Sigma(\tau)= \int_0^{\infty} q^2 dq \left[ | f_q(\tau)|^2 - {1 \over
{q\; a(\tau)^2}} + {{\Theta(q - 1)}\over {2 q^3}} \left(\frac{{\cal
M}^2(\tau)}{m^2_R}-{{{\cal{R}(\tau)}}\over{6 m^2_R}}\right)\right] \; .
\end{equation}
both for unbroken and broken symmetry.

The initial conditions for $\eta(\tau)$ will be specified later. 
An important point to notice is that the equation of
motion for the $q=0$ mode coincides with that of the zero mode
(\ref{modcr}). Furthermore, for $\eta(\tau \rightarrow \infty) \neq
0$, a stationary (equilibrium) solution of the eq.(\ref{modcr})  
is obtained for broken symmetry
when the sum rule\cite{us1,mink,De Sitter,frw2}
\begin{equation}
-1+\eta^2(\infty)+g\Sigma(\infty) = 0 \label{sumrule}
\end{equation}
is fulfilled. 

This sum rule is nothing but a proof that Goldstone's
theorem holds here out of thermal equilibrium. In addition, it
is a result of the fact that the large $ N $ approximation 
satisfies the Ward identities associated with the $ O(N) $ symmetry, since
the term  $ -1+\eta^2(\tau)+g\Sigma(\tau) $ is seen to be the
effective mass of the modes transverse to the symmetry breaking
direction, i.e. the Goldstone modes in the broken symmetry phase.

The renormalized dimensionless evolution equations in the Hartree approximation
are very similar to eqs.(\ref{modknr})-(\ref{modkr}). They can be obtained
 just dividing by three the $ \eta^2(\tau) $ term in the zero mode equation
\cite{eri97}. 

\bigskip

In terms
of the zero mode $\eta(\tau)$ and the quantum mode function given
by eq.(\ref{modcr}) we find that the Friedmann equation for the dynamics
of the scale factor in dimensionless variables is given by
\begin{equation}
 h^2(\tau)    =   4h^2_0 \; \epsilon_R(\tau) \quad  ;   \quad h^2_0 =
 \frac{4\pi N m^2_R}{3M^2_{Pl}\lambda_R} \; .\label{eif} 
 \end{equation}
and the renormalized energy and pressure are given by: 
\begin{eqnarray}\label{hubble}
 \epsilon_R(\tau) &  =  &  
\frac{1}{2}\dot{\eta}^2+\frac{1}{4}\left(-1+\eta^2+g\Sigma \right)^2 \\ 
& + &   \frac{g}{2}\int q^2 dq \left[|\dot{f_q}|^2-
{\cal S}^{(1)}(q,\tau)
+\frac{q^2}{a^2}\left(|f_q|^2-\Theta(q - 1)\; {\cal
S}^{(2)}(q,\tau)\right) \right]\; , \nonumber \\
 (p+\varepsilon)_R(\tau)  & = & 
\dot{\eta}^2  
+   g \int q^2 dq \left[|\dot{f_q}|^2- {\cal S}^{(1)}(q,\tau)
+\frac{q^2}{3a^2}\left(|f_q|^2-\Theta(q - 1)\;{\cal S}^{(2)}(q,\tau)\right)
\right] \;, \nonumber
\end{eqnarray}
where the subtractions $ {\cal S}^{(1)} $ and $ {\cal S}^{(2)} $ are given
by 
\begin{eqnarray}
{\cal S}^{(1)} &=&
\frac{k}{a^4(t)}+\frac{1}{2ka^4(t)}\left[B(t)+2\dot{a}^2 \right] \cr \cr  
& + & {1 \over {8 k^3 \; a^4(t) }}\left[ - B(t)^2 - a(t)^2 {\ddot
B}(t) + 3 a(t) {\dot a}(t) {\dot B}(t) - 4 {\dot a}^2(t) B(t) \right]\cr \cr 
{\cal S}^{(2)}&=& \frac{1}{ka^2(t)}- \frac{1}{2k^3
a^2(t)}\;B(t)  + {1 \over {8 k^5 \; a^2(t) }}\left\{  3 B(t)^2 + a(t)
\frac{d}{dt} \left[ a(t) {\dot B}(t) \right]  \right\}\nonumber
\end{eqnarray}
The renormalized energy and pressure are covariantly conserved:
\begin{equation}\label{concov}
{\dot  \epsilon}_R(\tau) + 3 \, h(\tau)\, (p+\varepsilon)_R(\tau) = 0 \; .
\end{equation}
From the evolution of the mode functions that determine the quantum
fluctuations, we can study the growth of correlated domains with the equal time
correlation function,

\begin{eqnarray}\label{correlator}
S(\vec{x},t) & = & \langle\pi(\vec{x},t)\pi(\vec{0},t)\rangle
=  \int\frac{d^3k}{2(2\pi)^3}\; e^{i\vec{k}\cdot\vec{x}}\; |f_k(t)|^2\; ,
\end{eqnarray}
which can be written in terms of the  power spectrum of quantum fluctuations,
$ |f_q(\tau)|^2 $. It is convenient to define the dimensionless
correlation function, 
\begin{equation}
{\cal S}(\rho,\tau) = \frac{S(|\vec x|, t)}{m^2_R}= \frac{1}{{4\pi^2 }\rho}
\int_0^{\infty} q\; dq \; \sin[q\rho] \; |f_q(\tau)|^2  \; ; \; \rho= m_R|\vec
x| \; .  \label{rhocorr} 
\end{equation}

We now have all the ingredients to study the particular cases of interest.

\section{Scalar Field Dynamics in a fixed FRW background}

We consider in this section the evolution of scalar fields in radiation or
matter dominated FRW cosmologies\cite{frw2}. The case for de Sitter
expansion will be discussed in sec. IX \cite{De Sitter}. 

We write the scale factor as $ a(\tau) = (\tau/\tau_0)^n $ with $ n = 1/2 $ and
$ n = 2/3 $ corresponding to radiation and matter dominated
backgrounds, respectively. The value of $ \tau_0 $ determines
the initial Hubble constant since
$$
h(\tau_0) = {{{\dot a}(\tau_0)} \over {a(\tau_0)}} = {n \over \tau_0} \; .
$$
We now solve the system of equations (\ref{evocos}) in the
large $N$ limit.  We begin by presenting an early time analysis of the
slow roll scenario.  We then undertake a thorough numerical investigation of
various cases of interest.  For the symmetry broken case, we also provide an
investigation of the late time behavior of the zero mode and the quantum
fluctuations. We use the dimensionless variables 
(\ref{dimvars1})-(\ref{dimvars3}).

We will assume minimal coupling to
the curvature, $\xi_r = 0$.  In the cases of interest, ${\cal R} \ll \mu^2$, so
that finite $\xi_r$ has little effect.

\subsection{Early Time Solutions for Slow Roll}

For early times in a slow roll scenario [$m^2=-\mu^2$, $\eta(\tau_0)
\ll 1$], we 
can neglect in eqs.(\ref{modkr}) both the
quadratic and cubic terms in $\eta(\tau)$ as well as 
the quantum fluctuations $\langle\pi^2(\tau)\rangle_r $ [recall that
$\langle\pi^2(\tau_0)\rangle_r = 0 $]. Thus, the differential equations for the
zero mode and the mode functions (\ref{evocos})
become linear equations. In terms of the scaled variables
introduced above, with $ a(\tau)= \tau^n $ ($ n=2/3 $ for a matter dominated
cosmology while $ n=1/2 $ for a radiation dominated cosmology) we have:
\begin{eqnarray}
\ddot{\eta}(\tau)+\frac{3n}{\tau} \;\dot{\eta}(\tau)-\eta(\tau) & = & 0 \; ,
\label{earlyeta} \\ \nonumber \\
\left[\frac{d^2}{d\tau^2}+\frac{3n}{\tau}\frac{d}{d\tau}
+k^2\;\left(\frac{\tau_0}{\tau}\right)^{2n}-1\right]U_k(\tau) & = & 0
\;. \label{earlyuk} 
\end{eqnarray}

The solutions to the zero mode equation (\ref{earlyeta}) are
\begin{equation}
\eta(\tau)=c\; \tau^{-\nu}I_{\nu}(\tau)+d\; \tau^{-\nu}K_{\nu}(\tau) \; ,
\label{earlyetasoln}
\end{equation}
where $ \nu \equiv (3n-1)/2 $, and $ I_{\nu}(\tau) $ and $ K_{\nu}(\tau)
$ are modified 
Bessel functions.  The coefficients, $ c $ and $ d $, are determined
by the initial conditions on $ \eta $.  For $ \eta(\tau_0)=\eta_0 $ and
$ \dot{\eta}(\tau_0)=0 $, we have: 
\begin{eqnarray}
&&c =  \eta_0  \; \tau_0^{\nu+1} \left[
\dot{K}_{\nu}(\tau_0)-\frac{\nu}{\tau_0}K_{\nu}(\tau_0)\right] \quad ,\quad 
d  =  -\eta_0  \; \tau_0^{\nu+1}
\left[\dot{I}_{\nu}(\tau_0)-\frac{\nu}{\tau_0}I_{\nu}(\tau_0)\right] \; .
\nonumber
\end{eqnarray}
Taking the asymptotic forms of the modified Bessel functions, we find that for
intermediate times $\eta(\tau)$ grows as
\begin{equation}
\eta(\tau) \stackrel{\tau \gg 1}{=} {c \over {\sqrt{2 \pi}}}
\; \tau^{-3n/2}\; e^\tau\left[1-\frac{9n^2-6n}{8\tau}+
{\cal O}({1\over{\tau^2}})\right].
\label{asymeta}
\end{equation}
We see that $\eta(\tau)$ grows very quickly in time, and the approximations
(\ref{earlyeta}) and (\ref{earlyuk}) will quickly break down.  For the case
shown in fig.1 (with $n=2/3$, $\eta(\tau_0)=10^{-7}$, and
$\dot{\eta}(\tau_0)=0$), we find that this approximation is valid up
to $\tau-\tau_0 \simeq 10$. 

The equations for the mode functions ({\ref{earlyuk}) can be solved in closed
form for the modes in the case of a radiation dominated cosmology with $n=1/2$.
The solutions are
\begin{equation}
U_k(\tau) = c_k \; e^{-\tau}\, U\left(\frac34-\frac{k^2}{2},\frac32,2\tau\right) +d_k \;
e^{-\tau}\, M\left(\frac34-\frac{k^2}{2},\frac32,2\tau\right).
\label{earlyuksoln}
\end{equation}
Here, $U(\cdot)$ and $M(\cdot)$ are confluent hypergeometric functions
\cite{aands} (in another common notation, $M(\cdot) \equiv\; _1F_1(\cdot)$),
and the $c_k$ and $d_k$ are coefficients determined by the initial conditions
(\ref{condini}) on the modes.  The solutions can also be written in terms of
parabolic cylinder functions.

For large $ \tau $ we have the asymptotic form
\begin{equation}
U_k(\tau) \stackrel{\tau \gg 1}{=} d_k \; e^\tau \; (2\tau)^{-(3/4+k^2 \tau_0 /2)} \;
\frac{\sqrt{\pi}}{2\;\Gamma\left(\frac34-\frac{k^2 \tau_0 }{2}\right)}
\left[1+{\cal O}({1\over{\tau}})\right]
+ c_k \; e^{-\tau} \;  (2\tau)^{(-3/4+k^2 \tau_0 /2)}\left[1+{\cal
O}({1\over{\tau}})\right]\; . \label{asymuk}
\end{equation}
Again, these expressions only apply for intermediate times before the
nonlinearities have grown significantly.

\subsection{Numerical Analysis}

We now present the numerical analysis of the dynamical evolution of
scalar fields in time dependent, matter and radiation dominated cosmological
backgrounds.  We use initial values of the Hubble constant such that 
$ h(\tau_0) \geq 0.1 $.  For expansion
rates much less than this value the evolution will look similar to
Minkowski space, which has been studied in great detail elsewhere
\cite{us1,mink,simrota}.  As will be seen, the equation of state found
numerically is, in the majority of cases, that of cold matter.  We therefore
use matter dominated expansion for the evolution in much of the analysis that
follows.  The evolution in radiation
dominated universes remains largely unchanged, although there is greater
initial growth of quantum fluctuations due to the scale factor growing more
slowly in time.  Using the large $N$ approximation to study 
theories with continuous and discrete symmetries respectively, we treat three
important cases.  They are 1) $ m^2<0, \; \eta(\tau_0)\ll 1 ; \; 2)\; m^2<0,
\; \eta(\tau_0)\gg 1; \; 3) \; m^2>0, \; \eta(\tau_0)\gg 1$.

In presenting the figures, we have shifted the origin of time such that
$\tau \to \tau'=\tau-\tau_0$.  This places the initial time, $\tau_0$, at
the origin. 
In these shifted coordinates, the scale factor is given by 
$$
a(t)=\left(\frac{\tau+\tau_0}{\tau_0}\right)^n,
$$ 
where, once again, $n=2/3$ and $n=1/2$ in matter and radiation dominated
backgrounds respectively, and the value of $\tau$ is determined by the 
Hubble constant at the initial time:
$$
h(\tau_0)=\frac{n}{\tau_0}.
$$

{\bf Case 1: $m^2<0$, $\eta(\tau_0)\ll 1$}.  This is the case of an
early universe 
phase transition in which there is little or no biasing in the initial
configuration (by biasing we mean that the initial conditions break the $ \eta
\to -\eta $ symmetry).  The transition occurs from an initial temperature above
the critical temperature, $T>T_c$, which is quenched at $t_0$ to the
temperature $T_f \ll T_c$.  This change in temperature due to the rapid
expansion of the universe is modeled here by an instantaneous change in the
mass from an initial value $m_i^2=T^2/T_c^2-1$ to a final value
$m_f^2=-1$.  We will use the value $m_i^2=1$ in what follows.
This quench approximation is necessary since the low momentum
frequencies (\ref{frec}) appearing in our initial conditions (\ref{condini}) 
are complex for negative mass squared and small $\eta(t_0)$.  An alternative
choice is to use initial frequencies given by
$$
\omega_k(\tau_0)=\left[k^2+{\cal{M}}^2(\tau_0)\tanh\left(
\frac{k^2+{\cal{M}}^2(\tau_0)}{|{\cal{M}}^2(\tau_0)|}\right)\right]^{1/2}.
$$
These frequencies have the attractive feature that they match the conformal
adiabatic frequencies given by eq.(\ref{frec}) for large values of $k$ while 
remaining positive for small $k$.  We find that such a choice of initial
conditions changes the quantitative value of the particle number by a few 
percent, but leaves the qualitative results unchanged.

We plot the the zero mode $ \eta(\tau) $, the equal time correlator $
g\Sigma(\tau) $, the total number of produced particles $ gN(\tau) $
(see sec. VI for a discussion of our definition of particles), the
number of particles $ gN_k(\tau) $ as a function of wavenumber for
both intermediate and late times, and the ratio of the pressure
and energy densities $ p(\tau)/\varepsilon(\tau) $ (giving the
equation of state). 

Figs. 1a-e shows these quantities in the large $N$ approximation for a matter
dominated cosmology with an initial condition on the zero mode given by
$ \eta(\tau_0)=10^{-7}, \; \dot{\eta}(\tau_0)=0$ and for an initial
expansion rate of $ h(\tau_0)=0.1 $.  This choice for
the initial value of $\eta$ stems from the fact that the quantum fluctuations
only have time to grow significantly for initial values satisfying
$ \eta(\tau_0) \ll  \sqrt{g} $; for values $ \eta(\tau_0) \gg \sqrt{g} $
the evolution is essentially classical.  This result is clear from the
intermediate time dependence of the zero mode and the low momentum
mode functions given by the expressions (\ref{asymeta}) and
(\ref{asymuk}) respectively. 

\begin{figure}
\epsfig{file=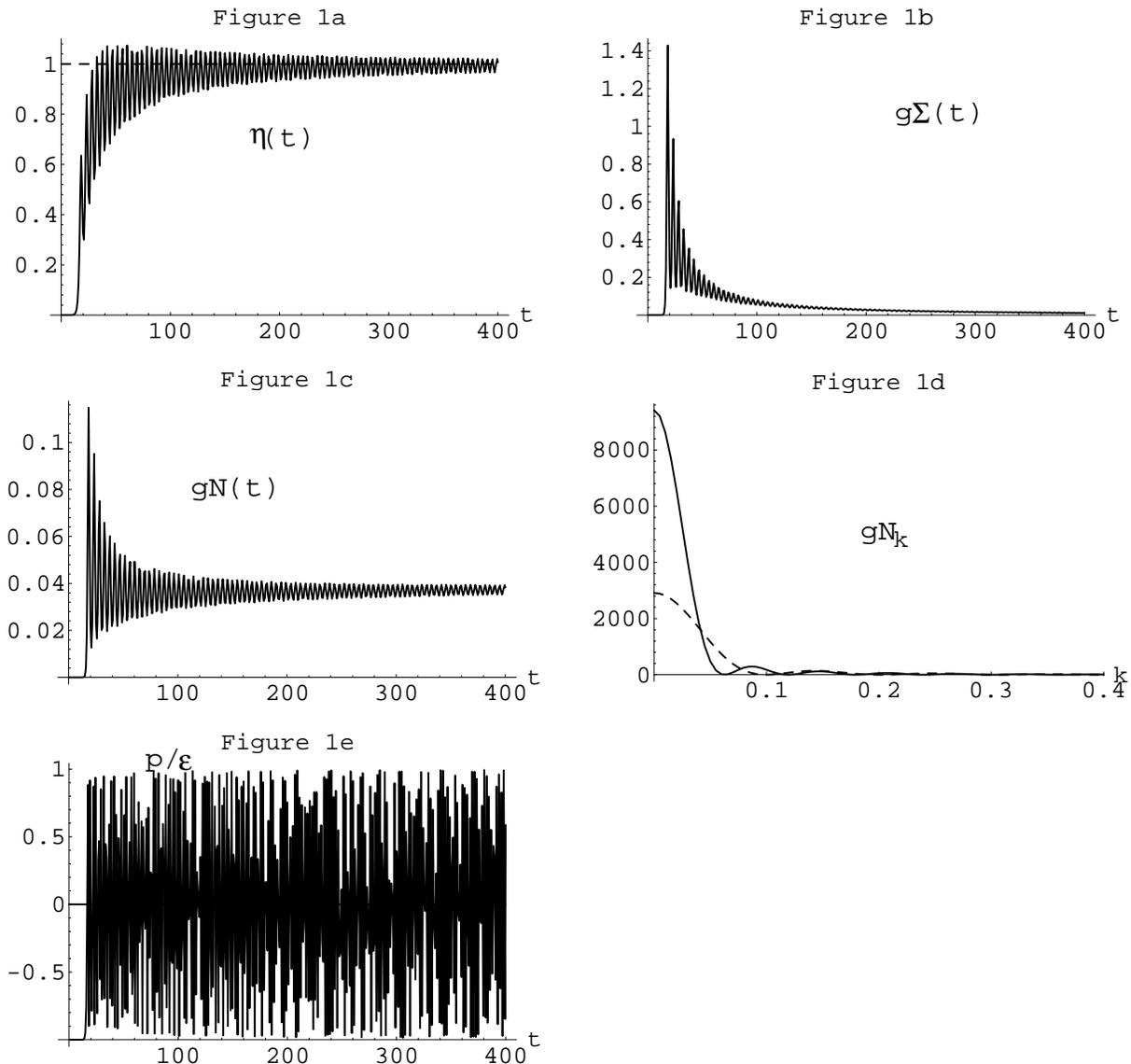}
\caption{Symmetry broken, slow roll, large $N$, matter dominated
evolution of (a) the zero mode $\eta(\tau)$ vs. $\tau$, (b) the
quantum fluctuation 
operator $ g\Sigma(\tau) $ vs. $ \tau $, (c) the number of particles
$ gN(\tau) $ vs. $ \tau $, (d) the particle distribution $ gN_k(\tau)
$ vs. $ k $ at $ \tau=149.1 $ (dashed line)
and $ \tau=398.2 $ (solid line),  and (e) the ratio of the pressure and
energy density $ p(\tau)/\varepsilon(\tau) $ vs. $ t $ for the  values,
$ \eta(\tau_0) = 10^{-7}, \; \dot{\eta}(\tau_0)=0, \; g = 10^{-12},\;
h(\tau_0) = 0.1 $. \label{fig1}} 
\end{figure}

After the initial growth of the fluctuation $ g\Sigma(\tau) $ (fig.1b)
we see that the zero mode 
(fig.1a) approaches the value given by the minimum of the tree level
potential, $ \eta=1 $, while $ g\Sigma(\tau) $ decays for late times as
$$
g\Sigma(\tau) \simeq {{\cal C}\over{a^2(\tau)}} ={{\cal
C}\over{\tau^{4/3}}} \; . 
$$
For these late times, the Ward identity corresponding to the $O(N)$ symmetry of
the field theory is satisfied, enforcing the condition
\begin{equation}
-1 + \eta^2(\tau) + g\Sigma(\tau) = 0.
\label{ward}
\end{equation} 
Hence, the zero mode approaches the classical minimum as
$$
\eta^2(\tau) \simeq 1 -{{\cal C}\over{a^2(\tau)}} \; .
$$ 

Figure 1c depicts the number of particles produced.  After an initial burst of 
particle production, the number of particles settles down to a relatively
constant value.  Notice that the number of particles produced is approximately
of order $ 1/g $. In fig.1d, we show the number of particles as a
function of the wavenumber, $ k $.  For intermediate times we see the
simple structure depicted by the dashed line in the figure, while for
late times this quantity becomes concentrated more at low values of
the momentum $ k $.   

Finally, fig.1e shows that the field begins with a de Sitter equation of 
state $p=-\varepsilon$ but evolves quickly to a state dominated by
ordinary matter,  
with an equation of state (averaged over the oscillation timescale) $p=0$.
This last result is a bit surprising as one expects from the condition
(\ref{ward}) that the particles produced in the final state are massless
Goldstone bosons  which should have the equation of state of radiation.
However, as shown in fig.1d, the produced particles are of low momentum,
$ q \ll 1 $, and while the effective mass of the particles is zero to
very high  
accuracy when averaged over the oscillation timescale, the effective mass 
makes small oscillations about zero so that the dispersion relation for these
particles differs from that of radiation.  In addition, since the produced
particles have little energy, the contribution to the energy density from
the zero mode, which contributes to a cold matter equation of state, remains
significant.

Finally, we show the special case in which
there is no initial biasing in the field, $\eta(\tau_0)=0, \;
\dot{\eta}(\tau_0)=0 $, and $ h(\tau_0)=0.1 $ in figs.
2a-d.  The zero mode remains zero for all time, so that the quantity
$ g\Sigma(\tau) $ (fig.2a) satisfies the sum rule (\ref{ward}) by reaching 
the value one without decaying for late times. 
Notice that many more particles are produced in this case (fig 2b); the growth
of the particle number for late times is due to the expansion of 
the universe.  The particle distribution (fig.2c) is similar to that of the 
slow roll case in fig.1.  The equation of state (fig.2d) is likewise
similar.

In each of these cases of slow roll dynamics, increasing the Hubble constant
has the effect of slowing the growth of both $\eta$ and $g\Sigma(t)$.
The equation of state will be that of a de Sitter universe for a
longer period before moving to a matter dominated equation of state.
Otherwise, the dynamics is much the same as in figs. 1-3.

{\bf Case 2: $ m^2<0$, $ \eta(\tau_0)\gg 1 $}.  We now examine the case of 
 a chaotic inflationary scenario with a symmetry broken potential.
In chaotic inflation, the zero mode begins with a value $
 \eta(\tau)\gg 1 $.  During
the de Sitter phase, $ h \gg 1 $, and the field initially evolves classically,
dominated by the first order derivative term appearing in the zero mode
equation [see eq.(\ref{evocos})].  
Eventually, the zero mode
rolls down the potential, ending the de Sitter phase and beginning the
FRW phase.  We consider the field dynamics in the FRW universe after
the end of inflation. We thus take the initial temperature to be zero,
 $ T=0 $.

\begin{figure}
\epsfig{file=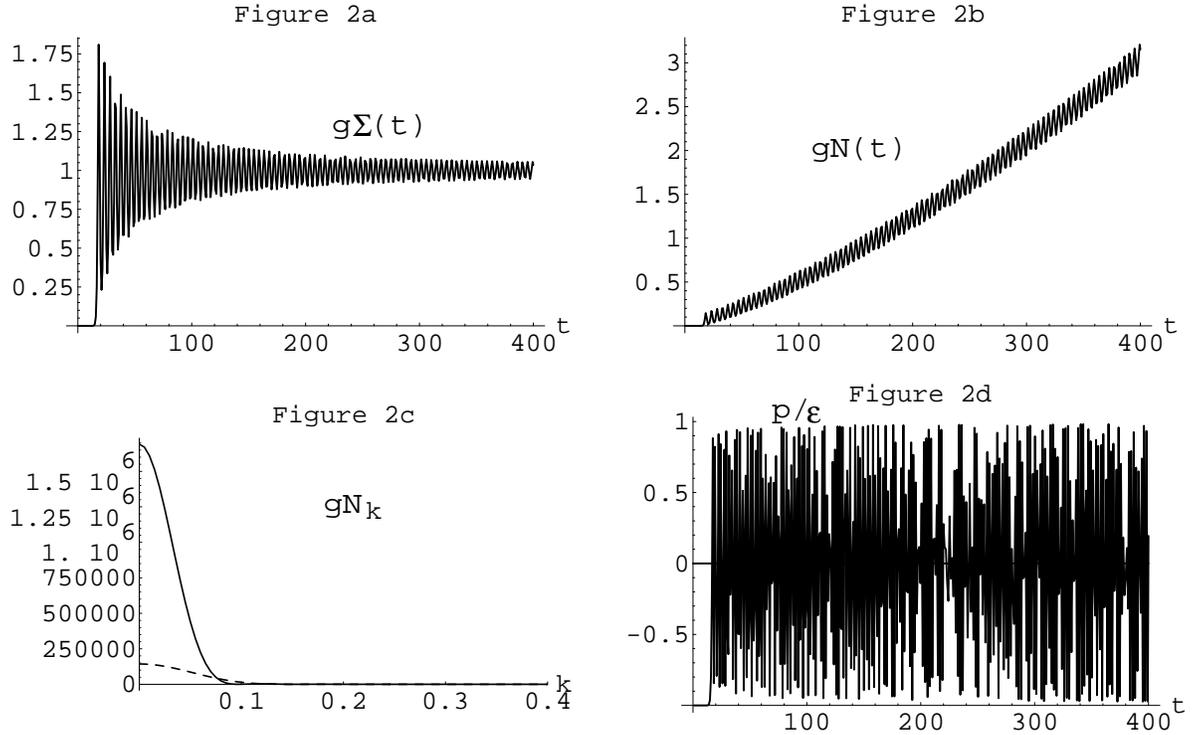}
\caption{Symmetry broken, $ \eta(\tau) \equiv 0 $, matter dominated
evolution of (a) the quantum fluctuation operator $g\Sigma(\tau)$ vs. $\tau$, 
(b) the number of particles $ gN(\tau) $ vs. $\tau$,
(c) the particle distribution $ gN_k(\tau) $ vs. $k$ at $ \tau=150.1 $
(dashed line) and $ \tau =397.1 $ (solid line),  and (d)
the ratio of the pressure and energy density
$p(\tau)/\varepsilon(\tau)$ vs. $\tau $ for
the  values $ \eta(\tau_0) = 0, \; \dot{\eta}(\tau_0)=0, \; g =
10^{-12}, \; h(\tau_0) = 0.1 $. \label{fig3}}
\end{figure}

Figure 4 shows our results for the quantities, $ \eta(\tau) , \;
g\Sigma(\tau), \; gN(\tau), \; gN_k(\tau) $ and $
p(\tau)/\varepsilon(\tau) $ for the evolution in  
the large $ N $ approximation within a {\em radiation} dominated gravitational
background with $ h(\tau_0)=0.1 $.  The initial condition on the zero mode
is chosen to have the representative value $ \eta(\tau_0)=4 $ with 
$ \dot{\eta}(\tau_0)=0 $.  
Initial values of the zero mode much smaller than this will not produce
significant growth of quantum fluctuations; initial values larger than this
produces qualitatively similar results, although the resulting number of 
particles will be greater and the time it takes for the zero mode to settle
into its asymptotic state will be longer. 

We see from fig.3a that the zero
mode oscillates rapidly, while the amplitude of the oscillation decreases due
to the expansion of the universe.

This oscillation induces particle production
through the process of parametric amplification (fig.3c) and causes the
fluctuation $g\Sigma(\tau)$ to grow (fig.3b).  Eventually, the zero mode loses
enough energy that it is restricted to one of the two minima of the tree level
effective potential.  The subsequent evolution closely follows that of Case 1
above with $g\Sigma(\tau)$ decaying in time as $ 1/a^2(\tau) \sim 1/\tau$ 
with $\eta(\tau)$ given by
the sum rule (\ref{ward}).  The spectrum (fig.3d) indicates a single unstable
band of particle production dominated by the modes $k=1/2$ to about $k=3$ 
for late times.  The structure within this band becomes more complex with time 
and shifts somewhat toward lower momentum modes.  
\begin{figure}
\epsfig{file=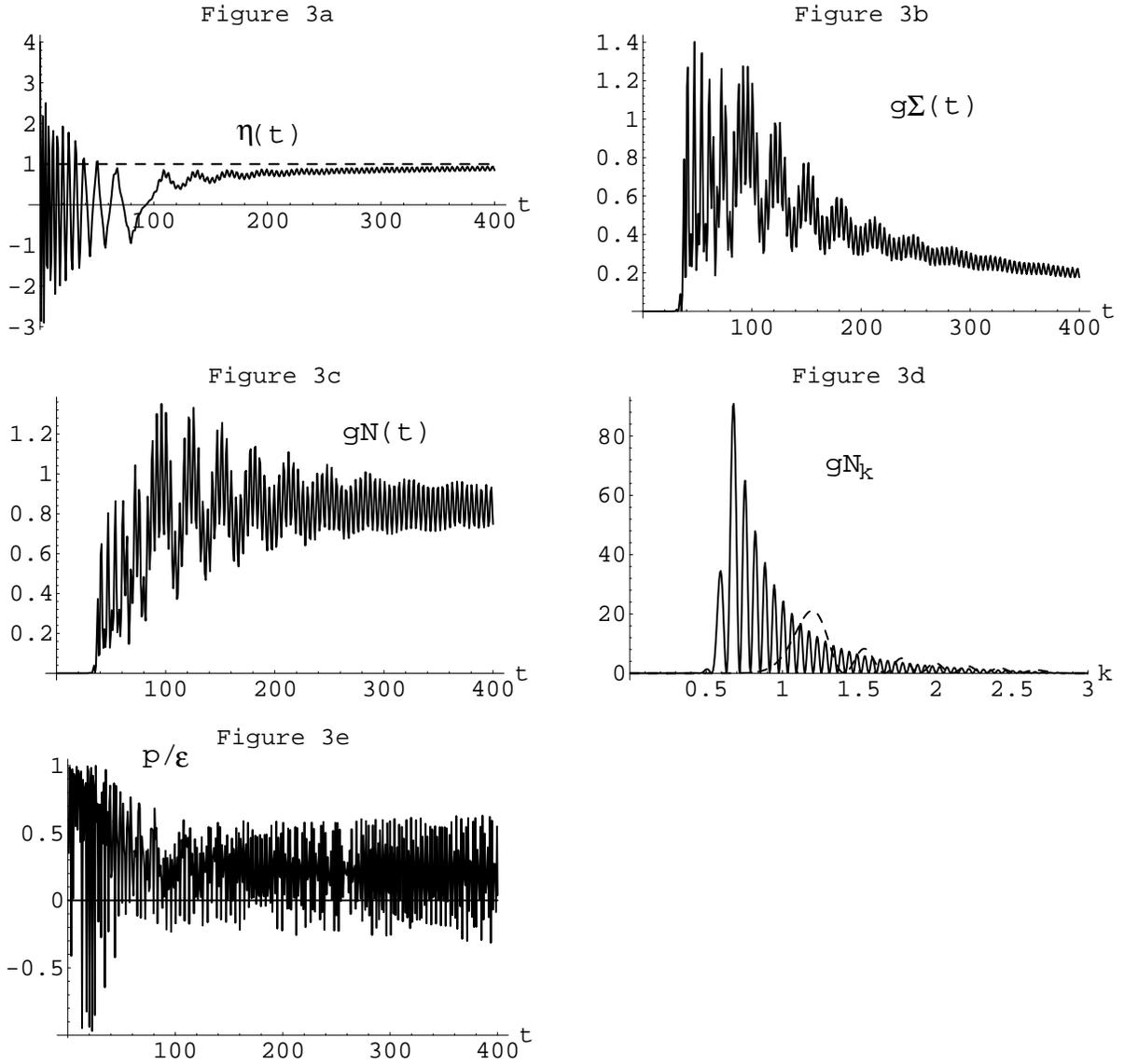}
\caption{Symmetry broken, chaotic, large $N$, radiation dominated
evolution of (a) the zero mode $ \eta(\tau) $ vs. $ \tau $, (b) the
quantum fluctuation 
operator $ g\Sigma(\tau) $ vs. $ \tau $, (c) the number of particles $
gN(\tau)$ vs. $ \tau $, 
(d) the particle distribution $ gN_k(\tau) $ vs. $ k $ at $ \tau=76.4
$ (dashed line) 
and $ \tau=392.8 $ (solid line),  and (e) the ratio of the pressure and energy
density $ p(\tau)/\varepsilon(\tau) $ vs. $ \tau $ for the  values
$ \eta(\tau_0) = 4, \; \dot{\eta}(\tau_0)=0, \; g = 10^{-12}, h(\tau_0) =
0.1 $. \label{fig4}}
\end{figure}

Such a shift is also  observed in Minkowski spacetimes \cite{us1,mink}. 
Figure 4e shows the equation of
state which we see to be somewhere between the relations for matter and
radiation for times out as far as $t=400$, but slowly moving to a matter
equation of state.  Since matter redshifts as $1/a^3(t)$ while radiation
redshifts as $1/a^4(t)$, the equation of state should eventually become matter
dominated.  Given the equation of state indicated by fig.3e, we estimate that
this occurs for times of order $t=10^4$.  The reason the equation of state
in this case differs from that of cold matter as was seen in figs. 1-3 is 
that the particle distribution 

\begin{figure}
\epsfig{file=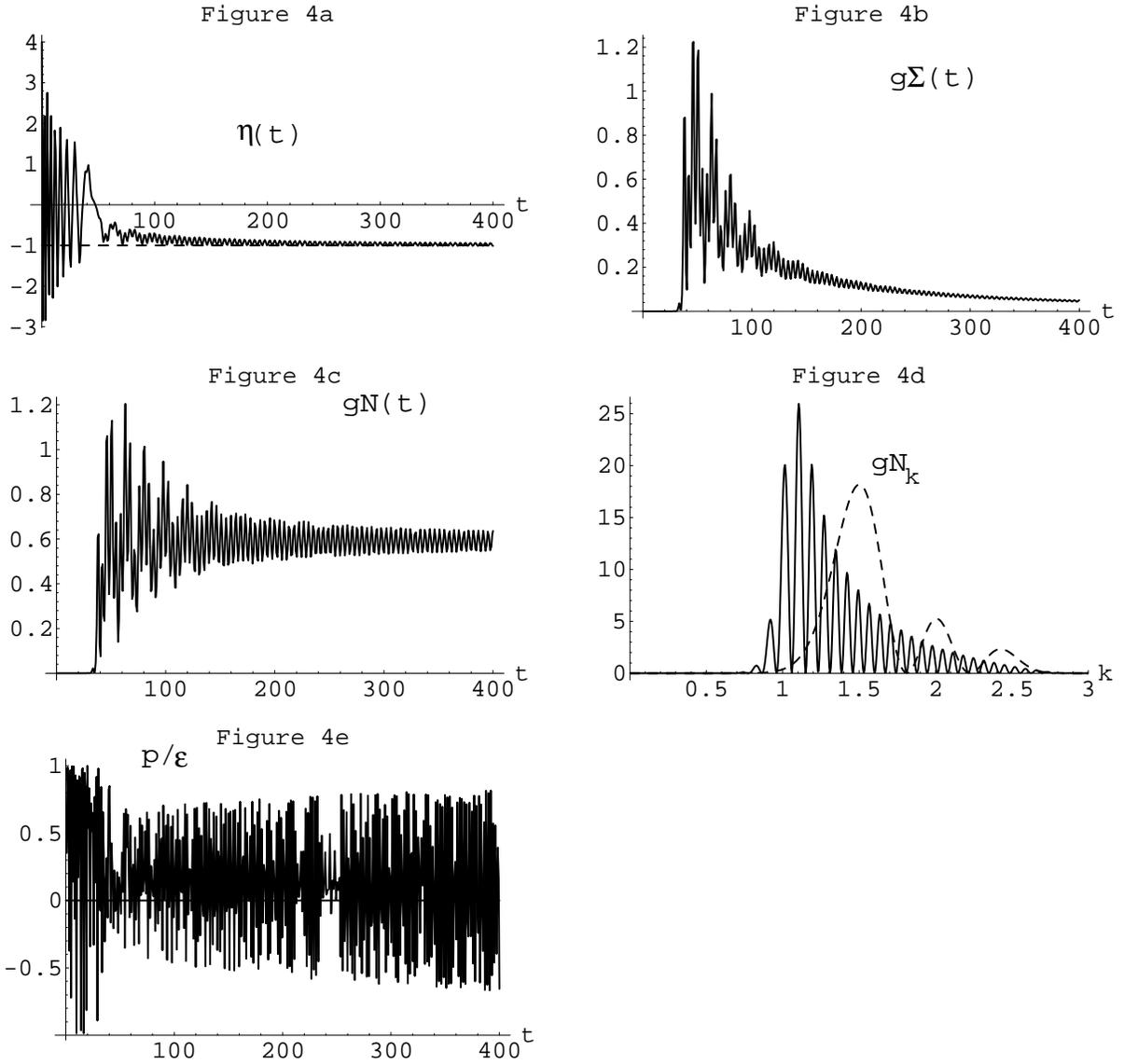}
\caption{Symmetry broken, chaotic, large $N$, matter dominated
evolution of (a) the zero mode $\eta(\tau)$ vs. $\tau$, (b) the
quantum fluctuation 
operator $g\Sigma(\tau)$ vs. $\tau$, (c) the number of particles
$gN(\tau)$ vs. $\tau$, 
(d) the particle distribution $gN_k(\tau)$ vs. $k$ at $t=50.8$ (dashed line)
and $\tau=399.4$ (solid line),  and (e) the ratio of the pressure and
energy density $ p(\tau)/\varepsilon(\tau) $ vs. $\tau$ for the parameter
values $ m^2=-1$, $\eta(\tau_0) = 4 $, $ \dot{\eta}(\tau_0)=0 $, $ g =
10^{-12} $, $ h(\tau_0) = 0.1 $. \label{fig5}}
\end{figure}
\noindent
produced by parametric amplification is 
concentrated at higher momenta, $k \simeq 1$. 

Figure 4 shows the corresponding case with a matter dominated background.  The
results are qualitatively very similar to those described for fig.3 above.
Due to the faster expansion, the zero mode (fig.4a) finds one of the two wells
more quickly and slightly less particles are produced.  For late times, the
fluctuation $g\Sigma(t)$ (fig.4b) decays as $1/a^2(t) \propto 1/t^{4/3}$.  
Again we see an equation of state (figs. 4e) which evolves from a
state between that of pure radiation or matter toward one of cold matter.

A larger Hubble constant prevents significant particle production unless the 
initial amplitude of the zero mode is likewise increased such that the relation
$\eta(t_0) \gg h(t_0)$ is satisfied. For very large amplitude
$\eta(t_0) \gg 1$,  to the extent that the mass term can be neglected
and while the quantum fluctuation term has not grown to be large, the
equations of motion (\ref{modcer}) are scale invariant with the
scaling $\eta \to \mu 
\eta$, $H \to \mu H$, $t \to t/\mu$, and $k \to \mu k$, where $\mu$ is
an arbitrary scale. 

{\bf Case 3: $m^2>0$, $\eta(t_0)\gg 1$}.  The final case we examine is that of
a simple chaotic scenario with a positive mass term in the Lagrangian.  Again,
the FRW stage occurs after the inflationary expansion; this allows us
to take  zero  initial temperature. 

Figure 5 shows this situation in the large $N$ approximation for a matter
dominated cosmology.  The zero mode, $\eta(\tau)$, oscillates in time while
decaying in amplitude from its initial value of $\eta(\tau_0\! =\! 0)=5$,
$\dot{\eta}(\tau_0\! =\! 0)=0$ (fig.5a), while the quantum fluctuation,
$g\Sigma(t)$, grows rapidly for early times due to parametric resonance
(figs. 5b).  We choose here an initial condition on the zero mode which
differs from that of figs 2-3 above since there is no significant growth
of quantum fluctuations for smaller initial values.  From fig.5d, we
see that there exists a single unstable band at values of roughly $k=1$ to
$k=3$, although careful examination reveals that the unstable band 
extends all the way to $k=0$.  The equation of state is depicted by the 
quantity $p(\tau)/\varepsilon(\tau)$ in fig.5e. As expected in this massive 
theory, the equation of state is matter dominated.

First, we note that, for early times when $g\Sigma(\tau) \ll 1$, the zero mode 
is well fit by the function $\eta(\tau)=\eta_0 f(\tau)/a(\tau)$ where
$ f(\tau) $ is an  
oscillatory function taking on values from $-1$ to $1$.  This is clearly seen
from the envelope function $\eta_0/a(\tau)$ shown in fig.6a (recall that 
$g\Sigma(\tau) \ll 1$ during the entire evolution in this case).  Second, the 
momentum that appears in the equations for the modes (\ref{evocos}) is the
{\em physical} momentum $k/a(\tau)$.  We therefore write the approximate 
expressions for the locations of the forbidden bands in FRW by using the
Minkowski results of \cite{mink} with the substitutions 
$\eta_0^2 \to \gamma\eta_0^2/a^2(\tau)$ (where the factor of $\gamma$ accounts
for the difference in the definition of the non-linear coupling between
this study and \cite{mink}) and $q^2 \to k^2/a^2(\tau)$.  

Making these substitutions, we find for the location in comoving momentum $k$ 
of the forbidden band in the large $N$ (fig.5-6)  case:
\begin{eqnarray}\label{hartband}
0 \leq & k^2 & \leq \frac{\eta_0^2}{2}  \; .
\end{eqnarray}
The important feature to notice is that while the location of the unstable
band (to a first approximation) in the case of the continuous $O(N)$ theory
is the same as in Minkowski and does not change in time. Again, the
qualitative dynamics remains largely unchanged from the case of a
smaller Hubble constant. 

As in the symmetry broken case of figs. 1-4, the equations of motion for
large amplitude and relatively early times are approximately scale invariant.
In fig.6 we show the case of the large $N$ evolution in a radiation
dominated universe with initial Hubble constant of $H(\tau_0)=2$ 
 with appropriately scaled initial value of the zero mode of
 $\eta(\tau_0)=16$.

\begin{figure}
\epsfig{file=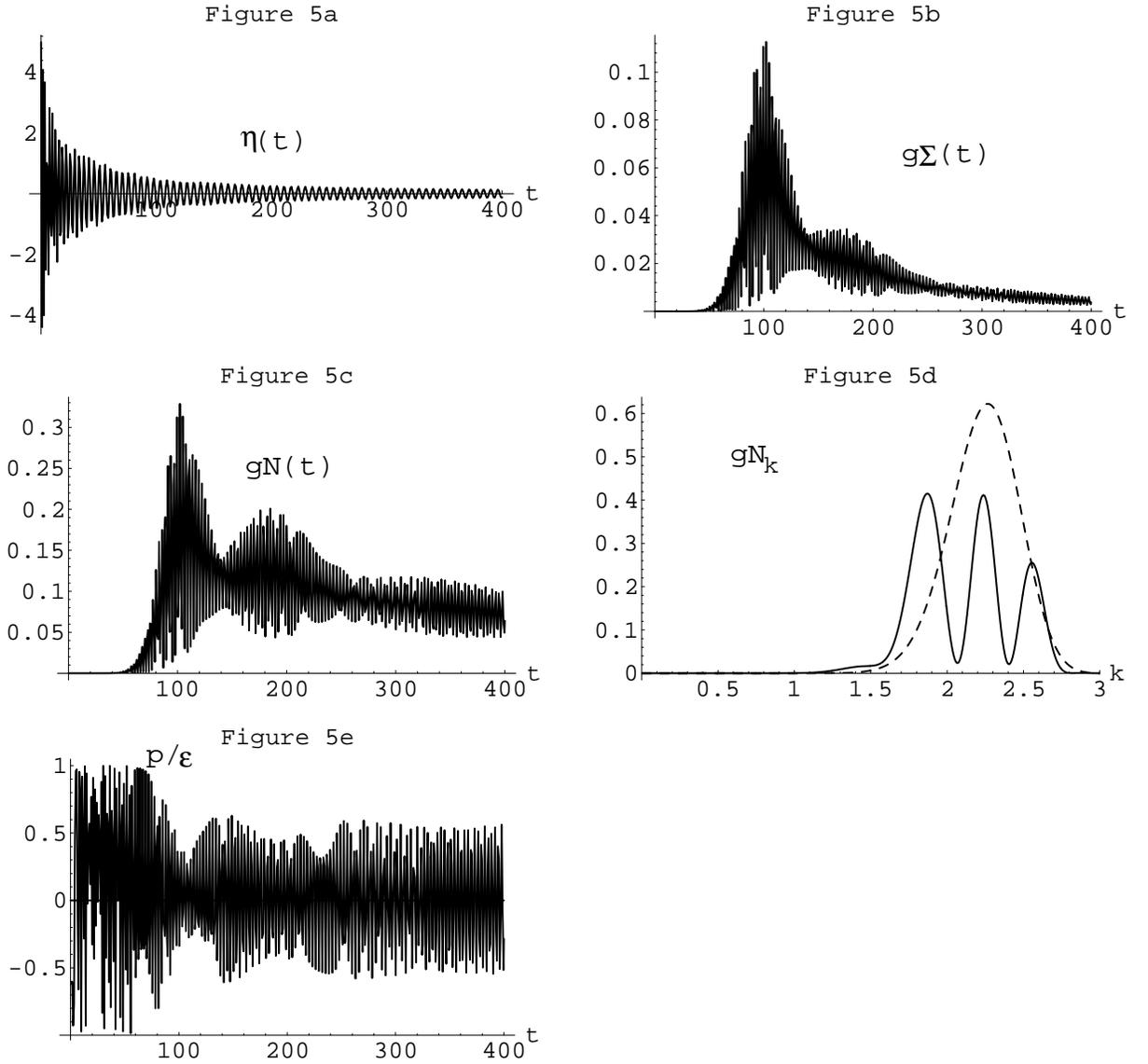}
\caption{Symmetry unbroken, chaotic, large $N$, matter dominated
evolution of (a) the zero mode $\eta(\tau)$ vs. $\tau$, (b) the
quantum fluctuation 
operator $g\Sigma(\tau)$ vs. $\tau$, (c) the number of particles
$gN(\tau)$ vs. $\tau$, 
(d) the particle distribution $gN_k(\tau)$ vs. $k$ at $\tau=77.4$ (dashed line)
and $\tau=399.7$ (solid line),  and (e) the ratio of the pressure and
energy density  
$p(\tau)/\varepsilon(\tau)$ vs. $\tau$ for the parameter values
$m^2=+1$, $\eta(\tau_0) = 
5$, $\dot{\eta}(\tau_0)=0$, $g = 10^{-12}$, $H(\tau_0) = 0.1$. \label{fig7}}
\end{figure}

\subsection{Late Time Behavior}
We see clearly from the numerical evolution that in the case of a symmetry
broken potential, the late time large $N$ solutions obey the sum rule
(\ref{sumrule}). This sum rule is a consequence of the late time Ward
identities which enforce Goldstone's Theorem.  Because of this sum
rule, we can write down the analytical expressions for the late time
behavior of  the fluctuations and the zero mode.  Using
eq.(\ref{sumrule}), the mode equation (\ref{evocos}) becomes 

\begin{figure}
\epsfig{file=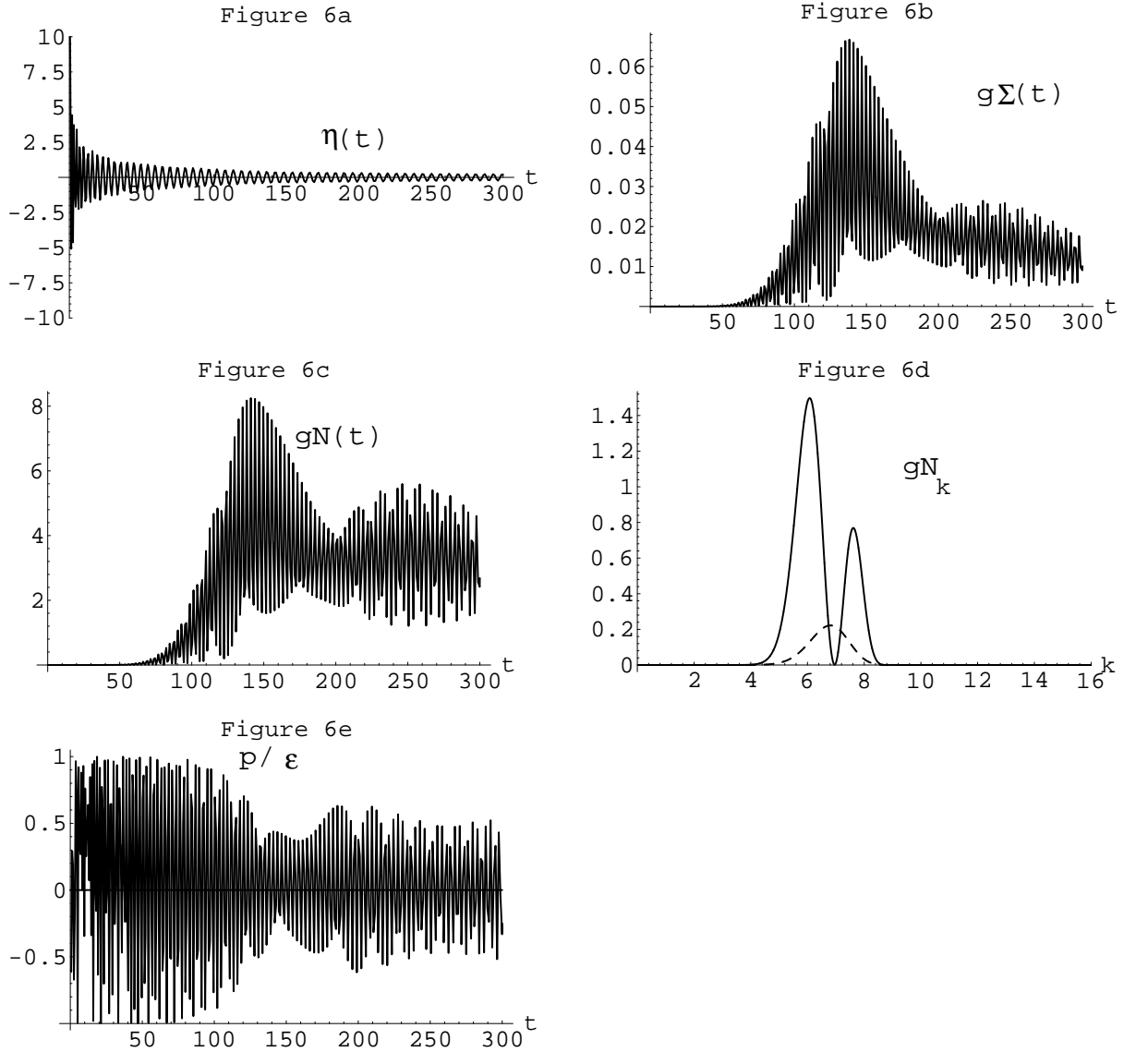}
\caption{Symmetry unbroken, chaotic, large $N$, radiation dominated
evolution of (a) the zero mode $\eta(\tau)$ vs. $\tau$, (b) the
quantum fluctuation operator $g\Sigma(\tau)$ vs. $\tau$, (c) the
number of particles $gN(\tau)$ vs. $\tau$, 
(d) the particle distribution $gN_k(\tau)$ vs. $k$ at $\tau=102.1$
(dashed line) and $\tau=251.6$ (solid line),  and (e) the ratio of the
pressure and energy density $ p(\tau)/\varepsilon(\tau)$ vs. $\tau$
for the parameter values $m^2=+1$, $\eta(\tau_0) = 
16$, $\dot{\eta}(\tau_0)=0$, $g = 10^{-12}, \, h(\tau_0) = 2.0$. \label{fig12}}
\end{figure} 
\begin{equation}
\left[\frac{d^2}{d\tau^2}+3\frac{\dot{a}(\tau)}{a(\tau)}
\frac{d}{d\tau}+\frac{k^2}{a^2(\tau)} \right]U_k(\tau) = 0 \; . \label{lateuk}
\end{equation}
This equation can be solved exactly if we assume a power law dependence for the
scale factor $ a(\tau) = (\tau/\tau_0)^n $ with solution
\begin{equation}
U_k(\tau) = c_k \; \tau^{(1-3n)/2} \;
	J_{\frac{1-3n}{2-2n}}\left(\frac{k \tau_0^n 
	\tau^{1-n}}{n-1}\right) + 
	d_k \; \tau^{(1-3n)/2} \; Y_{\frac{1-3n}{2-2n}}\left(\frac{k \tau_0^n
	\tau^{1-n}}{n-1}\right), 
\label{latesoln}
\end{equation}
where $J_{\nu}$ and $Y_{\nu}$ are Bessel and Neumann functions respectively, 
and the constants $c_k$ and $d_k$ 
carry dependence on the initial conditions and on the dynamics up to the point
at which the sum rule is satisfied.  

These functions have several important properties.  In particular, in radiation
or matter dominated universes, $n<1$, and for values of wavenumber satisfying
$k \gg \tau^{-(1-n)}/\tau_0^n$, the mode functions decay in time as 
$1/a(\tau) \sim \tau^{-n}$.  Since
the sum rule applies for late times, $\tau-\tau_0 \gg 1$ in dimensionless units, we see
that all values of $k$ except a very small band about $k=0$ redshift as
$1/a(\tau)$.  The $k=0$ mode for any scale factor $ a(\tau) $ takes the form
$$
U_0(\tau) = C_1 + C_2 \; \int^\tau {d\tau' \over a^3(\tau')} \; .
$$
where we used eq.(\ref{lateuk}) and $ C_1 $ and $ C_2 $ are constants
depending on the initial conditions.

We see that the $k=0$ mode {\bf freezes out} for late times tending to
a constant. This explains the support evidenced in the numerical
results for values of small $k$ (see figs. 1,3).  

These results mean that the quantum fluctuation has a late time
dependence of $\langle \pi^2(\tau)\rangle_r \sim 1/a^2(\tau)$.  The late time
dependence of the zero mode is given by this expression combined with the sum
rule (\ref{sumrule}).  These results are accurately reproduced by our numerical
analysis.  Note that qualitatively this late time dependence is independent of
the choice of initial conditions for the zero mode, except that there is no
growth of modes near $k=0$ in the case in which particles are produced via
parametric amplification (figs. 4,5).

For the radiation $ n= \frac12 $ and matter dominated $ n = \frac23 $
universes, eq.(\ref{latesoln}) reduces to elementary functions:
\begin{eqnarray}
a(\tau) \; U_k(\tau) &=&  c_k \; e^{2ik \tau_0^{1/2} \tau^{1/2} } + 
d_k\; e^{-2ik \tau_0^{1/2} \tau^{1/2} }  
\;  \; \mbox{(RD) } \; ,\cr \cr
a(\tau) \; U_k(\tau) &=&  c_k \; e^{3ik \tau_0^{2/3} \tau^{1/3} } \; 
\left[ 1 + {i \over {3k \tau_0^{2/3}
\tau^{1/3}}}\right] + d_k\; e^{-3ik \tau_0^{2/3} \tau^{1/3}}\left[ 1 - {i \over {3k
\tau_0^{2/3} \tau^{1/3}}}\right] \; \; \mbox{ (MD) }.
\end{eqnarray}

It is also of interest to examine the $n>1$ case.  Here, the modes of interest
satisfy the condition $k \ll \tau^{n-1}/t_0^n$ for late times.  
These modes are constant
in time and one sees that the modes are {\bf frozen}.  In the case of a de
Sitter universe, we can formally take the limit $n \to \infty$ and we see that
{\bf all} modes become frozen at late times.   This case is detailed in
sec. VII \cite{De Sitter}.

\subsection{Discussion, Conclusions and Further Results for the FRW background}

We have shown that there can be significant particle production through quantum
fluctuations after inflation\cite{frw2}. However, this production is somewhat 
sensitive 
to the expansion of the universe. From our analysis of the equation of state,
we see that the late time dynamics is given by a matter dominated cosmology.
We have also shown that the quantum fluctuations of the inflaton decay for late
times as $1/a^2(t)$, while in the case of a symmetry broken inflationary model,
the inflaton field moves to the minimum of its tree level potential.  The
exception to this behavior is the case when the inflaton begins exactly at the
unstable extremum of its potential for which the fluctuations grow out to the
minimum of the potential and do not decay.  Initial production of particles due
to parametric amplification is significantly greater in chaotic scenarios with
symmetry broken potentials than in the corresponding theories with positive
mass terms in the Lagrangian, given similar initial conditions on the zero mode
of the inflaton.

In ref.\cite{frw3} we further investigate a symmetry breaking phase
transition triggered by the lowering of the temperature as $ 1/a(t) $ both
in radiation dominated and matter dominated FRW spacetimes. We
identify  three different time scales:  
an early regime dominated by linear instabilities and the exponential
growth of long-wavelength fluctuations, an intermediate scale when the
field fluctuations probe the broken symmetry states and an asymptotic
 scale wherein a {\bf scaling regime} emerges for  modes of
wavelength comparable to or larger than the horizon. The scaling
regime is characterized by a dynamical physical correlation length 
$ \xi_{phys} = d_H(t) $ with $ d_H(t) $ the size of the causal horizon,
thus there is one correlated region per causal horizon. Inside these
correlated regions the field fluctuations sample the broken symmetry
states. The amplitude of the long-wavelength fluctuations becomes
non-perturbatively large due to  the early times instabilities 
and a semiclassical but {\em stochastic} description emerges in the 
asymptotic regime. In the scaling regime, 
the power spectrum is peaked at zero momentum revealing the onset of a
Bose-Einstein condensate. The scaling solution results in 
that the  equation of state of the scalar fields is the {\em same as
that of the background fluid}. This implies a Harrison-Zeldovich spectrum of 
scalar density perturbations for long-wavelengths.
We discuss the corrections to scaling  as well as
the universality of the scaling solution and the differences and
similarities with the classical non-linear sigma model.

\section{Scalar Field Dynamics in a fixed Inflationary Background 
(the de Sitter Universe)}

We describe in this section the scalar field  evolution in a fixed de Sitter
background using the large $ N $ approximation, postponing the evolution
with a dynamical background to sec. VII.

We consider an initial state at a non-zero temperature $ T_i $. This
change on the initial conditions does not affect the initial values of
the mode functions in eqs.(\ref{modknr}) or (\ref{modkr}). 

Only the expression for the quantum fluctuations $ \Sigma(\tau) $
changes due to the 
fact that the expectation value of the product of a creation and an
annihilation operator is temperature dependent\cite{frw}. We now have
instead of eq.(\ref{sigre}) 
\begin{equation}\label{sigreT}
\Sigma(\tau)= \int_0^{\infty} q^2 dq \left[ | f_q(\tau)|^2 \coth\left(
{\omega_q \over 2 T_i } \right) - {1 \over
{q\; a(\tau)^2}} + {{\Theta(q - 1)}\over {2 q^3}} \left(\frac{{\cal
M}^2(\tau)}{m^2_R}-{{{\cal{R}(\tau)}}\over{6 m^2_R}}\right)\right] \; .
\end{equation}

We consider the case $ m^2 < 0 $ with a critical temperature $ T_c
$ such that $ T_i \gtrsim T_c \gg |m| $\cite{De Sitter}. The symmetry
is initially 
unbroken and through the expansion of the universe the effective
temperature decreases as $  T_i/a(t) $. Therefore, there is a symmetry
breaking phase transition after a few e-folds of inflation.  When the
temperature falls below the critical value, the effective mass becomes
negative.  As will be seen 
explicitly below, when this occurs, long-wavelength modes become unstable and
grow. Local thermodynamic equilibrium will set in again if the contribution
from the quantum fluctuations can grow and adjust to compensate for the
negative mass terms on the same time scales as that in which the temperature
drops. However, as discussed below, for very weak coupling the important time
scales for the non-equilibrium fluctuations are of the order of $ [H/m^2_R]
\ln[1/\lambda] $, which are much longer than the time it takes for the
temperature to drop well below the critical value to practically zero. Thus,
the non-equilibrium dynamics will proceed as if the phase transition occured
via a quench, that is with an effective mass term,
\begin{equation}
m^2_{eff}(t)= m^2_i \; \theta(t_i-t)-m^2_R \; \theta(t-t_i)
\quad ; \quad m^2_i=  m^2_R \; \left[\frac{T^2_i}{T^2_c}-1\right] >0 \, . 
 \label{quench}
\end{equation}
Therefore,  we choose the initial conditions on the mode functions at
$t_i=0$ to be given in terms of the effective mass,
\begin{equation}
M^2_0= m^2_R\left[ \frac{T^2_i}{T^2_c}-1\right]+\frac{\lambda_R}{2} \;
\phi^2(0) de\quad ; \quad \frac{T_i}{T_c} >1 \; .
\label{minitial}
\end{equation}
\subsection{Evolution for $\phi(0)= \dot{\phi}(0)=0$. Analytical Results}

We begin by considering the broken symmetry 
situation in which the expectation value of the
inflaton field sits atop the potential hill with zero initial velocity. This
situation is expected to arise if the system is initially in local
thermodynamic equilibrium  an initial
temperature larger than the critical temperature and cools down
through the critical temperature in the absence of an external field or bias.

The order parameter and its time derivative vanish in the
local equilibrium high
temperature phase, and this condition is a fixed point of the evolution
equation for the zero mode of the inflaton.  There is no rolling of the
inflaton zero mode in this case, although the fluctuations will grow
and will be responsible for the dynamics. 

We can understand the early stages of the dynamics analytically as follows. For
very weak coupling and early time we can neglect the backreation in the mode
equations, which become,

\begin{equation} 
\left[\frac{d^2}{d\tau^2}+3h
\frac{d}{d\tau}+\frac{q^2}{a^2(\tau)}-1\right]f_q(\tau)=0\;, \label{earlytime}
\end{equation}
\begin{equation}
f_q(0)=\frac{1}{\sqrt{\omega_q}} \quad ; \quad \dot{f}_q(\tau) = -i
\sqrt{\omega_q} \quad ; \quad \omega_q= \sqrt{q^2+r^2-1} \quad ; \quad
r = \frac{T_i}{T_c}\; . 
\end{equation}
The solutions are of the form,
\begin{equation}
f_q(\tau) = \exp[-\frac{3}{2}h\tau] \left\{a(q)\; J_{\nu}(z)+b(q)\;
J_{-\nu}(z) \right\} \; ; \;
z=\frac{q}{h}\exp[-h\tau] 
\; ; \; \nu = \sqrt{\frac{1}{h^2}+\frac{9}{4}}\; , \label{bessel}
\end{equation}
where the coefficients $a(q)$ and $b(q)$ are determined by the initial
conditions:
\begin{equation}
b(q) = - {{\pi \, q}\over { 2 h \, \sin{\nu \pi} }} \; \left[ {{i \omega_q -
\frac32 \, h }\over q} \; J_{\nu}\left(\frac{q}{h}\right) - 
J'_{\nu}\left(\frac{q}{h}\right) \right]\; ,
\label{coefb}
\end{equation}
\begin{equation}
a(q) =   {{\pi \, q}\over { 2 h \, \sin{\nu \pi} }} \; \left[ {{i \omega_q -
\frac32 \, h }\over q} \;  J_{-\nu}\left(\frac{q}{h}\right) -  
J'_{-\nu}\left(\frac{q}{h}\right)\right] \;. \label{coefa}
\end{equation}

For long times, $e^{h\tau}\geq q/h$, these mode functions grow exponentially,
\begin{equation}\label{Uasi}
f_q(\tau) \simeq b(q) \; J_{-\nu}(z) \simeq {{b(q)} \over {\Gamma(1-\nu)}}
\; \left( {{2h\, }\over q}\right)^{\nu}e^{(\nu-3/2)h \tau}   \;.
\end{equation}

The Bessel functions appearing in the expression for the modes $ f_q(\tau) $
can be approximated by their series expansion,
\begin{equation}
f_q(\tau) = \frac12 \left[ 1 + \frac{1}{\nu} \; \left( \frac32 -{{q^2}\over{4
h^2}} - i {{\omega_q}\over h} \right) + {\cal O}\left(\frac1{\nu^2}\right) \right] \; 
e^{(\nu - 3/2) h\tau} \; .
\end{equation}
This is an expansion in powers of $ q^2/(\nu h^2) $ and we conclude that $
g\Sigma(\tau) $ is dominated by the modes with $q \leq \sqrt{h}$.

The integral for $g\Sigma(\tau)$ can be approximated by keeping only the modes
$ q \leq f \sqrt{h} $, where $ f $ is a number of order one, and by neglecting
the subtraction term which will cancel the contributions from high
momenta. Numerically, even with the backreaction taken into account, the
integral is dominated by modes $q \leq f \approx 10-20$ in all of the cases
that we studied (see ref. \cite{De Sitter}).

The contribution to the fluctuations from these unstable modes is:
\begin{equation}\label{estim}
 g\Sigma(\tau) \simeq \sqrt{{g}\over 3}\, {{f^3\,h^{3/2} \, r m^2_R}
\over { 2 \pi \, \, M_0^2 }}\; \left(1 + {{ M_0^2}\over { m^2_R}}\right) \;
e^{(2\nu-3)h\tau} \; ,
\end{equation}
where again, we have taken the high temperature limit, $ T_i \sim T_c \gg m_R
$.

From this equation, we can estimate the value of $\tau_s$, the `spinodal
time', at which the contribution of the quantum fluctuations becomes
comparable to the contribution from the tree level terms in the equations of
motion. This time scale is obtained
 from the condition $g\Sigma(\tau_s) = {\cal O}(1)$:
\begin{equation}
\tau_s \simeq -\frac{1}{(2\nu-3)h}\ln\left[\sqrt{g\over 3}
{{f^3\,h^{3/2}}\over { 2 \pi \, m_R M_0^2 }}\;\frac{T_i}{T_c} 
\left(1 + {{ M_0^2}\over { m_R^2}}\right)\right]\label{spinoest} ,
\end{equation}
which is in good agreement with our numerical results (see
ref. \cite{De Sitter}). For values of $h \geq 1$, which, as argued
below, lead to the most interesting case, an estimate for the spinodal time is,
\begin{equation} \label{tslargeh}
\tau_s \simeq  \frac{3h}{2} \ln[1 / \sqrt{g}] + {\cal{O}}(1)
\label{spinotime}
\end{equation}
which is consistent with our numerical results (see \cite{De Sitter}).

For $\tau > \tau_s$, the effects of backreaction become very important, and the
contribution from the quantum fluctuations competes with the tree level terms
in the equations of motion, shutting-off the instabilities. Beyond $\tau_s$,
only a full numerical analysis will capture the correct dynamics.

It is worth mentioning that had we chosen zero temperature initial conditions,
then the coupling $\bar{g} \rightarrow g $ (see fig.\ref{gsigma}) 
and the estimate for the spinodal
time would have been,
\begin{equation}
\tau_s \simeq \frac{3h}{2} \ln[1/{g}] + {\cal{O}}(1), \label{spinotimeT0}
\end{equation}
that is, roughly a factor 2 larger than the estimate for which the de Sitter
stage began at a temperature above the critical value. Therefore,
eq.(\ref{spinotime}) represents an {\em underestimate} of the spinodal
time scale at which fluctuations become comparable to tree level contributions.

The number of e-folds occurring during the stage of growth of spinodal
fluctuations is therefore,
$$
{\cal{N}}_e \approx \frac{3h^2}{2} \ln[1/ \sqrt{g}]\;
\mbox{(high~temperature)~~and~~} {\cal{N}}_e \approx \frac{3h^2}{2} \ln[1/ g]
\; \; \mbox{(zero temperature)}
$$
It is a factor 2 larger for zero temperature.  Thus, it becomes clear
that with $ g \approx 10^{-12}$ and $h \geq 2$, a required number of
e-folds, ${\cal{N}}_e \approx 100$ can easily be accommodated before
the fluctuations become large, modifying the dynamics and the equation
of state. 

The implications of these estimates are important.  The first conclusion drawn
from these estimates is that a `quench' approximation is well justified (see
\cite{De Sitter}). While the temperature drops from an initial value of a few
times the critical temperature to below critical in just a few e-folds, the
contribution of the quantum fluctuations needs a large number of e-folds
to grow to compensate for the tree-level terms and overcome the
instabilities. Only for 
a strongly coupled theory is the time scale for the quantum fluctuations to
grow short enough to restore local thermodynamic
equilibrium during the transition.

The second conclusion is that most of the growth of spinodal fluctuations
occurs during the inflationary stage, and with $ g \approx 10^{-12}$ and
$H \geq m_R$, the quantum fluctuations become of the order of the tree-level
contributions to the equations of motion within the number of e-folds necessary
to solve the horizon and flatness problems. Since the fluctuations
grow to become of the order of the tree level contributions at times of the
order of this time scale, for larger times they will
modify the equation of state substantially and will be shown in
sec. IX to provide a graceful exit from the inflationary phase within
an acceptable number of e-folds.

\subsection{\bf The late time limit}

For late  the dynamics freezes out. The
fluctuation, $g\Sigma(\tau) =1$, and the mode functions effectively describe
free, minimally coupled, massless particles.  The sum rule,
\begin{equation}
-1+g\Sigma(\infty) =0, \label{sumrule1}
\end{equation}
is obeyed exactly in the large $N$ limit as in the Minkowski
case\cite{us1,mink}.

We now show that this value is a self-consistent solution of the
equations of motion for the mode functions, and the {\em only} stationary
solution for asymptotically long times.

In the late time limit, the effective time dependent mass term,
$-1+\eta^2+g\Sigma$, in the equation for the mode functions,
(\ref{modcr}), vanishes (in this case with $\eta =0$).  Therefore,
these mode equations asymptotically become,
\begin{equation} 
\left[\frac{d^2}{d \tau^2}+3h \frac{d}{d \tau}+\frac{q^2}{a^2(\tau)}
\right]f_q(\tau)=0 \; . 
\end{equation}
The general solutions are given by,
\begin{equation}\label{asi}
f_q^{asy}(\tau)= \exp\left[-\frac{3}{2}h \tau \right] \left[c_+(q) \;
J_{3/2}\left( \frac{q}{h}e^{-h\tau}\right) -i c_-(q) \; N_{3/2}\left(
\frac{q}{h}e^{-h\tau}\right)\right] \; ,
\end{equation}
where $ J_{3/2}(z) $ and $ N_{3/2}(z) $ are the Bessel and Neumann functions,
respectively.  The coefficients, $ c_{\pm}(q) $  can be computed for
large $ q $ by matching $ f_q^{asy}(\tau) $ with the WKB approximation to the
exact mode functions $ f_q(\tau) $ that obey the initial conditions
(\ref{modkr}). The WKB approximation to $f_q(\tau) $ has been computed
in ref.\cite{frw}, and we find for large $q $,
\begin{equation}\label{abdek}
c_{\pm}(q) = \sqrt{{\pi\, q}\over {2\, h}}\; \left[ 1 - {i\over q}(h +
\Delta) + 
{\cal O}(q^{-2}) \right]\;e^{-iq/h} \pm \sqrt{{\pi\, h}\over {8\,
q}}\; \left[ 1 
+ {\cal O}\left({1\over q}\right) \right]\;e^{iq/h} \; ,
\end{equation}
where
\begin{equation}
\Delta \equiv \int_0^{\infty} d\tau \; e^{h\tau} \; M^2(\tau) \; .
\end{equation}

In the $ \tau \to \infty $ limit, we have for fixed $ q $,
\begin{equation}\label{uasi}
f_q^{asy}(\tau) \buildrel{ \tau \to \infty}\over=  i \sqrt{2 \over
{\pi}}\left({h \over q }\right)^{3/2}\; c_-(q) \; \; .
\end{equation}
which are independent of time asymptotically, and explains why the power
spectrum of quantum fluctuations freezes at times larger than the
spinodal. This behavior is confirmed numerically [see ref. \cite{De
Sitter}]. Clearly at early
times the mode functions grow exponentially, and at times of the order of
$\tau_s$, when $g\Sigma(\tau) \approx 1$ the mode functions freeze-out and
become independent of time.  Notice that the largest $q$ modes have grown the
least, explaining why the
integral is dominated by $q \leq 10-20$.

For asymptotically large times, $g\Sigma(\tau)$ is given by,
\begin{equation}
g\Sigma(\infty) =g \; h^2 \; \int_0^{+\infty} {{dq}\over q } \;
\coth\left(\frac{ \omega_q}{2  T} \right) \; \left[ {{2h}\over {\pi }}
\mid c_-(q) \mid^2 - \, q \right] \;, \label{asinto}
\end{equation}
where only one term in the UV subtraction survived in the $\tau =\infty $
limit.  The factor $ \coth\left(\frac{ \omega_q}{2 T} \right) $ in 
eq.(\ref{asinto}) takes into account the nonzero initial temperature $ T $.

For consistency, this integral must converge and be equal to $1$ as
given by the sum rule.  For this to be the case and to avoid the potential
infrared divergence in (\ref{asinto}), the coefficients $ c_-(q) $ must vanish at
$ q = 0 $. The mode functions are finite in the $ q \to 0 $ limit provided,
\begin{equation}
c_-(q) \buildrel{ q \to 0 }\over= {\cal C} \; q^{3/2} \; ,
\end{equation}
where $ {\cal C} $ is a constant.

The numerical analysis clearly shows that the mode functions
remain finite as $q \rightarrow 0$, and the coefficient ${\cal C}$ can be read
off from these figures.  This is a remarkable result. It is well known that for
{\em free} massless minimally coupled fields in de Sitter space-time with
Bunch-Davies boundary conditions, the fluctuation contribution $\langle
\pi^2(\vec x, t) \rangle$ grows linearly in time as a consequence of the
logarithmic divergence in the
integrals\cite{linde2}. However, in our case, although the
asymptotic mode functions are free, the coefficients that multiply the Bessel
functions of order $3/2$ have all the information of the interaction and
initial conditions and must lead to the consistency of the sum rule. Clearly
the sum rule and the initial conditions for the mode functions prevent the
coefficients $ c_{\pm}(q) $ from describing the Bunch-Davies vacuum. These
coefficients are completely determined by the initial conditions and the
dynamics.  This is the reason why the fluctuation freezes at long times unlike
in the free case in which they grow linearly\cite{linde2}.

It is easy to see from eqs.(\ref{hubble}) and (\ref{uasi})
that the energy and pressure vanish for $\tau \to \infty$.

\subsection{Discussion and Conclusions for the de Sitter background}

We have identified analytically and numerically two distinct regimes for the
dynamics determined by the initial condition on the expectation value of the
zero mode of the inflaton \cite{De Sitter}.

\begin{enumerate}
\item{When $\eta(0) << g^{1/4}$ (or $ g^{1/2}$ for $T_i=0$), the
dynamics is driven by quantum (and thermal) fluctuations. Spinodal
instabilities grow and eventually compete with tree level terms at a time
scale, $\tau_s \geq -3h\ln[g]/2$.  The growth of spinodal fluctuations
translates into the growth of spatially correlated domains which attain a
maximum correlation length (domain size) of the order of the horizon.  For very
weak coupling and $h \geq 1$ this time scale can easily accommodate enough
e-folds for inflation to solve the flatness and horizon problems. The quantum
fluctuations modify the equation of state dramatically providing a
means for a graceful exit to the inflationary stage without slow-roll.

This non-perturbative description of the non-equilibrium effects in this
regime in which quantum (and thermal) fluctuations are most important
provides a reliable understanding of the relevant 
non-perturbative, non-equilibrium effects of the fluctuations that have
not been revealed before in this setting\cite{us1} - \cite{din}.

These initial conditions are rather natural if the de Sitter era arises during
a phase transition from a radiation dominated high temperature phase in local
thermodynamic equilibrium, in which the order parameter and its time derivative
vanish.}

\item{When $\eta(0) >> g^{1/4}$ (or $ g^{1/2} $ for $T_i=0$), the
dynamics is driven solely by the classical evolution of the inflaton zero
mode. The quantum and thermal fluctuations are always perturbatively small
(after renormalization), and their contribution to the dynamics is negligible
for weak couplings. The de Sitter era will end when the kinetic contribution to
the energy becomes of the same order as the `vacuum' term. This is the realm of
the slow-roll analysis whose characteristics and consequences have been
analyzed in the literature at length. These initial conditions, however,
necessarily imply some initial state either with a biasing field that favors a
non-zero initial expectation value, or that in the radiation dominated stage,
prior to the phase transition, the state was strongly out of equilibrium with
an expectation value of the zero mode different from zero. Although such a
state cannot be ruled out and would naturally arise in chaotic scenarios, the
description of the phase transition in this case requires further input on the
nature of the state prior to the phase transition.}
\end{enumerate}

\section{Self-consistent  Evolution of Matter Fields with a dynamical
cosmological background}

We present in this section the full self-consistent matter-geometry 
dynamics\cite{din}. That is,
the scale factor $ a(t) $ is here a dynamical variable determined by the
Einstein-Friedman eq.(\ref{eif})-(\ref{hubble}) coupled with the
scalar field evolution eqs.(\ref{modcr})-(\ref{modkr}).

\bigskip

In order to provide the full
solution we now must provide the values of $\eta(0)$, $\dot{\eta}(0)$,
and $h_0$. Assuming that the 
inflationary epoch is associated with a phase transition at the GUT scale,
this requires that $ N m^4_R/g \approx 
(10^{15}\mbox{ Gev })^4 $ and assuming the bound on the scalar
self-coupling $ g \approx 10^{-12}-10^{-14}$ (this will be seen
later 
to be a compatible requirement), we find that $h_0 \approx N^{1/4}$ which
we will take to be reasonably given by $h_0 \approx 1-10$ (for example
in popular GUT's $ N \approx 20 $ depending on particular representations). 

We will begin by studying the case of most interest from the point of view
of describing the phase transition: $\eta(0)=0$ and $\dot{\eta}(0)=0$,
which are the initial conditions that led to puzzling questions. With
these initial conditions, the evolution equation for the zero mode
eq.(\ref{modcr}) determines that $\eta(\tau) = 0$ by symmetry.

\subsection{Early time dynamics:}
Before engaging in the numerical study, it proves illuminating to
obtain an estimate of the relevant time scales and an intuitive idea of
the main features of the dynamics. Because the coupling is so weak 
($ g \sim 10^{-12} \ll 1 $) and after
renormalization the contribution from the quantum fluctuations to 
the equations of motion is finite, we can neglect  all the terms
proportional to $ g $ in eqs.(\ref{hubble}) and (\ref{modcr}). 

For the case where we choose $\eta(\tau) = 0$ and
the evolution equations for the 
mode functions are those for an inverted oscillator in De Sitter space-time,
which have been studied in sec. VII \cite{guthpi}. One
obtains the approximate solutions (\ref{bessel})-(\ref{coefb}).

After the physical wavevectors cross the horizon, i.e. when $q e^{-h_0
\tau}/h_0 \ll 1$ we find that the mode functions factorize: 
\begin{equation}
f_q(\tau)\buildrel {q e^{-h_0 \tau}\ll h_0 } \over =   {{B_q} \over
{\Gamma(1-\nu)}}  \; \left( {{2h_0\, }\over
q}\right)^{\nu}e^{(\nu-3/2)h_0 \tau}. \label{factor} 
\end{equation}
This result reveals a very
important feature: because of the negative mass squared
term in the matter Lagrangian leading to symmetry breaking (and $\nu > 3/2$), we see
that all of the mode functions {\em grow exponentially} after horizon
crossing (for positive mass squared $ \nu < 3/2 $, and  
they would {\em decrease
exponentially} after horizon crossing). This exponential growth is a 
consequence of the spinodal instabilities which 
is a hallmark of the process of phase separation that
occurs to complete the phase transition. 
 We note, in addition that the time 
dependence is exactly given by that of the $ q=0 $ mode, i.e. the zero 
mode, which is a consequence of the redshifting of the wavevectors and 
the fact that after horizon crossing the contribution of the term
$q^2/a^2(\tau)$ in the equations of motion become negligible.
 We clearly  see that the quantum fluctuations grow exponentially and
they will begin to be of the order of the tree level terms in the
equations of motion when $g\Sigma(\tau) \approx 1$. At large
times substantially before the end of inflation
$$
\Sigma(\tau)  \approx {\cal
F}^2(h_0) \; h_0^2 \; e^{(2\nu-3)h_0 \tau} \; ,   
$$
with $ {\cal F}(h_0) $ a finite constant that depends on the initial
conditions and is found numerically to be of ${\cal O}(1)$ [see
fig.\ref{fofh}].   

In terms of the initial dimensionful variables, the
 condition  $ g\Sigma(\tau) \approx 1$ translates
to $ <\pi^2(\vec x,t)>_R \approx 2m^2_R/g $, i.e. the quantum
fluctuations sample the minima of the (renormalized) tree level potential.
We find that the
time at which the contribution of the 
quantum fluctuations becomes of the same order as the tree level terms is
estimated to be\cite{De Sitter}
\begin{equation}
\tau_s \approx \frac{1}{(2\nu-3)h_0}
\ln\left[\frac{1}{g\, h_0^2\,  {\cal F}^2(h_0)}\right] 
= \frac32 h_0 \ln\left[\frac{1}{g\, h_0^2\,  {\cal F}^2(h_0)}\right]
+ {\cal O}(1/h_0).
\label{spinodaltime}
\end{equation}
At this time, the contribution of the quantum fluctuations makes the
back reaction very important and, as will be seen numerically, this
translates into the fact that $\tau_s$ also determines the end of the
De Sitter era and the end of inflation. The total number of e-folds during
the stage of exponential expansion of the scale factor (constant
$h_0$) is  given by   
\begin{equation}
N_e \approx \frac{1}{2\nu-3}\;
\ln\left[\frac{1}{g\; h_0^2\; {\cal F}^2(h_0)}\right] 
= \frac32\; h_0^2\; \ln\left[\frac{1}{g\;  h_0^2 \;{\cal
F}^2(h_0)}\right]  
+ {\cal O}(1)\label{efolds}
\end{equation}
For large $h_0$ we see that the number of e-folds scales as $h^2_0$ as well
as with the logarithm of the inverse coupling. 
These results (\ref{factor}-\ref{efolds}) will be
confirmed numerically below and will be of paramount importance for the
interpretation of the main consequences of the dynamical evolution. 

As discussed in sec. VII.C, the early time dynamics is dominated by classical
or quantum effects depending on the ratio between the time scales
 $ \tau_c $ and $ \tau_s $.

If $ \tau_c $ is much smaller than the spinodal time $ \tau_s $ given
by eq.(\ref{spinodaltime}) then the {\em classical} evolution of the
zero mode will dominate the dynamics and the quantum fluctuations will
not 
become very large, although they will still undergo spinodal growth. On 
 the other hand, if $\tau_c \gg \tau_s$ the quantum fluctuations will
grow to be very large well before the zero mode reaches the non-linear
regime. In this case the dynamics will be determined completely by
the quantum fluctuations. Then the criterion for the classical or quantum
dynamics is given by
\begin{eqnarray} 
\eta(0) & \gg & \sqrt{g}\;h_0 \Longrightarrow \mbox{ classical dynamics }
\nonumber \\
\eta(0) & \ll & \sqrt{g}\;h_0 \Longrightarrow \mbox{ quantum dynamics }
\label{classquandyn} 
\end{eqnarray}
or in terms of dimensionful variables $\phi(0) \gg H_0$ leads to 
{\em classical dynamics} and $\phi(0) \ll H_0$ leads to 
{\em quantum dynamics}. 

However, even when the classical evolution of the
zero mode dominates the dynamics, the quantum fluctuations grow
exponentially after horizon crossing unless the value of $\phi(t)$ is
very close to the minimum of the tree level potential. In the large $
N $ approximation the spinodal line, that is the values of $\phi(t)$ for  
which there are spinodal instabilities, reaches all the way to the minimum
of the tree level potential as can be seen from the equations of motion for
the mode functions. 
Therefore even in the
classical case one must understand how to deal with quantum fluctuations
that grow after horizon crossing.  

\subsection{Numerics}
The time evolution is carried out by means of a fourth order Runge-Kutta
routine with adaptive step-sizing while the momentum 
integrals are carried out using an 11-point Newton-Cotes integrator.  
The relative errors in both
the differential equation and the integration are of order $10^{-8}$.
We find that the energy is covariantly conserved throughout the evolution
to better than a part in a thousand. Figs. \ref{gsigma}--\ref{modu} 
show $g\Sigma(\tau)$ vs. $\tau$,
$h(\tau)$ vs. $\tau$ and $ \ln|f_q(\tau)|^2 $ vs. $\tau$ for several values
of $q$ with larger $q's$ corresponding to successively lower curves. 
Figs. \ref{povere},\ref{hinverse} show $p(\tau)/\varepsilon(\tau)$ 
and the horizon size $h^{-1}(\tau)$ for 
$g = 10^{-14} \; ; \; \eta(0)=0 \; ; \; \dot{\eta}(0)=0$
and we have chosen the representative value $h_0=2.0$.

\begin{figure}
\epsfig{file=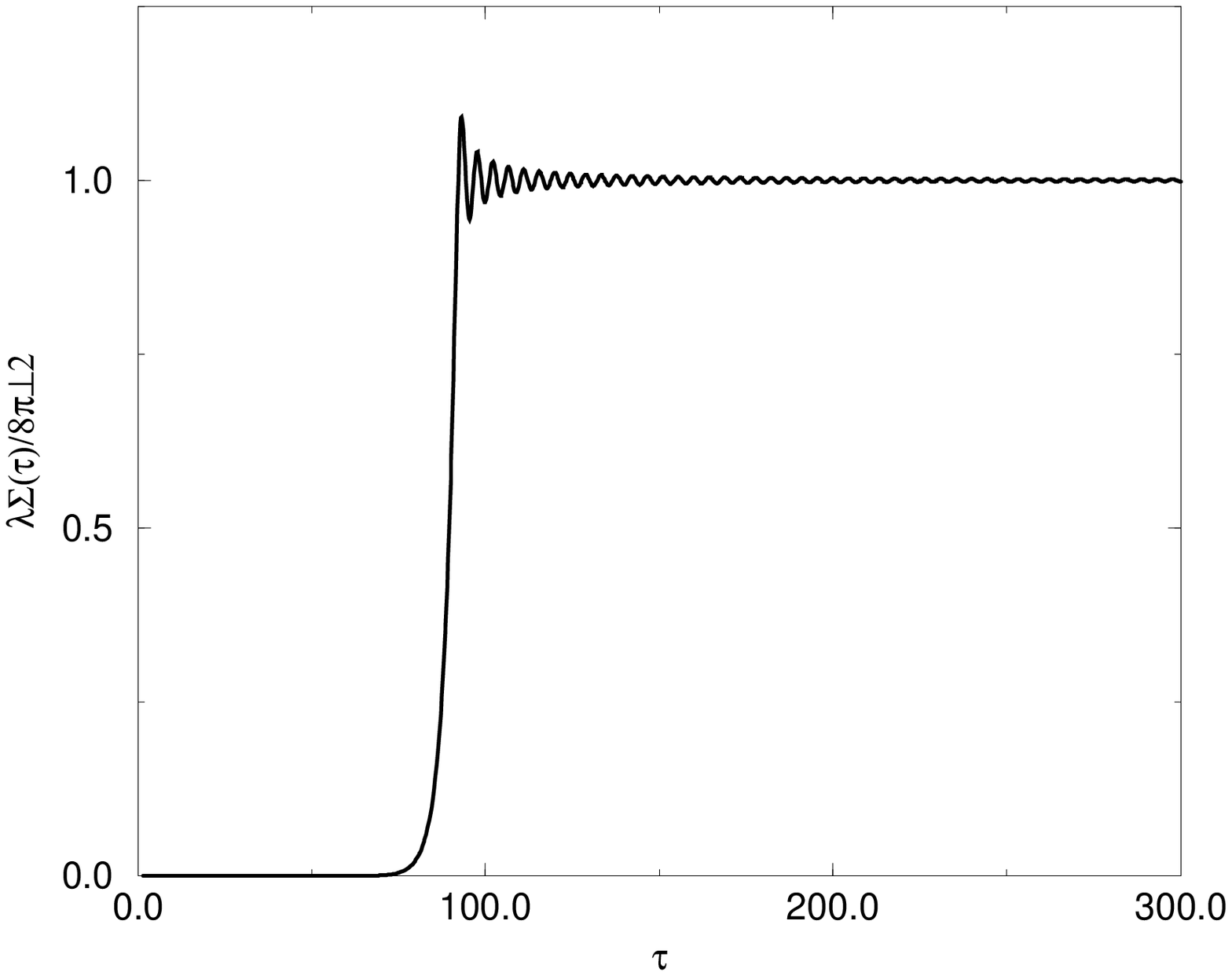,width=10.5cm,height=5.5cm}
\caption{ $ g\Sigma(\tau) $ vs. $\tau$, for $\eta(0)=0, \dot{\eta}(0)=0,\, 
g = 10^{-14},\,  h_0 = 2.0$. }
\label{gsigma}
\epsfig{file=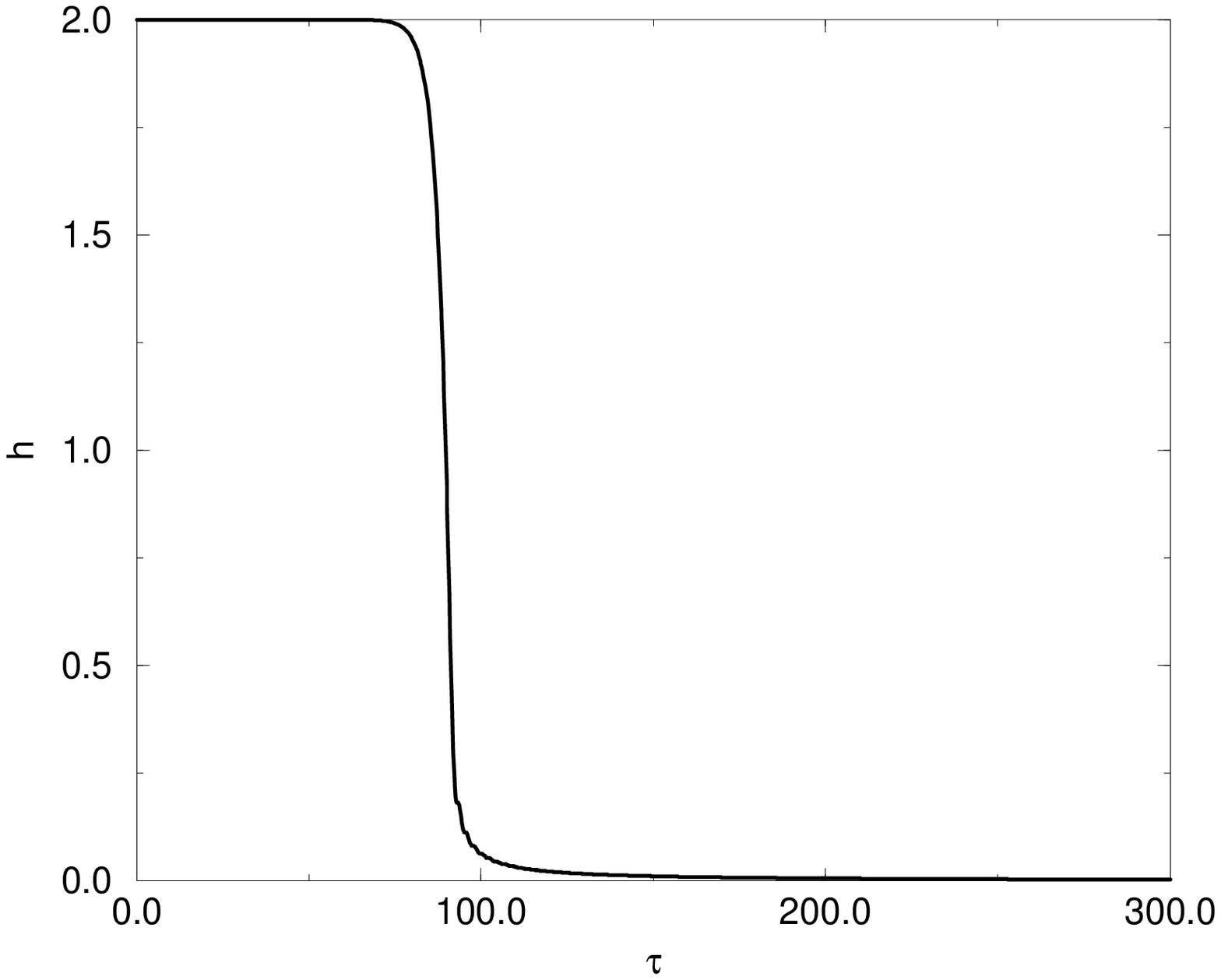,width=10.5cm,height=5.5cm}
\caption{$ h(\tau) $ vs. $ \tau $, for $ \eta(0)=0,\,  \dot{\eta}(0)=0, \, 
g = 10^{-14},\,  h_0 = 2.0 $. }
\label{hubblefig}
\end{figure}

Figs. \ref{gsigma} and \ref{hubblefig} show clearly that 
when the contribution of the quantum
fluctuations $ g\Sigma(\tau) $ becomes of order 1 inflation ends,
and the time scale for $ g\Sigma(\tau) $ to reach ${\cal O}(1)$ is very well
described by  the estimate (\ref{spinodaltime}). From
fig.\ref{gsigma} we see that this happens for $ \tau =\tau_s\approx
90 $, leading to a number of e-folds $ N_e \approx 180 $ which is
correctly estimated by  eqs. (\ref{spinodaltime})-(\ref{efolds}).  

Fig. \ref{modu} shows clearly the factorization of the modes after they
cross the horizon as described by eq.(\ref{factor}).
 The slopes of all the curves after they become
straight lines in fig.\ref{modu} is given exactly by $(2\nu-3)$,
whereas the 
intercept depends on the initial condition on the mode function and
the larger the value of $ q $ the smaller the intercept because the
amplitude of the mode function is smaller initially. Although the
intercept depends on the initial conditions on the long-wavelength
modes, the slope is independent of the value of $q$ and is the same as
what would be obtained in the linear approximation for the {\em
square} of the zero mode at times 
long enough that the decaying solution can be neglected but short enough
that the effect of the non-linearities is very small.
 Notice from the figure that when inflation ends and
the non-linearities become important all of the modes effectively saturate.
This is also what one would expect from the solution of the zero mode:
exponential growth in early-intermediate times (neglecting the
decaying solution), with a growth exponent
given by $(\nu - 3/2)$ and an asymptotic behavior of small oscillations
around the equilibrium position, which for the zero mode is $\eta =1$, but
for the $q \neq 0$ modes depends on the initial conditions. 
All of the mode functions have this behavior once they cross the horizon.
We have also studied the phases of the mode functions and we found that 
they freeze after horizon crossing in the sense that they become independent
of time. This is natural since both the
real and imaginary parts of $ f_q(\tau) $ obey the same equation but
with different 
boundary conditions. After the physical wavelength crosses the horizon, the
dynamics is insensitive to the value of $q$ for real and imaginary parts and
the phases become independent of time. Again, this is a consequence of the
factorization of the modes. 

\begin{figure}
\epsfig{file=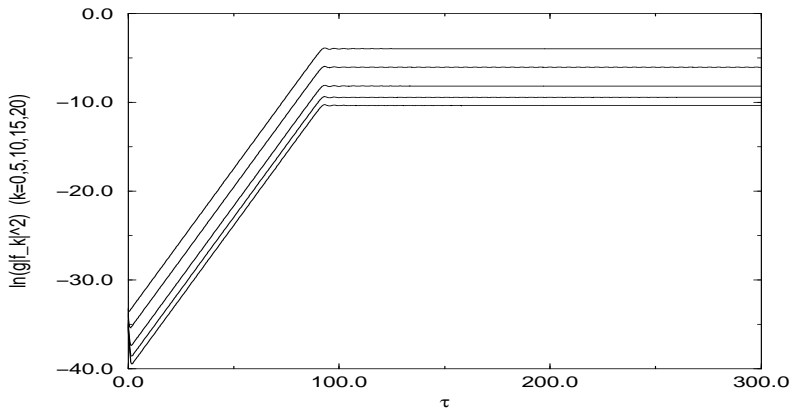,width=10.5cm,height=5.5cm}
\caption{$ \ln|f_q(\tau)|^2 $ vs. $\tau$, for $\eta(0)=0,
\dot{\eta}(0)=0,  g = 10^{-14}, h_0=2.0$ for
$q=0.0,5,10,15,20$ with smaller 
$q$ corresponding to larger values of $\ln|f_q(\tau)|^2 $.}
\label{modu}
\end{figure}

The growth of the quantum fluctuations is sufficient to end inflation
at a time given by $\tau_s$ in eq.(\ref{spinodaltime}). Furthermore fig.
\ref{povere} shows that during the inflationary epoch
$p(\tau)/\varepsilon(\tau)  
\approx -1$ and the end of inflation is rather sharp at $\tau_s$ with
$p(\tau)/\varepsilon(\tau)$ oscillating between $\pm 1$ with zero average
over the cycles, resulting in matter domination. Fig. \ref{hinverse}
shows this  
feature very clearly; $h(\tau)$ is constant during the de Sitter epoch and
becomes matter dominated after the end of inflation with $h^{-1}(\tau) 
\approx \frac32 (\tau -\tau_s)$. There are small oscillations around
this value because both $p(\tau)$ and $\varepsilon(\tau)$
oscillate. These oscillations 
are a result of small oscillations of the mode functions after they 
saturate, and are also a
feature of the solution for a zero mode. 

All of these features hold for a variety of initial conditions.  As an
example, we show in ref.\cite{din} the
case of an initial Hubble parameter of $h_0=10$.

\begin{figure}
\epsfig{file=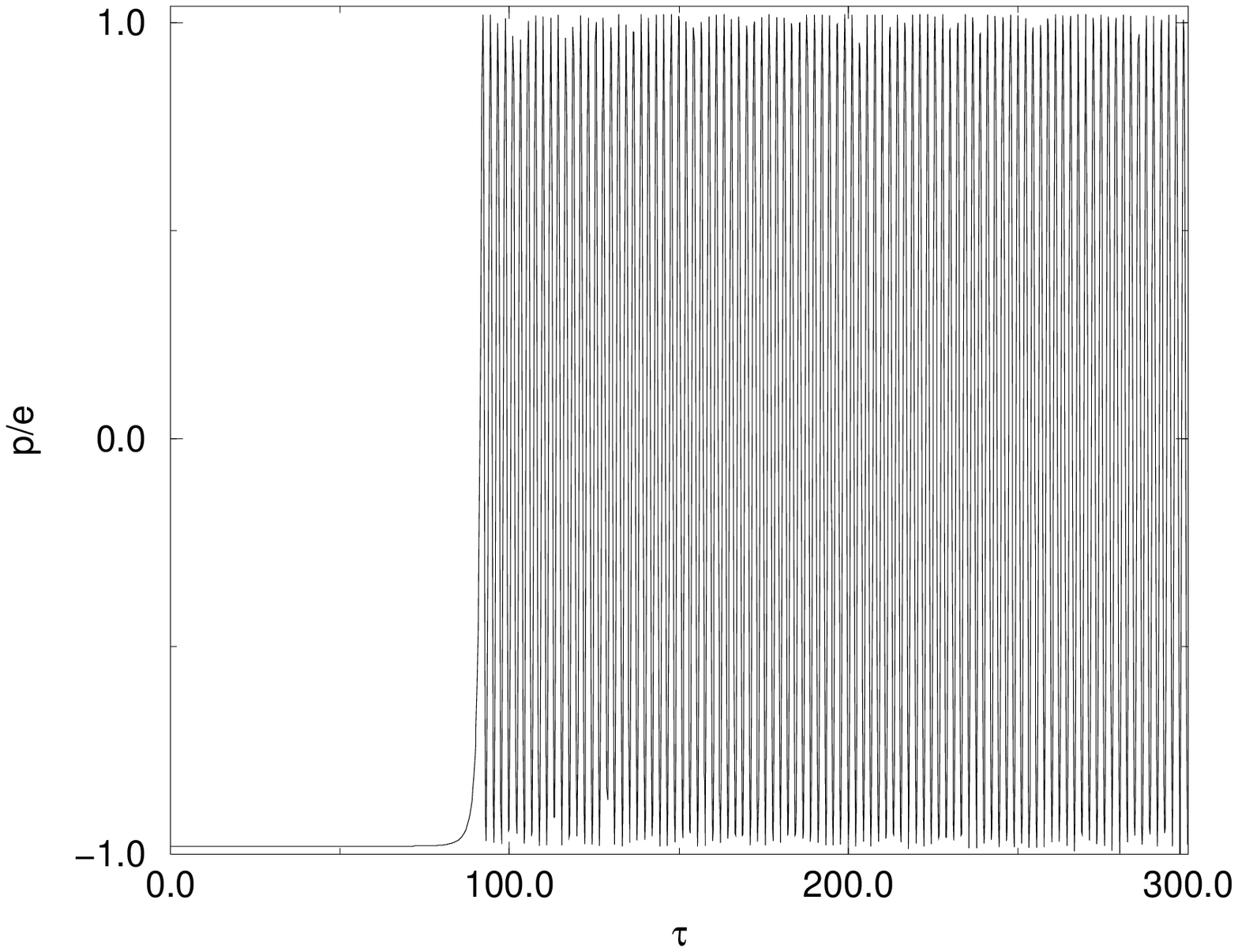,width=10.5cm,height=5.5cm}
\caption{$p/\varepsilon$ vs. $\tau$, for $\eta(0)=0, \, \dot{\eta}(0)=0,\, 
g = 10^{-14}, h_0=2.0$.}
\label{povere}
\epsfig{file=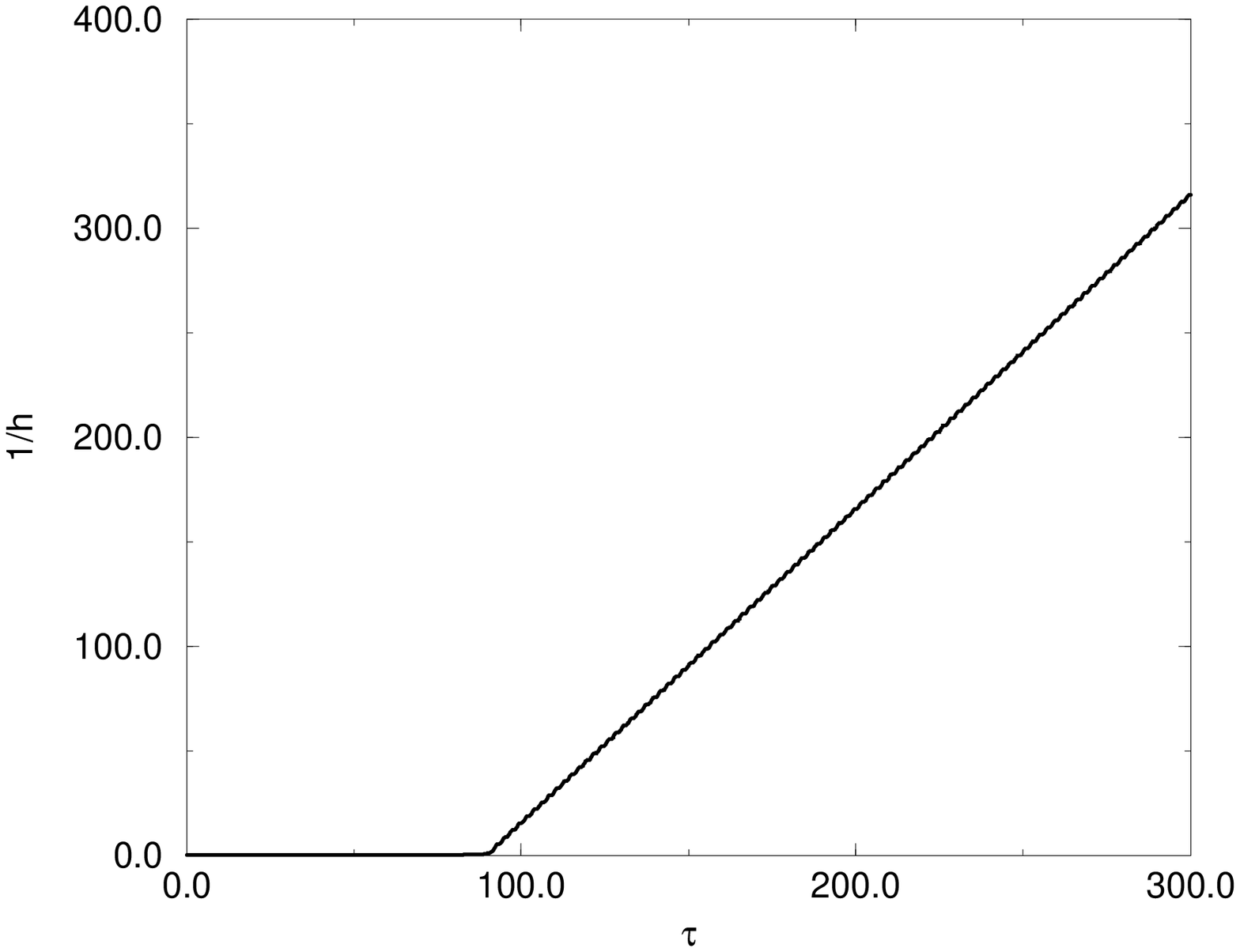,width=10.5cm,height=5.5cm}
\caption{$1/h(\tau)$ vs. $\tau$, for $\eta(0)=0,\,  \dot{\eta}(0)=0,
g = 10^{-14},\,  h_0=2.0$. }
\label{hinverse}
\end{figure}

\subsection{Zero Mode Assembly:}
This remarkable feature of factorization of the mode functions after
horizon crossing can be elegantly summarized as
\begin{equation}
f_k(t)|_{k_{ph}(t) \ll H} = g(q,h)f_0(\tau),\label{factor2}
\end{equation}
with $k_{ph}(t) = k\,e^{-Ht}$ being the physical momentum, 
$ g(q,h)$ a complex constant, and $f_0(\tau)$ a {\em real} function
of time that satisfies the mode equation with $q=0$ and real initial
conditions which will be inferred later.

Then we consider the 
contribution of these modes to the  {\em renormalized} quantum
fluctuations a long time after the beginning of inflation (so as to
neglect the decaying solutions), we find that 
$$g\Sigma(\tau) \approx {\cal C}e^{(2\nu-3)h \tau} + \mbox{ small} \; , 
$$
where 
`small' stands for
the contribution of mode functions associated with momenta that have not
yet crossed the horizon at time $\tau$, which give a perturbatively
small (of order $ g $) contribution. We find that several e-folds
after the beginning of inflation and till inflation ends, this
factorization of superhorizon modes implies the following:
\begin{eqnarray}
&&g\int q^2 dq \; |f^2_q(\tau)|  \approx |C_0|^2 f^2_0(\tau)\quad , \quad
g\int q^2 dq \; |\dot{f}^2_q(\tau)|  \approx  |C_0|^2
 \dot{f}^2_0(\tau)\; ,  \cr \cr
&&g\int \frac{q^4}{a^2(\tau)} dq \; |f^2_q(\tau)|  \approx 
 \frac{|C_1|^2 }{a^2(\tau)}f^2_0(\tau)\; , \label{int1}
\end{eqnarray}
where we have neglected the weak time dependence arising from the
perturbatively small 
contributions of the short-wavelength modes that have not yet crossed the
horizon, and the integrals above are to be understood as the fully
renormalized (subtracted), finite integrals. For $\eta = 0$, we note
that (\ref{int1}) and the fact that $f_0(\tau)$ obeys the equation of
motion for the mode with $q=0$ leads at once to the conclusion that  
in this regime $\left[g\Sigma(\tau)\right]^{\frac{1}{2}} =
|C_0|f_0(\tau)$ obeys the zero mode equation of motion 
\begin{equation}
\left[\frac{d^2}{d \tau^2}+ 3h \frac{d}{d\tau}-1+
(|C_0|f_0(\tau))^2\right]|C_0|f_0(\tau) = 0 \; . 
\label{zeromodeeff}
\end{equation}
It is clear that at a time $ \tau_A $ several e-folds after the
beginning of inflation, we can define an effective zero mode   as 
\begin{equation}
\eta^2_{eff}(\tau) \equiv g\Sigma(\tau), \mbox{ or in dimensionful
variables, } \phi_{eff}(t) \equiv \left[\langle \pi^2(\vec x, t)
\rangle_R \right]^{\frac{1}{2}} 
\label{effectivezeromode}
\end{equation}
Although this identification seems natural, we emphasize that it
is by no means a trivial or ad-hoc statement. There are several
important features that allow an {\em unambiguous} identification:
i) $\left[\langle \pi^2(\vec x, t) \rangle_R \right]$ is a fully 
renormalized operator product and hence finite, ii) because of  the
factorization of the superhorizon modes that enter in the evaluation of 
$\left[\langle \pi^2(\vec x, t) \rangle_R \right]$,  
$\phi_{eff}(t)$ (\ref{effectivezeromode}) 
{\em obeys the equation of motion for the zero mode}, iii) this identification 
is valid several e-folds after the beginning of inflation,
after the transient decaying solutions have died away and the integral
in $\langle \pi^2(\vec x,t) \rangle$
is dominated by the modes with wavevector $k$ that have crossed the horizon at 
$t(k) \ll t$.
Numerically we see that this identification holds throughout the
dynamics except for a very few e-folds at the beginning of inflation. This
factorization determines at once the initial conditions of the effective
zero mode that can be extracted numerically: after the first few e-folds and
long before the end of inflation we find
\begin{equation}
\phi_{eff}(t) \equiv \phi_{eff}(t_A)\; e^{(\nu-\frac{3}{2})H \,
(t-t_A)} \; \; ,   
\label{effzeromodein} 
\end{equation}
where we parameterized 
$$
\phi_{eff}(t_A ) \equiv \frac{H}{2\pi} \; {\cal F}(H/m)
$$
to make contact with the literature. As is shown in fig.~(\ref{fofh}),
we find numerically that $ {\cal F}(H/m)  \approx 
{\cal O}(1) $ for a large range of $ 0.1 \leq H/m \leq 50 $ and that
this quantity depends on the initial conditions of the long wavelength modes.  

Therefore, in summary, the effective composite zero mode obeys
\begin{equation}
\left[\frac{d^2}{d \tau^2}+ 3h \frac{d}{d\tau}-1+
\eta^2_{eff}(\tau)\right]\eta_{eff}(\tau) = 0 
\; ; \; \dot{\eta}_{eff}(\tau_A) = \left(\nu -
\frac{3}{2}\right) \; \eta_{eff}(\tau_A) \; , \label{effzeromode}
\end{equation}
where $ \tau_A $ is the time at which the composite mode 
becomes effective, $ \eta_{eff}(\tau_A) \equiv
{{\sqrt{\lambda_R/2}}\over {m_R}} \; \phi_{eff}(t_A) $  is obtained
numerically for a given $ h_0 $ by fitting the intermediate time
behavior of  $ g\Sigma(\tau) $ with the growing zero 
mode solution. Recall that $ \lambda_R = 8\pi^2 \, g $.
\begin{figure}
\epsfig{file=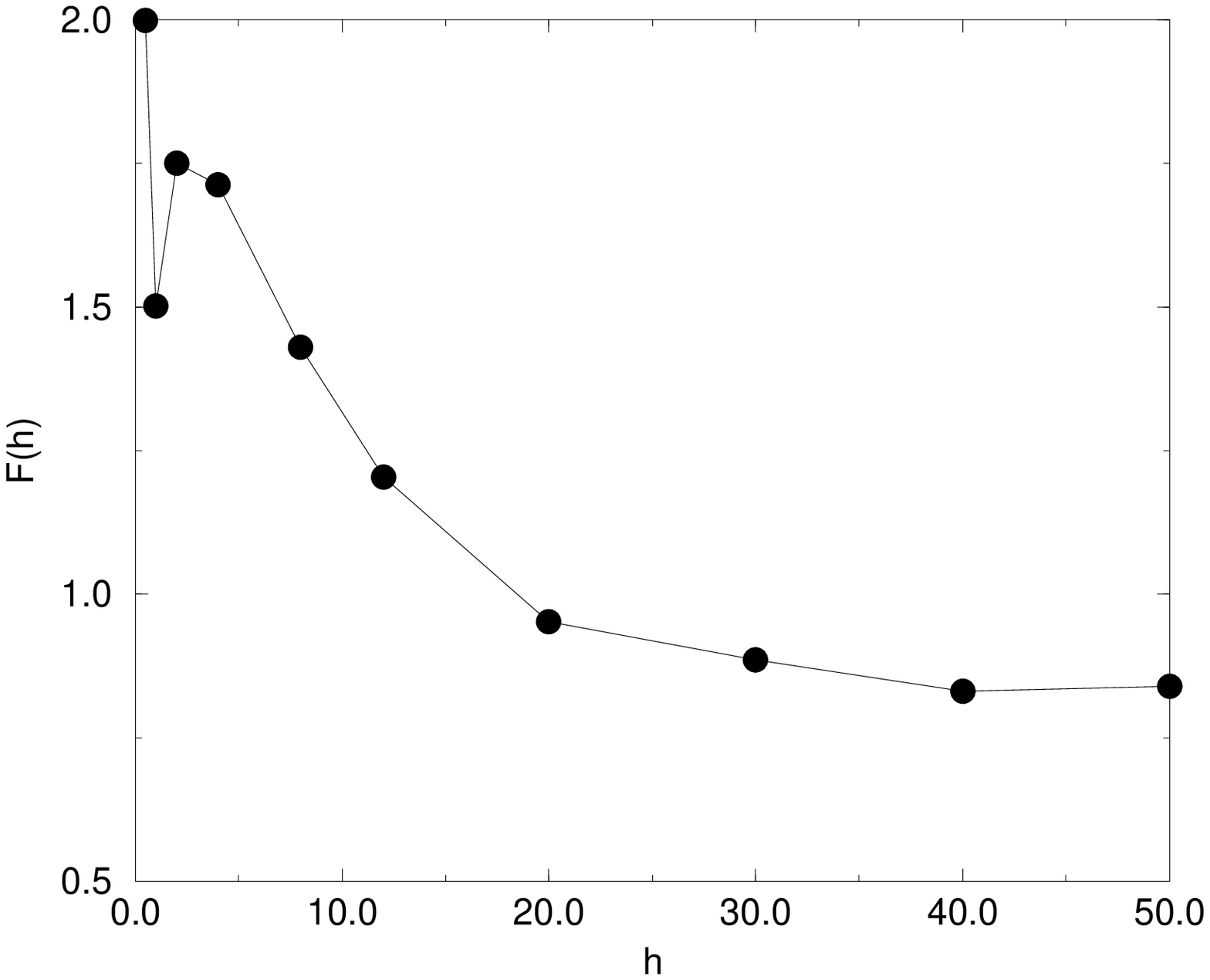,width=10.5cm,height=5.5cm}
\caption{${\cal F}(H/m)$ vs. $h$, where ${\cal F}(H/m)$
is defined by the relation
$\phi_{eff}(t_A) = (H/2\pi) {\cal F}(H/m)$ (see eqs.
(\ref{effectivezeromode}) and (\ref{effzeromodein})).}
\label{fofh}
\epsfig{file=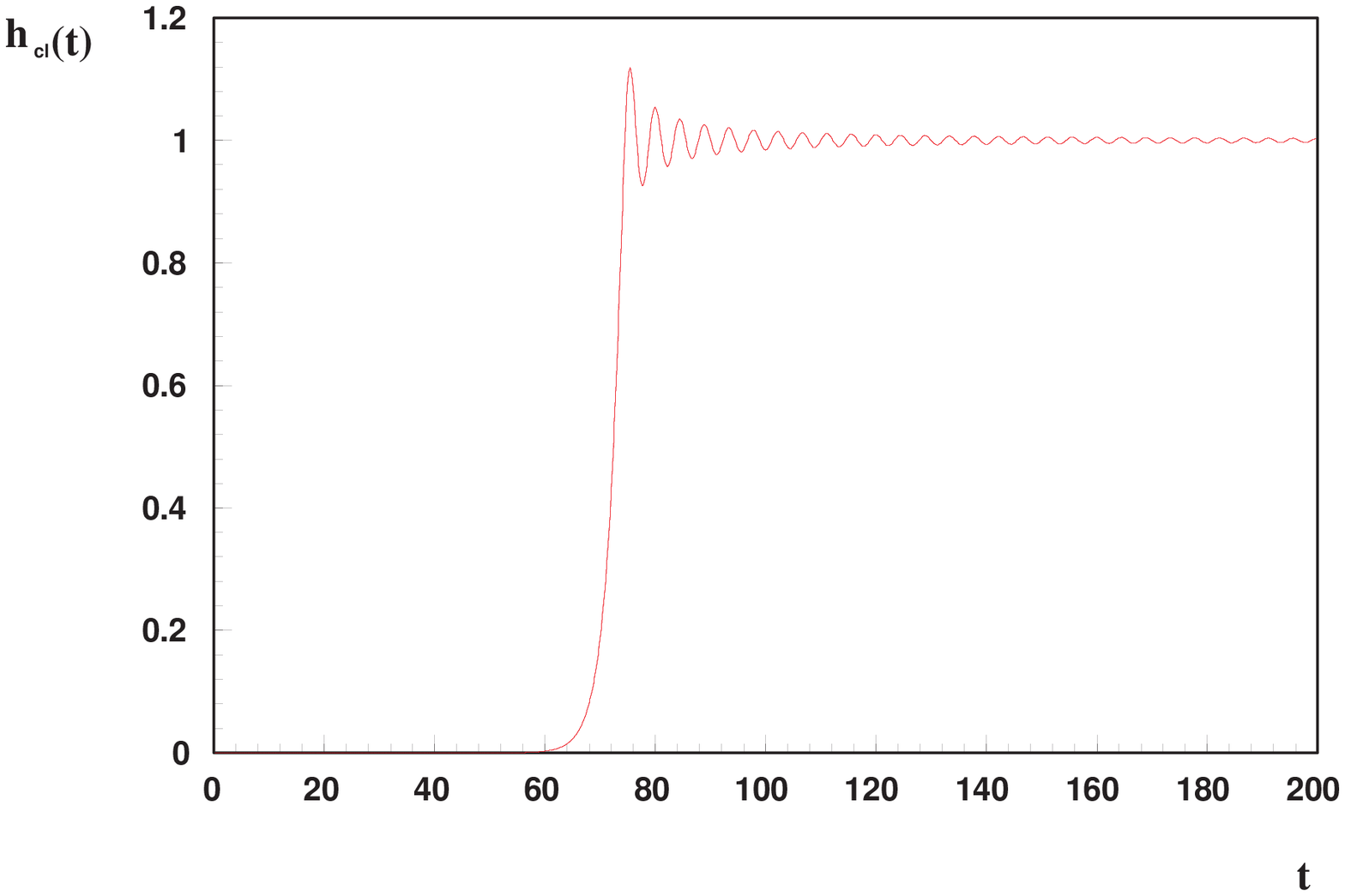,width=10.5cm,height=5.5cm}
\caption{ $\eta_{eff}^2(\tau)$ vs. $\tau$, for $\eta_{eff}(0)=3.94 \times
10^{-7},\,  \dot{\eta}_{eff}(0)=0.317\eta_{eff}(0),\, 
g = 10^{-14},\,  h_0 = 2.0$. The initial conditions were obtained by
fitting the intermediate time regime of $g\Sigma(\tau)$ in
fig.\ref{gsigma}. $\eta_{eff}(\tau)$ is the solution of
eq.(\ref{effzeromode}) 
with these initial conditions.}
\label{etaclas}
\epsfig{file=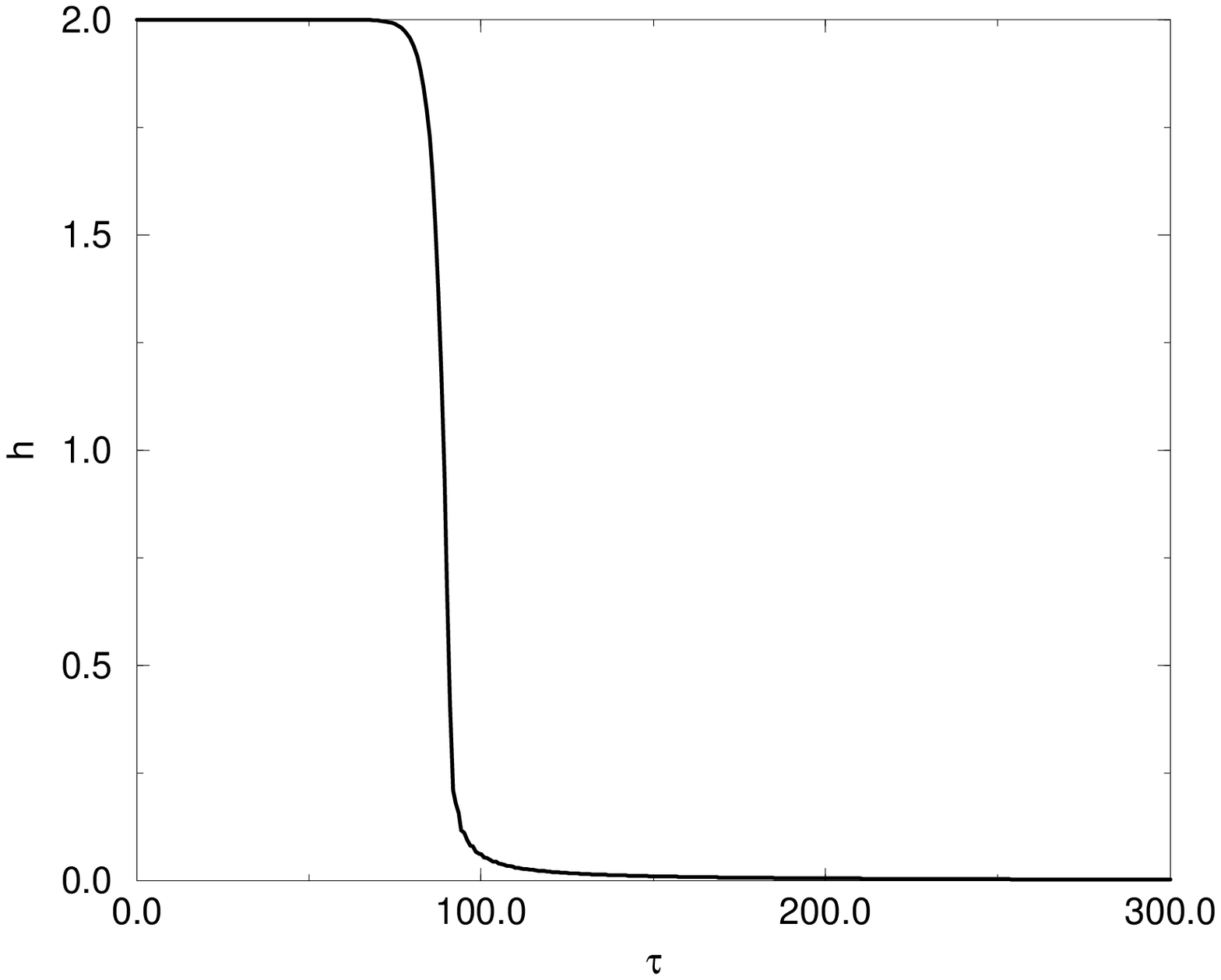,width=10.5cm,height=5.5cm}
\caption{$h(\tau)$ vs. $\tau$, obtained from the solution of
eqs. (\ref{effzeromode}) and (\ref{effscalefactor}) 
with the conditions of fig.\ref{etaclas}.}
\label{hubclas}
\end{figure}
Furthermore, this analysis shows that in the case $\eta = 0$,  the
renormalized energy and pressure in this regime in which the
renormalized integrals are dominated by the superhorizon modes are given by  
\begin{eqnarray}\label{effenergy}
&& \varepsilon_R(\tau)   =  \frac{2Nm^4_R}{\lambda_R} \left\{
\frac{1}{2}\dot{\eta}^2_{eff}+\frac{1}{4}\left(-1+\eta^2_{eff}\right)^2
\right\} \quad , \quad (p+\varepsilon)_R  =
\frac{2Nm^4_R}{\lambda_R}\left\{ \dot{\eta}^2_{eff}\right\} 
\end{eqnarray}
where we have neglected the contribution proportional to $1/a^2(\tau)$ 
because it is effectively red-shifted away after just a few e-folds.
We found numerically that this term is negligible after the interval
of time necessary for the superhorizon modes to dominate the contribution
to the integrals. 
Then the dynamics of the scale factor is given by 
\begin{equation}
h^2(\tau) = 4 h^2_0 \left\{
\frac{1}{2}\dot{\eta}^2_{eff}+\frac{1}{4}\left(-1+\eta^2_{eff}\right)^2
\right\}\; .
\label{effscalefactor}
\end{equation}

We have numerically evolved the set of effective equations
(\ref{effzeromode})-(\ref{effscalefactor}) by extracting the initial 
condition for the effective zero mode from the intermediate time behavior
of $g\Sigma(\tau)$. We found a remarkable agreement  between the
evolution of $\eta^2_{eff}$ and $g\Sigma(\tau)$ and between 
the dynamics of the scale factor in terms of the evolution of
$\eta_{eff}(\tau)$, and the {\em full}
dynamics of the scale factor and quantum fluctuations within our numerical
accuracy. Figs. \ref{etaclas} and \ref{hubclas} show the evolution
of $\eta^2_{eff}(\tau)$ and $h(\tau)$ respectively from the {\em
classical} evolution 
eqs. (\ref{effzeromode}) and (\ref{effscalefactor}) using the initial
condition  $ \eta_{eff}(\tau_A) $ extracted from the exponential fit of
$ g\Sigma(\tau) $ in the intermediate regime. These figures should be 
compared to figs. \ref{gsigma} and \ref{hubblefig}. We have also
numerically compared  
$p/\varepsilon$ given solely by the dynamics of the effective zero mode
and it is again numerically indistinguishable from that obtained with the
full evolution of the mode functions. 

This is one of the main results of our work\cite{din}. In summary: the
modes that 
become superhorizon sized and grow through the spinodal instabilities assemble
themselves into an effective composite zero mode a few e-folds after
the beginning of inflation. This effective zero mode drives the dynamics
of the FRW scale factor, terminating inflation when the non-linearities
become important. In terms of the underlying fluctuations, the spinodal
growth of superhorizon modes gives a non-perturbatively large contribution
to the energy momentum tensor that drives the dynamics of the scale factor.
Inflation terminates when the mean square root fluctuation probes the
equilibrium minima of the tree level potential. 

The extension of this analysis to the case for which $\eta(0) \neq 0$
is straightforward.  Since both $\eta(\tau)$ and 
$\sqrt{g\Sigma(\tau)} = |C_0|f_0(\tau)$ obey the equation for the
zero mode, eq.(\ref{modcr}), it is clear that we can generalize
our definition of the effective zero mode to be
\begin{equation}
\eta_{eff}(\tau) \equiv \sqrt{\eta^2(\tau)+g\Sigma(\tau)}\; .
\label{effeta}
\end{equation}
which obeys the equation of motion of a {\em classical} zero mode:
\begin{equation}
\left[\frac{d^2}{d\tau^2}+3h(\tau)\frac{d}{d\tau}-1+\eta_{eff}(\tau)^2\right]
\eta_{eff}(\tau) = 0 \; . \label{effzeroeqn}
\end{equation}
If this effective zero mode is to drive the FRW expansion, then the
additional condition
\begin{equation}\label{condeta}
\dot{\eta}_{eff}^2 \; f_0^2 - 2\;\eta_{eff}\;\dot{\eta}_{eff}\;f_0\;\dot{f_0} +
\eta_{eff}^2 \;\dot{f_0}^2 = \left[ \dot{\eta}_{eff}\; f_0 -
\eta_{eff}\;\dot{f_0} \right]^2 = 0 \; ,
\end{equation}
must also be satisfied.  One can easily show that this relation is indeed
satisfied if the mode functions factorize as in (\ref{factor2}) and if
the integrals in eqs.(\ref{int1}) are dominated by the 
contributions of the superhorizon mode functions.  This leads to the
conclusion that the gravitational dynamics is given by eqs. 
(\ref{effenergy}) -- (\ref{effscalefactor}) with $ \eta_{eff}(\tau) $ defined
by (\ref{effeta}). Eq.(\ref{condeta}) is just the vanishing of the
wronskian of $ \eta_{eff}(\tau) $ and $ f_0(\tau) $. Namely,  $
\eta_{eff}(\tau) $ and $ f_0(\tau) $ just differ in a constant
factor. 

We see that in {\em all} cases, the full large $N$ quantum dynamics in these 
models of inflationary phase transitions is well approximated by the
equivalent dynamics of a homogeneous, classical scalar field with initial
conditions on the effective field 
$\eta_{eff}(\tau_A) = \sqrt{g}\; h_0 \; {\cal F}(h_0)$.  
We have verified these
results numerically for the field and scale factor dynamics, finding that
the effective classical dynamics reproduces the results of the full
dynamics to within our numerical accuracy.  
We have also checked numerically
that the estimate for the classical to quantum crossover given by eq.(\ref{classquandyn}) is quantitatively correct. Thus in the classical case in
which $ \eta(0) \gg \sqrt{g}\; h_0$ we find that 
$\eta_{eff}(\tau) = \eta(\tau)$,
 whereas in the opposite, quantum case $\eta_{eff}(\tau) =
\sqrt{g\Sigma(\tau)}$. 

This remarkable feature of zero mode assembly of long-wavelength,
spinodally unstable modes is a consequence of the presence of the horizon.
It also explains why, despite the fact that asymptotically the
fluctuations sample the broken symmetry state, the equation of state is
that of matter. 
Since the excitations in the broken symmetry state are massless Goldstone
bosons one would expect radiation domination. However, the assembly
phenomenon, i.e. the redshifting of the wave vectors, makes these modes behave
exactly like zero momentum modes that give an equation of state of matter
(upon averaging over the small oscillations around the minimum).  

Subhorizon modes at the end of inflation with $ q > h_0 \, e^{h_0
\tau_s} $ do not participate in the zero mode assembly. The behavior of such
modes do depend on $ q $ after the end of inflation. Notice that these
modes have extremely large comoving $ q $ since $  h_0 \, e^{h_0
\tau_s} \geq 10^{26} $. As discussed in sec. VI such modes decrease
with time after inflation as $ \sim 1/a(\tau) $\cite{frw2}. 

\subsection{Making sense of small fluctuations:}
Having recognized the effective classical variable that can be interpreted
as the component of the field that drives the FRW background and rolls
down the classical potential hill, we want to recognize unambiguously
the small fluctuations. We have argued above that after horizon crossing,
all of the mode functions evolve proportionally to the zero mode,
and the question arises: which modes are assembled into the effective
zero mode whose dynamics drives the evolution of the FRW scale factor
and which modes are treated as perturbations? In principle every
$k\neq 0$ mode provides some spatial inhomogeneity, and assembling these
into an effective homogeneous zero mode seems in principle to do away with
the very inhomogeneities that one wants to study. However, scales of
cosmological importance today first crossed the horizon during the 
last 60 or so e-folds of inflation. Recently Grishchuk\cite{grishchuk}
 has argued that the
sensitivity of the measurements of $ \Delta T/T $ probe inhomogeneities on
scales $\approx 500$ times the size of the present horizon. Therefore scales
that are larger than these and that have first crossed the horizon
much earlier than the last 60 e-folds of inflation are unobservable
today and 
can be treated as an effective homogeneous component, whereas the scales that
can be probed experimentally via the CMB inhomogeneities today must be treated 
separately as part of the inhomogeneous perturbations of the CMB. 

Thus a consistent description of the dynamics in terms of an effective
zero mode plus `small' quantum fluctuations can be given provided
the following requirements are met:
a) the total number of e-folds $N_e \gg 60$, b) all the modes that have
crossed the horizon {\em before} the last 60-65 e-folds are assembled into
an effective {\em classical} zero mode via $\phi_{eff}(t) = 
\left[\phi^2_0(t)+ \langle \pi^2(\vec x,t) \rangle_R
\right]^{\frac{1}{2}}$, c) the modes that cross the horizon during the
last 60--65 e-folds are accounted as `small' perturbations. The reason
for the requirement a) is that in the separation 
$\phi(\vec x, t) = \phi_{eff}(t)+\delta \phi(\vec x,t)$ one requires that
$\delta \phi(\vec x,t)/\phi_{eff}(t) \ll 1$. As argued above, after the 
modes cross the horizon, the ratio of amplitudes of the mode functions remains
constant and given by $e^{(\nu - \frac{3}{2})\Delta N}$ with $\Delta N$ 
being the number of e-folds between the crossing of the smaller $ k $ and the
crossing of the larger $ k $. Then for $\delta \phi(\vec x, t)$ to be much
smaller than the effective zero mode, it must be that the Fourier components
of $\delta \phi$ correspond to very large $k$'s at the beginning of inflation,
so that the effective zero mode can grow for a long time before the components
of $\delta \phi$ begin to grow under the spinodal instabilities. 
In fact requirement a) is not very severe; in the
figs.\ref{gsigma}-\ref{hinverse} we have taken 
$h_0 = 2.0$ which is a very moderate value and yet for $ g = 10^{-12}$
the inflationary stage lasts for well over 100 e-folds, 
and as argued above, the
larger $h_0$ for fixed $ g $, the longer is the inflationary stage. 
Therefore under this set of conditions, the classical dynamics of the effective zero mode $\phi_{eff}(t)$  drives the FRW background, whereas
the inhomogeneous fluctuations $\delta \phi(\vec x,t)$, which are made up
of Fourier components with wavelengths that are much smaller than the
horizon at the beginning of inflation and that cross the horizon during
the last 60 e-folds, provide the inhomogeneities that seed density
perturbations.

\subsection{Scalar  Metric Perturbations:}
Having identified the effective zero mode and the `small perturbations',
we are now in position to provide an estimate for the amplitude and spectrum
of scalar metric perturbations. We use the clear formulation in
ref.\cite{mukhanov} in terms of gauge invariant
variables. In particular we focus on the dynamics of the Bardeen
potential\cite{bardeen}, which in longitudinal gauge is identified
with the Newtonian potential. The equation of motion for the Fourier
components (in terms of comoving wavevectors) for this variable in
terms of the effective zero mode is [eq.(4.48) in \cite{mukhanov}]
\begin{equation}
\ddot{\Phi}_k +
\left[H(t)-2\;\frac{\ddot{\phi}_{eff}(t)}{\dot{\phi}_{eff}(t)}\right] 
\dot{\Phi}_k+\left[\frac{k^2}{a^2(t)}+ 2\left(\dot{H}(t)-H(t)\;
\frac{\ddot{\phi}_{eff}(t)}{\dot{\phi}_{eff}(t)}\right)\right]\Phi_k =
0 .
\label{bardeen}
\end{equation}
We are interested in determining the dynamics of $\Phi_k$ for those
wavevectors that cross the horizon during the last 60 e-folds before the
end of inflation. During the inflationary stage and substantially
before its end the numerical analysis yields  to a very good approximation
\begin{equation}
H(t) = H_0  \; ; \; \phi_{eff}(t) = \phi_{eff}(t_A)\; e^{(\nu-
\frac{3}{2})H_0(t-t_A)}, \label{infla}
\end{equation}
where $H_0$ is the value of the Hubble constant during inflation, leading to 
\begin{equation}
\Phi_k(t) = e^{(\nu -2)H_0(t-t_A)}\left[a_k\;
H^{(1)}_{\nu-1}\left(\frac{ke^{-H_0(t-t_A)}}{H_0}\right) 
+b_k \; H^{(2)}_{\nu-1}\left(\frac{ke^{-H_0(t-t_A)}}{H_0}\right)\right]
\; \; .
\label{solbardeen}
\end{equation}
The coefficients $a_k,b_k$ are determined by the initial conditions
and $ \nu $ is given by eq.(\ref{bessel}).

Since we are interested in the wavevectors that cross the horizon during
the last 60 e-folds, the consistency for the zero mode assembly and
the interpretation of `small perturbations' requires that there must be
many e-folds before the {\em last} 60. We are then considering wavevectors
that were deep inside the horizon at the onset of
inflation. $\Phi_k(t)$ is related to the canonical `velocity field'
that determines scalar  perturbations 
of the metric and which is quantized with Bunch-Davies initial
conditions for the large $k$-mode functions. The relation between
$\Phi_k$ and $v$ and the  
initial conditions on $v$ lead at once to a determination of the
coefficients $a_k$ and $b_k$ for $k >> H_0$ [see eqs.(13.5) and (13.9)
in ref.\cite{mukhanov}],
\begin{equation}
a_k = { \dot{\phi}_{eff}(t_A) \over k \; \sqrt{2H_0} \; M^2_{Pl}} \;
e^{{i\pi \over 2} \left(\nu - \frac32\right)} \; .
\label{coeffs}
\end{equation} 
Thus we find that the amplitude of scalar metric perturbations after 
horizon crossing is given by
\begin{equation}\label{perts}
|\delta_k(t)| = k^{\frac{3}{2}}|\Phi_k(t)| =
{ \dot{\phi}_{eff}(t_A) \over \pi \; M^2_{Pl}} \, 
\Gamma(\nu-1) \;\left(\frac{2H_0}{k}\right)^{\nu -\frac{3}{2}}\;
e^{(2\nu-3)H_0(t-t_A)}\; .
\end{equation}
The power spectrum per logarithmic $k$  interval is given by
$ |\delta_k(t)|^2 $. The time dependence of $ |\delta_k(t)| $ displays the
unstable growth associated with the spinodal instabilities of
super-horizon modes and 
is a hallmark of the phase transition.
This time dependence can be also understood from the constraint equation
that relates the Bardeen potential to the gauge invariant field fluctuations\cite{mukhanov}, which in longitudinal gauge are identified
with $\delta \phi(\vec x,t)$. 
The constraint equation and the evolution equations for the
gauge invariant scalar field fluctuations are\cite{mukhanov}
\begin{equation}
\frac{d}{dt}(a \Phi_k) = \frac{4\pi}{M^2_{Pl}}\; a\; \delta \phi^{gi}_k
\; \dot{\phi}_{eff}\; ,
\label{constraint}
\end{equation}

\begin{equation}
\left[\frac{d^2}{dt^2}+3H \frac{d}{dt}+\frac{k^2}{a^2}+{\cal M}^2 \right]
\delta \phi^{gi}_k-4 \;
\dot{\phi}_{eff}\;\dot{\Phi}_k+2V'(\phi_{eff})\;\Phi_k=0\; . 
\label{gauginv}
\end{equation}

Since the right hand side of (\ref{constraint}) is proportional to
$\dot{\phi}_{eff}/M^2_{Pl} \ll 1 $ during the inflationary epoch in this
model,  we can neglect the terms proportional
to $\dot{\Phi}_k $ and $\Phi_k$ on the left hand side of (\ref{gauginv}),
in which case the equation for the gauge invariant scalar field
fluctuation is the same as for the mode functions. In fact, since $ 
\delta \phi^{gi}_k$ is gauge invariant we can evaluate it in the longitudinal gauge wherein it is identified with the mode functions
$f_k(t)$. Then absorbing a constant of integration in the initial
conditions for the Bardeen variable, we find
\begin{equation}
\Phi_k(t) = \frac{4\pi}{M^2_{Pl}\; a(t)}\int_{t_o}^t
a(t')\;\phi_{eff}(t')\; f_k(t')\; dt' + {\cal O}\left(\frac{1}{M^4_{Pl}}\right) ,
\label{bard}
\end{equation}
and using that $\phi(t) \propto e^{(\nu-3/2)H_0t} $ and that
 after horizon crossing $f_k(t) \propto e^{(\nu-3/2)H_0t}$, one obtains
at once the time dependence of the Bardeen variable after horizon
crossing. In particular the time dependence is found to be $\propto 
e^{(2\nu-3)H_0t}$. It is then clear that the time dependence is a reflection of
the spinodal (unstable) growth of the superhorizon field fluctuations. 

 To obtain the amplitude and spectrum
of density perturbations at {\em second} horizon crossing we use the
conservation law associated with the gauge invariant variable\cite{mukhanov}
\begin{equation}
\xi_k = \frac{2}{3}
\frac{\frac{\dot{\Phi}_k}{H}+\Phi_k}{1+p/\varepsilon} + \Phi_k
\; \; ; \; \; \dot{\xi}_k =0\; , \label{xivar}
\end{equation}
which is valid after horizon crossing of the mode with wavevector $ k $.
Although this conservation law is an exact statement of superhorizon mode
solutions of eq.(\ref{bardeen}), 
we have obtained solutions assuming that during
the inflationary stage $H$ is constant and have neglected the $\dot{H}$ term in
Eq. (\ref{bardeen}). Since during the inflationary stage,
\begin{equation}
\dot{H}(t) = -\frac{4\pi}{M^2_{Pl}}\, \dot{\phi}^2_{eff}(t) \propto
H^2_0 \; \left(\frac{d\eta_{eff}(\tau)}{d\tau}\right)^2\ll H^2_0 \label{Hdot}
\end{equation}
and $\ddot{\phi}/\dot{\phi} \approx H_0$, the above approximation is
justified. We then see that $\phi^2_{eff}(t) \propto
e^{(2\nu-3)H_0t}$ which is the same time dependence as that of
$\Phi_k(t)$. Thus the
term proportional to $1/(1+p/\varepsilon)$ in Eq. 
(\ref{xivar}) is indeed constant in time after horizon crossing. On the other
hand, the term that
does not have this denominator evolves in time but is of order 
$(1+p/\varepsilon) =
-2\dot{H}/3H^2 \ll 1$ with respect to the constant term and therefore can be
neglected. Thus, we confirm that the variable $\xi$ is conserved
up to the small term proportional to $(1+p/\varepsilon)\Phi_k$ which
is negligible during the inflationary stage. 
This small time dependence is consistent with the fact
that we neglected the $\dot{H}$ term in the equation of motion for $\Phi_k(t)$.
The validity of the conservation law has been recently studied and confirmed in
different contexts\cite{caldwell}. Notice that we do not have to assume
that $\dot{\Phi}_k$ vanishes, which in fact does not occur.

 However, upon second horizon crossing
it is straightforward to see that $\dot{\Phi}_k(t_f) \approx 0$. The 
reason for this assertion can be seen as follows: eq.(\ref{gauginv}) shows that
at long times, when the effective zero mode is oscillating around the minimum
of the potential with a very small amplitude and when the time dependence of
the fluctuations has saturated (see fig.\ref{modu}), $\Phi_k$ will redshift
as $\approx 1/a(t)$\cite{frw2} and its derivative becomes extremely small. 

Using this conservation law,  assuming matter domination at second horizon 
crossing,  and $\dot{\Phi}_k(t_f)\approx 0$\cite{mukhanov}, we find
\begin{equation}\label{amplitude}
|\delta_k(t_f)| = {3 \over 5 \; \pi} { \Gamma(\nu) \over
(\nu-\frac{3}{2})\,  {\cal F}(H_0/m)}
\left(\frac{2H_0}{k}\right)^{\nu-\frac{3}{2}} \; , 
\end{equation}
where ${\cal F}(H_0/m)$ determines the initial amplitude of the effective
zero mode (\ref{effzeromodein}). 
We can now read the power spectrum per logarithmic $k$ interval
\begin{equation}
{\cal P}_s(k) = |\delta_k|^2 \propto k^{-2(\nu-\frac{3}{2})}.
\end{equation}
leading to the index for scalar density perturbations
\begin{equation}
n_s = 1-2\left(\nu-\frac{3}{2}\right) \; . \label{index}
\end{equation}

For $H_0/m \gg 1$, we can expand $\nu-3/2$ as a series in $m^2/H_0^2$ in
eq.(\ref{amplitude}).  Given that the comoving wavenumber of the mode which 
crosses the horizon $n$ e-folds before the 
end of inflation is $k=H_0 \; e^{(N_e-n)}$ 
where $N_e$ is given by (\ref{efolds}), we arrive at the following expression
for the amplitude of fluctuations on the scale corresponding to $n$ 
in terms of the De Sitter Hubble constant and the coupling 
$ \lambda = 8\pi^2 \, g $:
\begin{equation}
|\delta_n(t_f)| \simeq 
\frac{9 \, H^3}{10 \, (2 \pi)^{3/2}\, m^3} \left(2e^n\right)^{m^2/3H_0^2}
\sqrt{\lambda} \left[1+\frac{2m^2}{3H_0^2} \left(\frac76 - 
\ln 2 - \frac{\gamma}{2}
\right) + {\cal O}\left( \frac{m^4}{H_0^4} \right) \right] \; .
\label{amplitude_n}
\end{equation}
Here, $\gamma$ is Euler's constant.  Note the explicit dependence of the 
amplitude of density perturbations on $\sqrt{g}$.  For $n \approx 60$,
the factor $\exp(nm^2/3H_0^2)$ is ${\cal O}(100)$ for $H_0/m = 2$, while
it is ${\cal O}(1)$ for $H_0/m \geq 4$.  Notice that for $H_0/m$ large,
the amplitude increases approximately as $(H_0/m)^3$, which will place 
strong restrictions on $ g $ in such models.

We remark that we have not included the small corrections to the dynamics
of the effective zero mode and the scale factor arising from the
non-linearities. We have found numerically that these nonlinearities 
are only significant for the
modes that cross about 60 e-folds before the end of inflation for
values of the Hubble parameter $H_0/m_R > 5$.  The effect of these
non-linearities in the large $N$ limit is to slow somewhat the exponential
growth of these modes, with the result of shifting the power spectrum
closer to an exact Harrison-Zeldovich spectrum with $n_s =1$.  Since
for $H_0/m_R > 5$ the power spectrum given by (\ref{index}) differs from
one by at most a few percent, the effects of the non-linearities are
expected to be observationally unimportant.
The spectrum given by (\ref{amplitude}) is
similar to that obtained in references\cite{turner,guthpi} although
the amplitude differs from that obtained there. In addition, we do not
assume slow roll for which $(\nu - \frac{3}{2})\ll 1$, although 
this would be the case if $N_e \gg 60$.

We emphasize an important
feature of the spectrum: it has more power at {\em long
wavelengths} because $\nu-3/2 > 0$. This is recognized to be a
consequence 
of the spinodal instabilities that result in the growth of long wavelength
modes and therefore in more power for these modes.  
This seems to be a robust prediction of new inflationary scenarios in
which the potential has negative second derivative in the region of field
space that produces inflation.  

 It is at this
stage that we recognize the consistency of our approach for separating
the composite effective zero mode from the small fluctuations. We have
argued above that many more than 60 e-folds are required for consistency,
and that the small fluctuations correspond to those modes that cross
the horizon during the last 60 e-folds of the inflationary stage. For these
modes $H_0/k = e^{-H_0 t^*(k)}$ where $t^*(k)$ is the time since the beginning 
of inflation of horizon crossing of the mode with wavevector $k$. 
The scale that  corresponds to the Hubble radius today $\lambda_0
=2\pi/k_0$ is the first to cross during the last 60 or so e-folds
before the end of 
inflation. Smaller scales today will correspond to $k > k_0$ at the
onset of inflation since they will cross the first horizon later and
therefore will reenter earlier. The bound on $|\delta_{k_0}| \propto  
\Delta T/ T \leq  10^{-5}$ on
these scales provides a lower bound on the number of e-folds required for
these type of models to be consistent:
\begin{equation}
N_e >
60+\frac{12}{\nu-\frac{3}{2}}-\frac{\ln(\nu-\frac{3}{2})}{\nu-\frac{3}{2}}\; ,
\label{numbofefolds}
\end{equation}
where we have written the total number of e-folds as $N_e=H_0\; t^*(k_0)+60$.
This in turn can be translated into a bound on the coupling constant using
the estimate given by eq.(\ref{efolds}).

The four year COBE  DMR Sky Map\cite{gorski} gives $n \approx 1.2 \pm 0.3$
thus providing an upper bound on $\nu$
\begin{equation}
0 \leq \nu-\frac{3}{2} \leq 0.05 \label{cobebound}
\end{equation}
corresponding to $h_0 \geq 2.6$. We then find that these values of $h_0$ and
$\lambda \approx 10^{-12}-10^{-14}$ provide sufficient e-folds to satisfy
the constraint for scalar density perturbations. 

\subsection{Tensor Metric Perturbations:} 
The scalar field does not couple to the tensor (gravitational wave)
modes directly, and the tensor perturbations are gauge invariant from
the beginning. Their dynamical evolution is completely determined by
the dynamics of the scale factor\cite{mukhanov,grishchuk2}. 
Having established numerically that the inflationary epoch is
characterized by $\dot{H}/H^2_0 \ll 1$ and that scales of cosmological
interest cross the 
horizon during the stage in which this approximation is excellent, we can
just borrow the known result for the power spectrum of gravitational waves
produced during inflation extrapolated to the matter
era\cite{mukhanov,grishchuk2} 
\begin{equation}
{\cal P}_T(k) \approx \frac{H^2_0}{M^2_{Pl}}k^0\; .
\end{equation}
Thus the spectrum to this order is scale invariant (Harrison-Zeldovich)
with an amplitude of the order $m^4/\lambda M^4_{Pl}$. Then, for values
of $m \approx 10^{12}-10^{14} \mbox{ Gev }$ and 
$\lambda \approx 10^{-12}-10^{-14}$
one finds that the amplitude is $\leq 10^{-10}$ which is much smaller than the
amplitude of scalar density perturbations. 
As usual the amplification of scalar perturbations
is a consequence of the equation of state during the inflationary epoch.

\section{Conclusions} 

Since there are a number of articles in the literature treating
related problems, it is useful to review the unique features of the
present work. First, we have treated the problem {\em dynamically},
without using the 
effective potential (an equilibrium construct) to determine the evolution.
Second, we have provided consistent non-perturbative calculations of the
non-linear quantum field evolution to bring out some of the most
relevant aspects of the late time 
behavior.  In particular, we found that the quantum backreaction naturally
inhibits catastrophic growth of fluctuations and provides a smooth transition
to the late time regime in which the quantum fluctuations decay as the zero
mode approaches its asymptotic state.  Third, the dynamics studied obeys the
constraint of covariant conservation of the energy momentum tensor.

It can be argued that the inflationary paradigm as currently understood is one
of the greatest applications of quantum field theory. The imprint of quantum
mechanics is everywhere, from the dynamics of the inflaton, to the
generation of 
metric perturbations, through to the reheating of the universe. It is clear
then that we need to understand the quantum mechanics of inflation in as deep a
manner as possible so as to be able to understand what we are actually testing
via the CMBR temperature anisotropies, say.

What we have found in our work is that the quantum mechanics of inflation is
extremely subtle. We now understand that it involves both non-equilibrium as
well as non-perturbative dynamics and that what you start from may {\it not} be
what you wind up with at the end!

In particular, we see now that the correct interpretation of the
non-perturbative growth of quantum fluctuations via spinodal decomposition is
that the background zero mode must be redefined through the process of zero
mode reassembly that we have discovered. When this is done (and {\it only}
when!) we can interpret inflation in terms of the usual slow-roll approach with
the now small quantum fluctuations around the redefined zero mode driving the
generation of metric perturbations. 

We have studied the non-equilibrium dynamics of a `new inflation' scenario in a
self-consistent, non-perturbative framework based on a large $ N $ expansion,
including the dynamics of the scale factor and backreaction of quantum
fluctuations. Quantum fluctuations associated with superhorizon modes grow
exponentially as a result of the spinodal instabilities and contribute to the
energy momentum tensor in such a way as to end inflation consistently.

Analytical and numerical estimates have been provided that establish the regime
of validity of the classical approach.  We find that these superhorizon modes
re-assemble into an effective zero mode and unambiguously identify the
composite field that can be used as an effective expectation value of the
inflaton field whose {\em classical} dynamics drives the evolution of the scale
factor. This identification also provides the initial condition for this
effective zero mode.

A consistent criterion is provided to extract small fluctuations that will
contribute to cosmological perturbations from large non-perturbative
spinodal fluctuations. This is an important ingredient for a consistent
calculation and interpretation of cosmological perturbations.  This criterion
requires that the model must provide many more than 60 e-folds to identify the
`small perturbations' that give rise to scalar metric (curvature)
perturbations. We then use this criterion combined with the gauge invariant
approach to obtain the dynamics of the Bardeen variable and the spectrum for
scalar perturbations.

We find that during the inflationary epoch, superhorizon modes of the Bardeen
potential grow exponentially in time reflecting the spinodal
instabilities. These long-wavelength instabilities are manifest in the spectrum
of scalar density perturbations and result in an index that is less than
one, i.e. a `red' power spectrum, providing more power at long wavelength.  We
argue that this red spectrum is a robust feature of potentials that lead to
spinodal instabilities in the region in field space associated with inflation
and can be interpreted as an imprint of the phase transition on the
cosmological background. Tensor perturbations on the other hand, are not
modified by these features, they have much smaller amplitude and a
Harrison-Zeldovich spectrum.

\section{Acknowledgements:} 
We thank F. Cao, D. Cormier, R. Holman, S. P. Kumar, J. Salgado,
A. Singh, M. Srednicki all of whom collaborated at different stages on
the works reviewed here. We thank J. Baacke,  C. Destri, A. Dolgov,
E. Kolb, E. Weinberg for conversations and discussions. D. B. thanks the
N.S.F for partial support through the grant awards: PHY-9605186 and INT-9815064
and LPTHE for warm hospitality.  We thank the CNRS-NSF
cooperation programme  for partial support.

\end{document}